\documentclass[preprint]{elsarticle}
\usepackage{amsmath}
\usepackage{amsfonts}
\usepackage{pgfplots}
\usepackage{amssymb}
\usepackage{amsthm}
\usepackage{appendix}
\usepackage{multirow}
\usepackage{multicol}
\usepackage{enumitem}
\usepackage[utf8]{inputenc}
\usepackage[english]{babel}
\biboptions{sort&compress}
\usepackage{tikz}
\usepackage{float}
\usepackage{subfig}
\usetikzlibrary{positioning}
\usetikzlibrary{matrix,arrows,decorations.pathmorphing}
\usetikzlibrary{patterns}
\usetikzlibrary{through,calc,3d}
\usetikzlibrary{plotmarks}
\usetikzlibrary{pgfplots.groupplots}
\usepackage{varioref}
\usepackage{comment}
\usepackage{empheq}
\usepackage{cancel}
\usepackage{graphicx}
\graphicspath{{Images/}{../Images}}
\usepackage[autostyle]{csquotes}
\usepackage{microtype}
\usepackage{mathtools}
\usepackage{mathrsfs}
\usepackage{mwe}
\usepackage{caption}
\usepackage{pdfpages}

\usepackage{blindtext}
\usepackage{import}

\newcommand{\KL}{Karhunen-Lo\'{e}ve }
\newcommand{\innerx}[2]{\Big< #1 \ , \  #2 \Big>}

\newlength{\tempheight}
\newlength{\tempwidth}
\newcommand\aug{\fboxsep=-\fboxrule\!\!\!\fbox{\strut}\!\!\!}
\newcommand{\rowname}[1]
{\rotatebox{90}{\makebox[\tempheight][c]{\text{#1}}}}

\newcommand{\columnname}[1]
{\makebox[\tempwidth][c]{\text{#1}}}

\makeatletter
\def\ps@pprintTitle{%
   \let\@oddhead\@empty
   \let\@evenhead\@empty
   \let\@oddfoot\@empty
   \let\@evenfoot\@oddfoot
}
\makeatother
\makeatletter
\def\hlinewd#1{%
\noalign{\ifnum0=`}\fi\hrule \@height #1 %
\futurelet\reserved@a\@xhline}
\makeatother

\hypersetup{
    colorlinks=true,
    linkcolor=blue,
    filecolor=blue,      
    urlcolor=blue
}

\usetikzlibrary{arrows}

\definecolor{Color1}{rgb}{1,0.89,0.77}
\definecolor{Color2}{rgb}{0.75,0.83,0.99}

\pgfplotsset{compat=1.14}

\newtheorem{theorem}{\bf Theorem}[section]
\newtheorem{lemma}{\bf Lemma}[section]
\newtheorem{remark}{\bf Remark}[section]

\newcommand{\pfrac}[2]{\ensuremath{\dfrac{\partial #1}{\partial #2}}}

\newcommand{\bm}[1]{\ensuremath{\mathbf{#1}}}
\newcommand{\ol}[1]{\ensuremath{\overline{#1}}}
\begin{document}
\title{Real-Time Reduced-Order Modeling of  Stochastic Partial Differential Equations via Time-Dependent Subspaces}
\begin{frontmatter}
\author[1]{Prerna Patil}
\address[1]{Department of Mechanical Engineering, University of Pittsburgh, Pittsburgh, PA-15206.}
\author[1]{Hessam Babaee\corref{cor1}}
\ead{h.babaee@pitt.edu}
\cortext[cor1]{Corresponding author}

\begin{abstract}
 We present a new methodology for the real-time reduced-order modeling of stochastic partial differential equations  called the \textit{dynamically/bi-orthonormal} (DBO) decomposition. In this method, the stochastic fields are approximated by a low-rank decomposition to spatial and stochastic subspaces. Each of these subspaces is  represented by a set of orthonormal time-dependent modes. We derive exact evolution equations of these time-dependent modes and the evolution of the factorization of the reduced covariance matrix. We show that DBO is equivalent to the dynamically orthogonal (DO) \cite{sapsis2009dynamically} and bi-orthogonal (BO) \cite{cheng2013dynamicallyI} decompositions via  linear and invertible transformation matrices that connect DBO to DO and BO. However, DBO shows several improvements compared to DO and BO: (i) DBO performs better than DO and BO for cases with ill-conditioned covariance matrix; (ii)  In contrast to BO, the issue of eigenvalue crossing is not present in the DBO formulation; (iii) In contrast to DO, the stochastic modes are orthonormal, which leads to more accurate representation of the stochastic subspace.  We study the convergence properties of the method and compare it to the DO and BO methods. For demonstration, we consider three cases: (i) stochastic linear advection equation,  (ii) stochastic Burgers' equation, and (iii) stochastic incompressible flow over a bump in a channel. Overall we observe improvements  in the numerical accuracy of DBO compared against DO and BO. 
\end{abstract}
\end{frontmatter}

\section{Introduction}\label{sec:Intro}
The pressing need of conducting verification and validation (V\&V) for realistic simulations in  scientific and engineering applications requires propagating uncertainty in these systems. These systems are often  subject to uncertainty that may come from imperfectly  known parameters  --- that can be modeled as random parameters --- or   random initial/boundary conditions, or by systems that are characterized by inherent stochastic dynamics, such as coarse grain models of multi-scale systems, in which the effects of unresolved scales are modeled as stochastic processes \cite{MM_DOEworks_11}. Uncertainty quantification (UQ) in such systems can disentangle the effects of different uncertain sources on the quantities of interest and it can guide the decision making process and ultimately lead to more reliable predictions and designs. 

One of the fundamental challenges in performing UQ in complex engineering and scientific systems is the computational cost associated with this task. These systems are often characterized by high-dimensional ordinary/partial differential equations, whose forward simulation can be computationally costly. There are a large number of techniques for performing UQ. These methods are primarily either sample based such as Monte Carlo (MC) method and its variants such as multi-level MC and quasi-MC (QMC) \cite{giles2008multilevel, barth2011multi, kuo2012quasi}, or are based on polynomial chaos expansion (PCE) \cite{ghanem2003stochastic, wan2006multi, xiu2005high, xiu2002wiener, foo2010multi, foo2008multi, babuvska2007stochastic, ganapathysubramanian2007sparse, yang2012adaptive, babaee2014effect, zhang2018stochastic}.

While PCE performs well for nearly elliptic problems or flow at low Reynolds numbers, solving highly transient stochastic ordinary/partial differential equations (SODE/SPDE)  is particularly challenging for this method. It was shown in \cite{wan2006long} that for the one dimensional advection equation with a uniform random transport velocity  the order of polynomial chaos must increase with time to maintain the error below a given value. PCE also loses its efficiency for nonlinear systems with intermittency and positive Lyapunov exponents \cite{branicki2013fundamental}. 


Reduced order modeling approaches are popular tools  for state prediction and control of  deterministic  evolutionary dynamical systems \cite{sirovich1987turbulenceI, sirovich1987turbulenceII, sirovich1987turbulenceIII,schmid2010dynamic,Alvergue:2015aa,rowley2010reduced,kutz2016dynamic,noack_2016}. With the recent developments in data-fusion and specifically  multi-fidelity modeling approaches \cite{PWG18,Perdikaris:2015aa,babaee2016}, in which imperfect predictions can be effectively utilized when combined with high-fidelity data,  reduced order modeling techniques will play a crucial role as a surrogate model that generates low-fidelity data at a low computational cost. In the context of SPDEs,  the dynamically orthogonal decomposition (DO) was introduced \cite{sapsis2009dynamically} as a stochastic reduced order modeling technique, in which the stochastic field $u(x,t;\omega)$ is approximated as:
\begin{equation*}
    u(x,t; \omega) = \bar{u}(x,t) + \sum_{i=1}^r u_i(x,t) y_i(t;\omega),
\end{equation*}
where $\bar{u}(x,t)$ is the mean, $u_i(x,t)$
are a set of deterministic time-dependent orthonormal modes in the spatial domain and $y_i(t;\omega)$ 
are zero-mean random processes in the stochastic domain and $r$ is the reduction order.  To remove the redundancy in time,  the evolution of the spatial subspace, i.e. $\partial u_i(x,t)/\partial t$, is chosen to be orthogonal to $u_j(x,t)$. By enforcing the above constraints, one can derive closed-form evolution equations for $\bar{u}(x,t)$, $u_i(x,t)$ and $y_i(t;\omega)$.
The imposed conditions on the above decomposition are not unique. Bi-orthogonal (BO) decomposition is one such variant, in which the spatial basis are orthogonal and the stochastic basis are orthonormal \cite{cheng2013dynamicallyII}. Recently, a non-intrusive DO formulation was introduced \cite{2019arXiv190409846B} and it was shown  that the DO evolution equations are the optimality conditions of a variational principle that seeks to minimize the distance between the rate of change of full-dimensional dynamics and that of the DO reduction. For linear parabolic SPDEs, the difference between the approximation error of $r$-term DO decomposition and $r$-term \KL (KL) decomposition  can be bounded \cite{MNZ15}.   Independently and prior to the development of DO/BO,  the idea of using time-dependent basis had been introduced in very different fields, namely chemistry and quantum mechanics for the approximation of the deterministic Schr\"{o}dinger equations by the Multi Configuration Time Dependent Hartree (MCTDH) method \cite{beck2000multiconfiguration, bardos2003mean}, and in deterministic settings \cite{koch2007dynamical}. 

It was shown in \cite{choi2014equivalence} that both DO and BO are equivalent: in both of these methods $u_i(x,t)$ and $y_i(t;\omega)$
span the same subspace and a linear invertible time-dependent matrix transforms one to the other. This matrix transformation amounts to an \emph{in-subspace} rotation and stretching for  $u_i(x,t)$ modes and    $y_i(t; \omega)$ coefficients. In contrast to PCE, BO/DO decompositions allow the stochastic coefficients evolve with time as opposed to time-invariant polynomial chaos basis. This relaxation allows BO/DO decompositions to ``follow" the transient dynamics. It was shown that in the limit of zero variance of $y_i(t;\omega)$, the subspace of $u_i(x,t)$
converges exponentially fast to the most unstable subspace of the dynamical system --- associated with the  $r$ most dominant eigendirections of the Cauchy–Green tensor  \cite{babaee2017reduced}. It was shown that the reduction based on the time-dependent basis and coefficients can capture the low-dimensional structure of the intermittent dynamics \cite{babaee2016minimization}.

Although both DO and BO are mathematically equivalent, they exhibit different numerical performance. When the eigenvalues of the reduced covariance matrix are close or cross  each other, the BO formulation becomes  numerically unstable. On the other hand,  the DO decomposition does not have the issue of eigenvalue crossing. However, when the eigenvalues of the reduced covariance matrix are not close, BO exhibits better numerical performance than DO \cite{choi2014equivalence}. This is mainly attributed to the orthonormality of $y_i(t;\omega)$ coefficients in the BO formulation, which maintains a well-conditioned representation of the stochastic subspace at all times. However, in the DO decomposition, the stochastic coefficients $y_i(t;\omega)$ could be highly correlated. This has inspired a hybrid DO/BO method where BO is the dominant solver, but near the eigenvalue crossing the solver switches to DO  \cite{babaee2017robust}.         

Both DO and BO decompositions  perform poorly when the covariance matrix is singular or near singular. In the case of DO, the covariance matrix is full, while in the case of BO the covariance matrix is diagonal. In DO the inverse of the covariance matrix is required for the evolution of the spatial basis and in BO the inverse of the diagonal covariance matrix are needed for the evolution of the stochastic basis.  The issue of singular  covariance matrix can commonly occur in DO/BO decompositions, since one has to resolve the stochastic system up to a small threshold eigenvalue. This necessitates adaptive DO/BO where modes are added and removed at the threshold eigenvalue \cite{choi2014equivalence}. This issue has motivated using pseudo-inverse of the covariance matrix \cite{babaee2017robust}, where the eigenvalue of the singular or near-singular mode below a threshold value is replaced with a minimum tolerable value.  This approach trades the stability of the DO/BO systems with introducing errors in the system of the order of the  minimum tolerable value. 

The motivation for this paper is to introduce a new decomposition that  resolves the aforementioned challenges in using DO and BO. To this end, we present a new methodology in which: (i)  the spatial and stochastic bases are represented by a set of time-dependent orthonormal modes; (ii) an additional equation for the evolution of a factorization of the covariance is derived; and (iii) the condition number of the decomposition is reduced to $\sqrt{\lambda_{max}(t)/\lambda_{min}(t)}$, where $\lambda_{min}(t)$  and $\lambda_{max}(t)$ are the minimum and maximum eigenvalues of the covariance matrix, respectively.

The structure of the paper is as follows: In Section \ref{sec:Method}, we review the formulation of the DBO representation, its evolution equations and prove the equivalence of this method to the DO and BO methods. In Section \ref{sec:DemCases}, we compare the performance of the presented method  with DO and BO via several benchmark problems: (i) Stochastic linear advection equation (ii) Stochastic Burgers' equation; and (iii) 2D stochastic incompressible Navier-Stokes equation for flow over a bump. In Section \ref{sec:Sum}, a brief summary of the present work is presented.

\section{Methodology}\label{sec:Method}
\subsection{Definitions and Notation}
We denote a random vector field by $u(x,t;\omega)$, where $x \in D$ is the spatial coordinate in the physical domain $D \subset \mathcal{R}^d$,  where d=1,2 or 3, and  $t>0$ is time and  $\omega \in \Omega$ is the random event in the sample space $\Omega$.  The inner product in the spatial domain between two random fields $u(x,t;\omega)$ and  $v(x,t;\omega)$ is then defined as: 
\begin{equation*}
    \left<u(x,t;\omega), v(x,t;\omega) \right> = \int_D u(x,t;\omega) v(x,t;\omega) dx, 
\end{equation*}
and the $L_2$ norm induced by the above inner product is:
\begin{equation*}
    \big\| u(x,t;\omega) \big\|_2 = \innerx{u(x,t;\omega)}{u(x,t;\omega)}^{1/2}. 
\end{equation*}
The expectation of the random field is defined as:
\begin{equation*}
  \bar{u}(x,t) = \mathbb{E}[u(x,t,\omega)] = \int_{\Omega} u(x,t;\omega) \rho(\omega) d\omega, 
\end{equation*}
 where $\rho(\omega)$ is the probability density function. The inner product in the random space is defined as the correlation between two random fields:
\begin{equation*}
    \mathbb{E}[u(x,t;\omega)v(x,t;\omega)] = \int_{\Omega} u(x,t;\omega) v(x,t;\omega) \rho(\omega) d\omega. 
\end{equation*}
The covariance operator between two random fields at time $t$ is then obtained from:
\begin{equation*}
    C(x, x',t) = \mathbb{E}\big[(u(x,t;\omega) - \bar{u}(x,t)) (v(x',t;\omega)- \bar{v}(x',t)) \big].
\end{equation*}
We introduce the \emph{quasimatrix} notation as defined in \cite{battles2004extension}, in which one of the dimensions is discrete as usual but the other dimension is continuous:  
\begin{equation*}
    U(x,t) = \begin{bmatrix}
    u_1(x,t) & \aug & u_2(x,t) & \aug & \cdots & \aug & u_r(x,t) \\ \end{bmatrix},
\end{equation*}
\begin{equation*}
    Y(t;\omega) = \begin{bmatrix}
    y_1(t;\omega) &\aug & y_2(t;\omega) & \aug & \cdots & \aug & y_r(t;\omega)
    \end{bmatrix},
\end{equation*}
where $U(x,t)$ and $Y(t;\omega)$ are quasimatrices of size $\infty \times r$. The inner product for two quasimatrices $U(x,t)=\begin{bmatrix}
    u_1(x,t) & \aug & u_2(x,t) & \aug & \cdots & \aug & u_{r_1}(x,t) \\ \end{bmatrix}$ and $V(x,t)=\begin{bmatrix}
    v_1(x,t) & \aug & v_2(x,t) & \aug & \cdots & \aug & v_{r_2}(x,t) \\ \end{bmatrix}$ is defined by a matrix A such that,
\begin{align*}
    A &= \left< U(x,t), V(x,t) \right>, 
 \end{align*}  
 where
 \begin{equation}\label{eq:quasiinner}
    A_{ij} = \left< u_i(x,t), v_j(x,t) \right>, \quad \quad \quad i=1,2,...,r_1, \quad j=1,2,..,r_2.
\end{equation}
A is a matrix of dimensions $r_1 \times r_2$. In general,  for the case of $r_1=r_2$, matrix $A$ is not symmetric.
\subsection{System of stochastic PDEs}
We consider the following stochastic partial differential equation (SPDE), which defines the system evolution:
\begin{subequations}\label{eq:SPDEall}
    \begin{align}
        \frac{\partial u(x,t;\omega)}{\partial t} &= \mathscr{F}(u(x,t; \omega)), && x \in D, \omega \in \Omega, \label{eq:SPDE}\\
        u(x, t_0; \omega) &= u_0(x; \omega),  && x \in D, \omega \in \Omega,\label{eq:SPDEIC}\\
        \mathscr{B}(u(x,t;\omega)) &= h(x,t), && x \in \partial D,\label{eq:SPDEBC}
    \end{align}
\end{subequations}
where $\mathscr{F}$ is, in general,  a  non-linear differential  operator, and $\mathscr{B}$ is, in general, a linear differential operator, and $\partial D$ denotes the boundary of the domain $D$. In this work we consider deterministic boundary conditions.  For an algorithm to treat  random boundary conditions for time-dependent subspaces, see reference \cite{MN18}.
\subsection{Dynamically bi-orthonormal decomposition}
We consider the following decomposition,
\begin{equation}\label{eq:DBODecomposition}
    u(x,t; \omega) = \bar{u}(x,t) + \sum_{j=1}^r \sum_{i=1}^r u_i(x,t) \Sigma_{ij} (t) y_j (\omega,t) + e(x,t;\omega),
\end{equation}
which is referred to as the \emph{dynamically bi-orthonormal (DBO)} decomposition.  In the above expression $u_i(x,t), i=1,2, \dots, r$ are a set of orthonormal spatial modes:
\begin{equation*}
  \langle u_i(x,t), u_j(x,t) \rangle = \delta_{ij},
\end{equation*}
and they constitute the spatial basis for the DBO decomposition, and $y_i(\omega,t), i=1,2, \dots, r$ are a set of orthonormal  stochastic modes:
\begin{equation*}
\mathbb{E}[ y_i(t;\omega) y_j(t;\omega) ]  = \delta_{ij},
\end{equation*}
that have zero mean i.e., $\mathbb{E}[y_i(t;\omega)]=0, i=1,2,\dots, r$, and $e(x,t;\omega)$ is the reduction error. Moreover, both the spatial and stochastic coefficients are dynamically orthogonal i.e., the rate of change of these subspaces is orthogonal to the space spanned by these modes:
\begin{align}
    \pfrac{U(x,t)}{t} \perp U(x,t) &\iff \left< \frac{\partial u_i(x, t)}{\partial t}, u_j(x,t) \right> = 0  &i,j=1,...,r,\\
    \frac{dY(t;\omega)}{dt} \perp Y(t;\omega) &\iff \mathbb{E}\left[\frac{d y_i(t;\omega)}{d t} y_j(t;\omega) \right] = 0 & i,j=1,...,r.
\end{align}
If the spatial and stochastic modes are orthonormal at $t=0$, imposing the above constraints ensures the orthonormality of the two bases for all time since: 
\begin{equation}
    \frac{d }{d t} \langle u_i(x,t), u_j(x,t) \rangle= \left< \frac{\partial u_i(x,t)}{\partial t}, u_j(x,t) \right> + \left<u_i(x,t), \frac{\partial u_j(x,t)}{\partial t} \right> = 0 \quad \quad \quad i,j =1,...,r,
\end{equation}
and similarly,
\begin{equation}
    \frac{d}{d t}\mathbb{E}[y_i(t;\omega)y_j(t;\omega)] = \mathbb{E}[\frac{d y_i(t;\omega)}{d t} y_j(t;\omega)] + \mathbb{E}[y_i(t;\omega) \frac{d y_j(t;\omega)}{d t}]=0, \quad \quad \quad i,j=1,...,r.
\end{equation}
We show in Section \ref{sec:RedTime}, that imposing the above constraints leads to a unique decomposition. 
The covariance operator is approximated from the DBO decomposition as in the following: 
\begin{align}
\mathcal{C}(x,x',t) &= \mathbb{E}[u_i(x,t)\Sigma_{ij}(t)y_j(t;\omega) u_m(x',t)\Sigma_{mn}(t) y_n(t;\omega) ]\nonumber \\
            & = u_i(x,t) u_m(x',t) \Sigma_{ij}(t) \Sigma_{mn}(t) \mathbb{E}[y_j(t;\omega) y_n(t;\omega)] \nonumber \\
            & = u_i(x,t) u_m(x',t) \Sigma_{ij}(t) \Sigma_{mn}(t) \delta_{jn} \nonumber \\
            & = u_i(x,t) u_m(x',t) \Sigma_{ij}(t) \Sigma_{mj}(t),
\end{align}
where we have used the orthonormality condition imposed on the stochastic basis.
The matrix $\Sigma(t) \in \mathbb{R}^{r\times r}$ is a factorization of  the reduced covariance matrix $C(t) \in \mathbb{R}^{r\times r}$ as in the following:
\begin{equation}\label{eq:CovDBO}
    C(t) = \Sigma(t) \Sigma(t)^T,
\end{equation}  
and it is related to the covariance matrix in the full-dimensional space with:
\begin{equation}
   \mathcal{C}(x,x',t)  = U(x,t)C(t)U^T(x',t).
\end{equation}

\subsection{DBO field equations}
In this section  we present closed-form evolution equations for $\bar{u}(x,t)$, $\Sigma(t)$, $Y(t;\omega)$ and $U(x,t)$ for the DBO decomposition.
\begin{theorem}\label{MainTh}
Let  Eq.(\ref{eq:DBODecomposition}) represent the DBO decomposition of the solution of SPDE given by Eq.(\ref{eq:SPDEall}).  
Then, under the assumptions of the DBO decomposition,  the closed-form evolution equations for the mean, covariance factorization, stochastic  and spatial bases are expressed by:
\begin{subequations}
    \begin{alignat}{4}
        \frac{\partial \bar{u}(x,t)}{\partial t} &= \mathbb{E}[\mathscr{F}(u(x, t; \omega))],  \label{subeq:DBOMean}\\
        \frac{d \Sigma_{ij}(t)}{d t} &= \left<u_i(x,t), \mathbb{E}[\mathscr{\tilde{F}}(u(x,t;\omega))y_j(t;\omega)] \right>,  \label{subeq:DBOSigma}\\
        \frac{dy_i(t;\omega)}{dt} &= \left[\left<u_j(x,t),\mathscr{ \tilde{F}}(u(x, t; \omega)) \right> - \left< u_j(x, t) , \mathbb{E}[\mathscr{\tilde{F}}(u(x, t;\omega)) y_k(t;\omega)] \right>y_k(t;\omega) \right] \Sigma_{ji}(t)^{-1},  \label{subeq:DBOY}\\
        \frac{\partial u_i(x,t)}{\partial t} &=  \left[ \mathbb{E}[\mathscr{\tilde{F}}(u(x, t;\omega)) y_j(t;\omega)] - u_k(x, t)\left< u_k(x, t), \mathbb{E}[\mathscr{\tilde{F}}(u(x,t;\omega)) y_j(t;\omega) ]\right>  \right] \Sigma_{ij}(t)^{-1},  \label{subeq:DBOU}
    \end{alignat}
\end{subequations}
where $\mathscr{\tilde{F}}(x,t;\omega)$ is a mean-subtracted quantity
    \begin{equation*}
      \mathscr{\tilde{F}}(u(x,t;\omega)) = \mathscr{F}(u(x, t;\omega) ) - \mathbb{E}[\mathscr{F}(u(x, t;\omega))]. 
    \end{equation*}
\noindent The associated boundary conditions are given by:
\begin{subequations}
    \begin{align}
        \mathscr{B}[\ol{u}(x,t)]&= h(x,t), & x \in \partial D,\\
        \mathscr{B}[u_i(x,t)] &= 0, & x \in \partial D.
    \end{align}
\end{subequations}
\end{theorem}
The proof for the above theorem is given in Appendix \ref{appendix:AppA}.
\subsection{Equivalence of DO, BO and DBO methods \label{equivalance}} 
Two decompositions are equivalent if they represent the same random fields for all times. The spatial subspaces of two equivalent decompositions  are identical and therefore, one can find  invertible transformation matrices that maps one subspace to the other. This amounts to an \emph{in-subspace} rotation.  The same is true for stochastic subspaces of two equivalent decompositions. The equivalence of DO and BO was first shown in \cite{choi2014equivalence}.  In this section, we show that DBO is equivalent to DO and BO. We first show that DBO is equivalent to DO and BO and then derive the equivalence relations. 
\begin{lemma}\label{lemma1}
Let DO and DBO be equivalent via the transformations: $U_{DO} = U_{DBO} R_u$ and $Y_{DO} = Y_{DBO} W_y$, where $R_u \in \mathbb{R}^{r \times r}$ and $W_y \in \mathbb{R}^{r \times r}$. Then: (i) $R_u$ is an orthogonal matrix (ii) $W_y = \Sigma_{DBO}^T R_u$, and (iii) $\dfrac{d R_u}{dt}=0$.
\end{lemma}
 The proof for  Lemma (\ref{lemma1}) is given in Appendix \ref{appendix:AppB}.
\begin{theorem}\label{DBOtoDO}
Let $U_{DO}(x,t)$, $Y_{DO}(t;\omega)$ represent the DO decomposition of SPDE in Eq.(\ref{eq:SPDEall}) and let $U_{DBO}(x,t)$,  $\Sigma_{DBO}(t)$ and      $Y_{DBO}(t;\omega)$ represent its DBO decomposition. Suppose that at $t=0$ the two bases are equivalent i.e., $U_{DO}(x,t_0) = U_{DBO}(x,t_0)R_u(t_0)$ and $Y_{DO}(t_0;\omega) = Y_{DBO}(t_0; \omega) W_y(t_0)$. Then the two subspaces remain equivalent for all $t>0$.  
\end{theorem}
The proof for  Theorem (\ref{DBOtoDO}) is given in Appendix \ref{appendix:AppB}.
\begin{lemma}\label{lemma2}
Let DBO and BO be equivalent via the transformations: $U_{DBO} = U_{BO} W_u$ and $Y_{DBO} = Y_{BO} R_y$,  where $W_u \in \mathbb{R}^{r \times r}$ and $R_y \in \mathbb{R}^{r \times r}$. Then: (i) $R_y$ is an orthogonal matrix (ii) $\Sigma_{DBO}=W_u^{-1} R_y$(iii) $ \frac{d W_u}{dt} = -(M+\Lambda^{-1} G) W_u$ (iv) $\dfrac{d R_y}{dt}=(S^T - G^T)\Lambda^{-1} R_y$; where $M=\mathbb{E}\left[ Y_{BO}^T \frac{d Y_{BO}}{dt}\right]$, $S=\left<U_{BO}, \pfrac{u_{BO}}{t} \right>$ and $G = \left<U_{BO},\mathbb{E}[\mathscr{\tilde{F}} Y_{BO}] \right>$.
\end{lemma}
The proof for  Lemma (\ref{lemma2}) is given in Appendix \ref{appendix:AppC}.
\begin{theorem}\label{DBOtoBO}
Let $U_{BO}(x,t)$, $Y_{BO}(t;\omega)$ represent the BO decomposition of SPDE in Eq.(\ref{eq:SPDEall}) and let $U_{DBO}(x,t)$, $\Sigma_{DBO}(t)$ and $Y_{DBO}(t;\omega)$ represent its DBO decomposition. Suppose that at $t=0$ the two bases are equivalent i.e., $U_{DBO}(x,t_0) = U_{BO}(x,t_0) W_u(t_0)$ and $Y_{DBO}(t_0;\omega) = Y_{BO}(t_0;\omega)R_y(t_0)$. Then the two subspaces remain equivalent for all $t>0$. 
\end{theorem}
 The proof for Theorem (\ref{DBOtoBO}) is given in Appendix \ref{appendix:AppC}.
 
\begin{figure}
    \centering
    \includegraphics[width=0.6\textwidth]{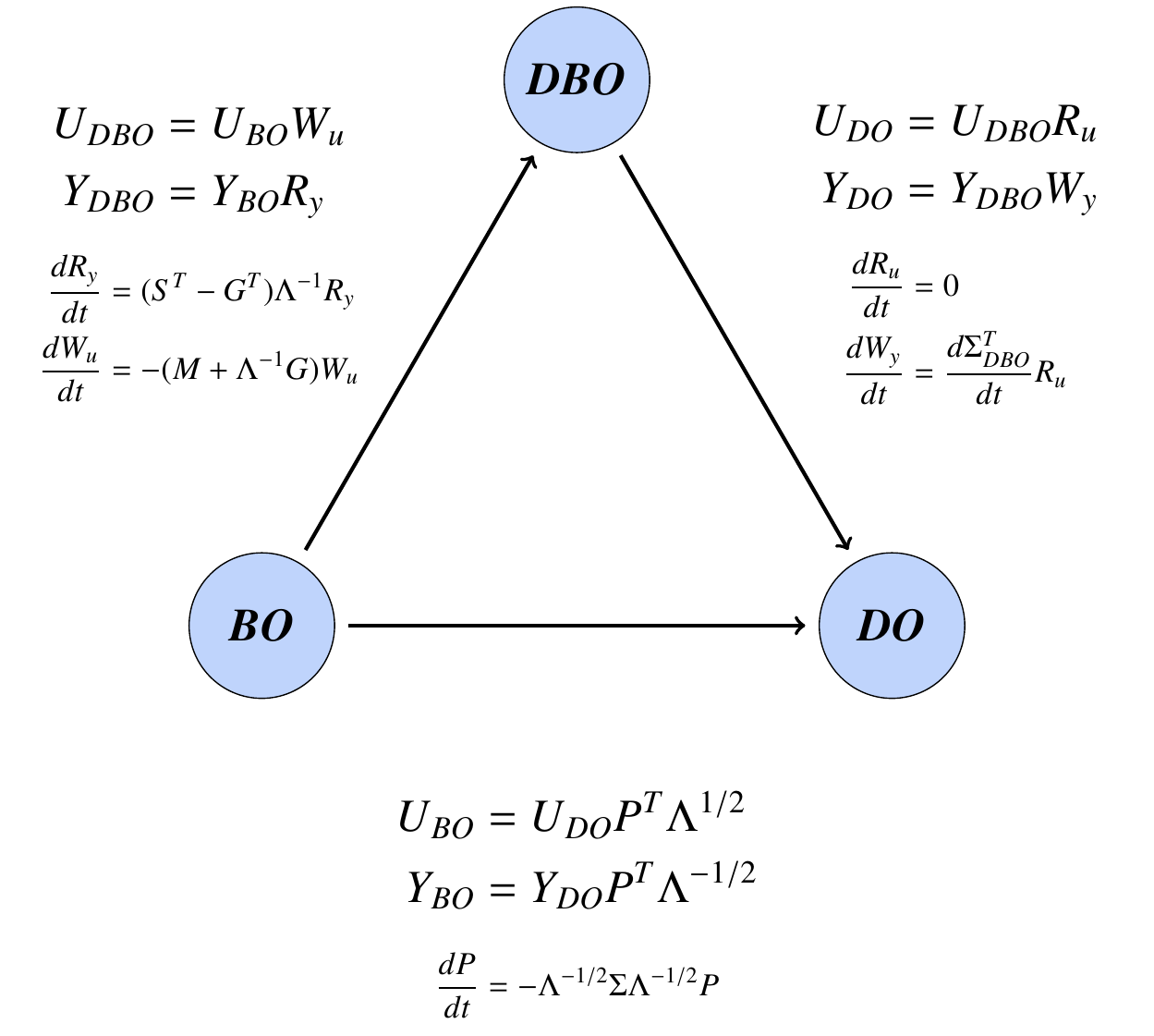}
    \caption{Equivalence relations between the three methods. The equivalence between DO and BO and the definitions of matrices $S$, $G$, $M$, $\Sigma$ are taken from reference \cite{choi2014equivalence}.}
    \label{fig:Equivalence}
\end{figure}

\begin{remark}
Based on the equivalence relation between BO and DBO, and that between DBO and DO; it can be easily shown that the equivalence between BO and DO obtained from \cite{choi2014equivalence} would be equal to $U_{DO} = U_{BO}W_u R_u$ and $Y_{DO} = Y_{BO} R_y W_y$.
\end{remark}
In Fig.(\ref{fig:Equivalence}) we summarize the equivalence relations between DBO, DO and BO. The equivalence relation between BO and DO and the definition of matrices: $M, G, S$ and $\Sigma$ are taken from \cite{choi2014equivalence}. 
\subsection{Mode ranking}
In this section, we determine the ranking of the modes in the stochastic and spatial subspace of DBO as performed in \cite{2019arXiv190409846B}. 
The spatial and stochastic DBO modes are ranked in the direction of the most energetic modes i.e., the modes are ranked based on the variance captured by each mode. To this end, we perform a singular value decomposition (SVD) of the $\Sigma_{DBO}$ matrix given by: 
\begin{equation*}
    \Sigma_{DBO}(t) = \Psi_{U}(t) \Lambda(t)^{1/2} \Psi_{Y}(t),
\end{equation*}
where $\Psi_{U}(t)$ and $\Psi_{Y}(t)$ are the left-singular vectors and the right-singular vectors of $\Sigma_{DBO}$, respectively. $\Lambda(t)$ is a diagonal matrix containing the eigenvalues of the covariance matrix. The eigenvalues are ranked such that $\lambda_1(t) \geq \lambda_2(t) \geq \cdots \geq \lambda_r(t) $. The ranked DBO modes based on the variance i.e., $\lambda_i(t)$, are obtained by an in-subspace rotation as in the following:
\begin{equation*}
    \widetilde{U}_{DBO}(t) = U_{DBO}(t)  \Psi_{U}(t),
\end{equation*}
\begin{equation*}
    \widetilde{Y}_{DBO}(t) = Y_{DBO}(t)  \Psi_{Y}(t).
\end{equation*}

\subsection{Redundancy in time}\label{sec:RedTime}
All  three components of the DBO decomposition i.e.,   $U(x,t)$, $Y(x,t)$ and $\Sigma(t)$  are time dependent. The issue of time redundancy also exists in both BO and DO decompositions.  We present a simple but insightful and unifying approach to clarify the \emph{constraints} and \emph{degrees of freedom} (DOF) in devising new time-dependent decompositions. For simplicity, we consider a finite-dimensional example. In particular, we consider the full-dimensional decomposition of a time-dependent matrix $A(t) \in \mathbb{R}^{n\times s}$. In this simplification $A(t)$ can be considered as a discrete representation of the mean subtracted random field, where $n$ is the number of discrete points in spatial domain and $s$ is the number of samples of the random field. In this section, we determine the degrees of freedom and the number of constraints imposed by each decomposition, and we show that in BO, DO and DBO decompositions the total number of constraints is equal to the number of degrees of freedom --- leading to  unique decompositions. In the following analysis we drop the explicit dependence on $t$ for brevity.

\subsubsection{BO}
We first consider the BO decomposition of matrix $A$ given by: $A= U Y^T$, where   $U \in \mathbb{R}^{n\times s}$ are the set of orthogonal spatial modes and $Y \in \mathbb{R}^{n\times s}$ are the set of orthonormal stochastic coefficients.  The total DOF is equal to the sum of number of entries in matrix $U$ i.e., $n\times s$  and entries in matrix $Y$ i.e., $s \times s$. Therefore,  the total DOF is given by: $N_{DOF}=n\times s + s\times s$. The  constraints imposed on the BO decomposition are as in the following: (i) The first set of constraints are the compatibility conditions, where $A_{ij} = U_{ik}Y_{jk}$, which imposes $N_{c_1}=n\times s$ constraints. (ii) The second set of constraints are imposed by the orthogonality of the spatial modes ($\left< u_i, u_j \right> = \delta_{ij}\lambda_j$), which impose $N_{c_2}=s(s-1)/2$ independent constraints. We take into account the number of $\left< u_i, u_j \right>=0, i=1,2, \dots, s$ for $j<i$. Note that for $j>i$ the constraints are equivalent to those of $i<j$, since $\left< u_i, u_j \right>=\left< u_j, u_i \right>$, and therefore they are not independent constraints and thus not counted. (iii) The third set of constraints are imposed by the orthonormality of the stochastic coefficients: $\mathbb{E}[y_i y_j]=\delta_{ij}, i=1,2, \dots, s$ and $j\leq i$, which imposes $N_{c_3}=s(s+1)/2$ independent constraints. Therefore, for the BO decomposition,  the total number of constraints is equal to total DOF, i.e.  $N_{DOF} = N_{c_1}+N_{c_2}+N_{c_3}=n\times s + s\times s$, leading to a fully determined unique decomposition.

\subsubsection{DO}
The DO decomposition is given by: $A= U Y^T$, where the spatial modes are a set of orthonormal vectors and $Y$ are the stochastic coefficients. 
 The total DOF of DO is the same as that of the BO for the same reasons mentioned above, $N_{DOF}=n\times s + s\times s$.  
The  constraints imposed on the DO decomposition are as in the following:  (i) Similar to the BO decomposition, there are $N_{c_1}= n\times s$ constraints imposed by the compatibility equations  $A_{ij} = U_{ik}Y_{jk}$. (ii) The orthonormality of the spatial modes ($\left< u_i, u_j \right> = \delta_{ij}$) imposes $N_{c_2}=s(s+1)/2$ independent constraints. (iii) The dynamically orthogonal condition $\left< \dot{u}_i, u_j \right> =0, i=1,2, \dots, s$ and $j<i$ imposes  $N_{c_3}=s(s-1)/2$ independent constraints. Note that $\left< \dot{u}_i, u_i \right> =0, i=1,2, \dots, s$ does not impose independent constraints as $\left< u_i, u_i \right>=1$ already enforces  this condition. This can be seen by taking the time derivative of the orthonormality constraints: 
\begin{equation*}
    \frac{d}{dt}\left< u_i, u_i \right> = \left< \dot{u}_i, u_i \right> + \left< u_i, \dot{u}_i \right> = 2 \left< \dot{u}_i, u_i \right>=0.
\end{equation*}
Therefore, similar to BO,  the DO decomposition leads to a fully determined decomposition as  the total number of DOF and constraints are equal, i.e.  $N_{DOF} = N_{c_1}+N_{c_2}+N_{c_3}=n\times s + s\times s$. 

\subsubsection{DBO}
Now, we consider the DBO decomposition, which is given by: $A = U \Sigma Y^T$, where the spatial modes and stochastic modes are a set of orthonormal bases. 
 The total DOF for DBO are given by the total number of elements in each of the matrices in the decomposition i.e., $n\times s $ entries in $U$ matrix, $s \times s $ entries in the $\Sigma$ matrix and $s \times s $ entries in the $Y$ matrix. Thus, the total DOF is: $N_{DOF} = n \times s + s \times s + s\times s$. 
The constraints imposed by the DBO decomposition are as in the following: (i) Similar to the BO and DO decompositions, there are $N_{c1}=n \times s$ constraints imposed by the compatibility conditions $A_{ij} = U_{ik}\Sigma_{km}Y_{jm}$. (ii) The orthonormality of stochastic and spatial modes ($\left< u_i, u_j \right> = \delta_{ij}$ and $\mathbb{E}[y_i y_j]= \delta_{ij}$) imposes $ s(s+1)/2$ constraints each, which in total imposes $N_{c2} = s(s+1)$. (iii) The dynamically orthogonal constraints for spatial and stochastic modes ($\left< \dot{u}_i, {u}_j \right> =0 $ and $\mathbb{E}[\dot{y}_i y_j]=0$) imposes $s(s-1)/2$ constraints each. Thus, the total constraints from the dynamically orthogonal condition are $N_{c3}= s(s-1)$. 

The total number of constraints for the DBO decomposition is $n \times s + s(s+1) + s(s-1)$, which is equal to the number of degrees of freedom, and this results in a fully determined DBO decomposition for matrix $A$. 

We  conclude that to obtain a unique time-dependent decomposition, the number of degrees of freedom and the number of constraints need to be equal. The summary of the constraints and degrees of freedom for BO, DO  and DBO are presented in Table \ref{tab:AllMethodsTab}. Introducing additional degrees of freedom requires additional constraints to keep the system fully determined and thus unique. In the light of the above analysis, DBO  allows for $s \times s$ additional degrees of freedom compared to DO by adding the matrix $\Sigma$ to the decomposition. These additional constraints are then utilized to enforce the orthonormality and dynamically orthogonal conditions on the stochastic coefficients $Y$.  The orthonormality  of $Y$ coefficients in  the DBO decomposition cannot be enforced in the DO decomposition.  As we will demonstrate this  loss of orthonormality of $Y$ in the DO decomposition can lead to degradation of accuracy in highly ill-conditioned problems.
\begin{center}
\begin{table}
\begin{tabular}{|c|c|c|c c|}
\hline
Method & Matrix Decomposition & Degrees of Freedom & \multicolumn{2}{c|}{Constraints}  \\ 
\hline \hline

\multirow{3}{*}{BO} & \multirow{3}{*}{$A_{n\times s} = U_{n\times s } Y^T_{s \times s} $ }& \multirow{3}{*}{$ns + s^2$} & $\left< U,U \right> = \Lambda$ &:$\frac{s(s-1)}{2}$ \\ 
&&& $\Lambda$ is diagonal matrix & \\ 
\cline{4-5}
&&& $\mathbb{E}[Y^T Y] = I $ &:$\frac{s(s+1)}{2}$\\
\cline{4-5}
\hlinewd{1.2pt}

\multirow{2}{*}{DO} & \multirow{2}{*}{$A_{n\times s} = U_{n\times s } Y^T_{s \times s} $} & \multirow{2}{*}{$ns + s^2$ }&  $\left< U, U \right> = I $ &: $\frac{s(s+1)}{2}$ \\ 
\cline{4-5}
&&& $\left< \dot{U}, U \right> = 0$ &: $\frac{s(s-1)}{2}$\\
\hlinewd{1.2pt}

\multirow{4}{*}{DBO} & \multirow{4}{*}{$A_{n\times s} = U_{n\times s} \Sigma_{s\times s} Y^T_{s \times s} $ }& \multirow{4}{*}{$ns + s^2 + s^2$} & $\left< U, U \right> = I$ &: $\frac{s(s+1)}{2}$  \\
\cline{4-5} 
&&& $\mathbb{E}[Y^T Y] = I$ &: $\frac{s(s+1)}{2}$  \\ 
\cline{4-5}
&&& $\left< \dot{U}, U \right> = 0$ &: $\frac{s(s-1)}{2}$ \\
\cline{4-5}
&&& $\mathbb{E}[\dot{Y}^T Y] = 0$  &: $\frac{s(s-1)}{2}$\\
\hline
\end{tabular}
\caption{Number of constraints and degrees of freedom for  BO, DO and DBO decompositions. Each decomposition imposes $n\times s$ compatibility constraints, which are not listed.}
\label{tab:AllMethodsTab}
\end{table}
\end{center}

\subsection{Error Analysis}
In Section \ref{sec:DemCases}, we compare the results of the DBO numerical solutions with the analytical solution using the following error calculations. To the end, we compute the  $L_2$ norm of the error of the mean $(\epsilon_{m}(t))$ as in the following:
\begin{equation}\label{eq:errormean}
    \epsilon_{m}(t) = \bigg(\int_D (\bar{u}(x,t)-\bar{u}_{DBO}(x,t))^2dx \bigg)^{1/2},
\end{equation}
where $\bar{u}(x,t)$ represents the mean of the analytical solution and $\bar{u}_{DBO}(x,t)$ represents the mean obtained from the DBO evolution equations. The error of the variance $(\epsilon_{v}(t))$   is calculated using the $L_2$-norm in both the spatial and stochastic dimensions:
\begin{subequations}\label{eq:errorvariance}
\begin{align}
    E(x,t;\omega) &=   u(x,t;\omega) -\bar{u}(x,t) - \sum_{j=1}^r \sum_{i=1}^r u_{DBO_i}(x,t) \Sigma_{DBO_{ij}} (t) y_{DBO_j} (\omega,t),\\
    \epsilon_{v}(t) &= \bigg(\int_D \mathbb{E}[E(x,t;\omega)^2] dx  \bigg)^{1/2},
\end{align}
\end{subequations}
where $u(x,t;\omega)$ represents the analytical stochastic field, $\bar{u}(x,t)$ represents the mean of the analytical stochastic flow field, whereas $u_{DBO_i}(x,t)$, $\Sigma_{DBO_{ij}}(t)$ and $y_{DBO_j}(\omega,t)$ represent the solutions of the components of the DBO decomposition obtained from the DBO evolution equations. 


\section{Demonstration cases}\label{sec:DemCases}
\subsection{Stochastic linear advection equation}
We consider linear advection governed by:
\begin{subequations}\label{eq:LinAd}
     \begin{align}
         \pfrac{u}{t} + V(\omega) \pfrac{u}{x} &= 0,   &&x \in [0, 2\pi] \quad \mbox{and} \quad  t\in[0,t_f],\\
          u(x,0) &= \sin(x),   &&x \in [0, 2\pi],
     \end{align}
\end{subequations}
with periodic boundary condition. The randomness in the system comes from the advection velocity $V( \omega)$. The random velocity is specified  by  $V(\omega) =\bar{v}+ \sigma \xi(\omega)$, where $\bar{v} =1.0$, $\sigma =1.0$ and $\xi(\omega)$  is a uniform random variable in the interval of $\xi  \sim \mathcal{U}[-1,1]$ with variance $1/3$. The physical domain  is discretized using the Fourier spectral method with $N_s =512$ Fourier modes. The random space is one dimensional and is discretized with the probabilistic collocation method (PCM) with $N_r =256 $ Legendre-Gauss points. The third-order Runge-Kutta scheme is used for the time integration with $\Delta t= 10^{-3}$. At $t=0$, the stochastic fluctuations are zero, and therefore, the simulation is initialized at $t=\Delta t$ to avoid singularity of the covariance matrix. The system is numerically evolved till $t_f=10$. 
The linear advection Eq.(\ref{eq:LinAd}) has a closed-form solution as follows:
 \begin{equation}\label{eq:LinAdSol}
     u(x,t;\omega) = g(x- V(\omega) t) = \sin(x-(\bar{v}+ \sigma \xi(\omega)) t).
 \end{equation}
This system can be expressed exactly with KL modes and the reduction order of $r=2$ as follows: 
  \begin{equation*}
     u(x,t;\omega) = \bar{u}(x,t) + \sum_{i=1}^r \sqrt{\lambda_i}(t) u_i(x,t) y_i(t,\omega),
 \end{equation*}
where,
\begin{align*}
    &\bar{u}(x,t) = \sin(x - \bar{v} t \pi) \frac{\sin(\sigma \pi t)}{\sigma \pi t }, \\
    &u_1(x,t) = \frac{1}{\sqrt{\pi}} \sin(x - \bar{v} \pi t), &&u_2(x,t) = \frac{-1}{\sqrt{\pi}} \cos(x - \bar{v} \pi t),\\
    &y_1(t;\omega) = \frac{\sqrt{\pi}}{\sqrt{\lambda_1(t)}} \left(\cos(\sigma \xi \pi t) - \frac{\sin(\sigma \pi t)}{\sigma \pi t} \right),   &&y_2(t;\omega) = \frac{\sqrt{\pi}}{\sqrt{\lambda_2(t)}} \sin(\sigma \xi \pi t),\\
     &\lambda_1(t) = 1- \frac{\sin(2\sigma \pi t) }{2 \sigma \pi t }, &&\lambda_2(t) = 1 + \frac{\sin(2 \sigma \pi t)}{2 \sigma \pi t} - \frac{2 \sin^2(\sigma \pi t)}{(\sigma \pi t)^2}.
\end{align*}
The mean, spatial and stochastic bases of the DBO decomposition are initialized with  KL modes given above. The covariance factorization is initialized by:
\begin{equation}
    \Sigma(t) = \begin{bmatrix}\sqrt{\lambda_1(t)} & 0 \\ 0 & \sqrt{\lambda_2(t)}\\
    \end{bmatrix}.
\end{equation}

\begin{figure}
     \centering
     \subfloat[Mean error]{\includegraphics[width=0.48\textwidth]{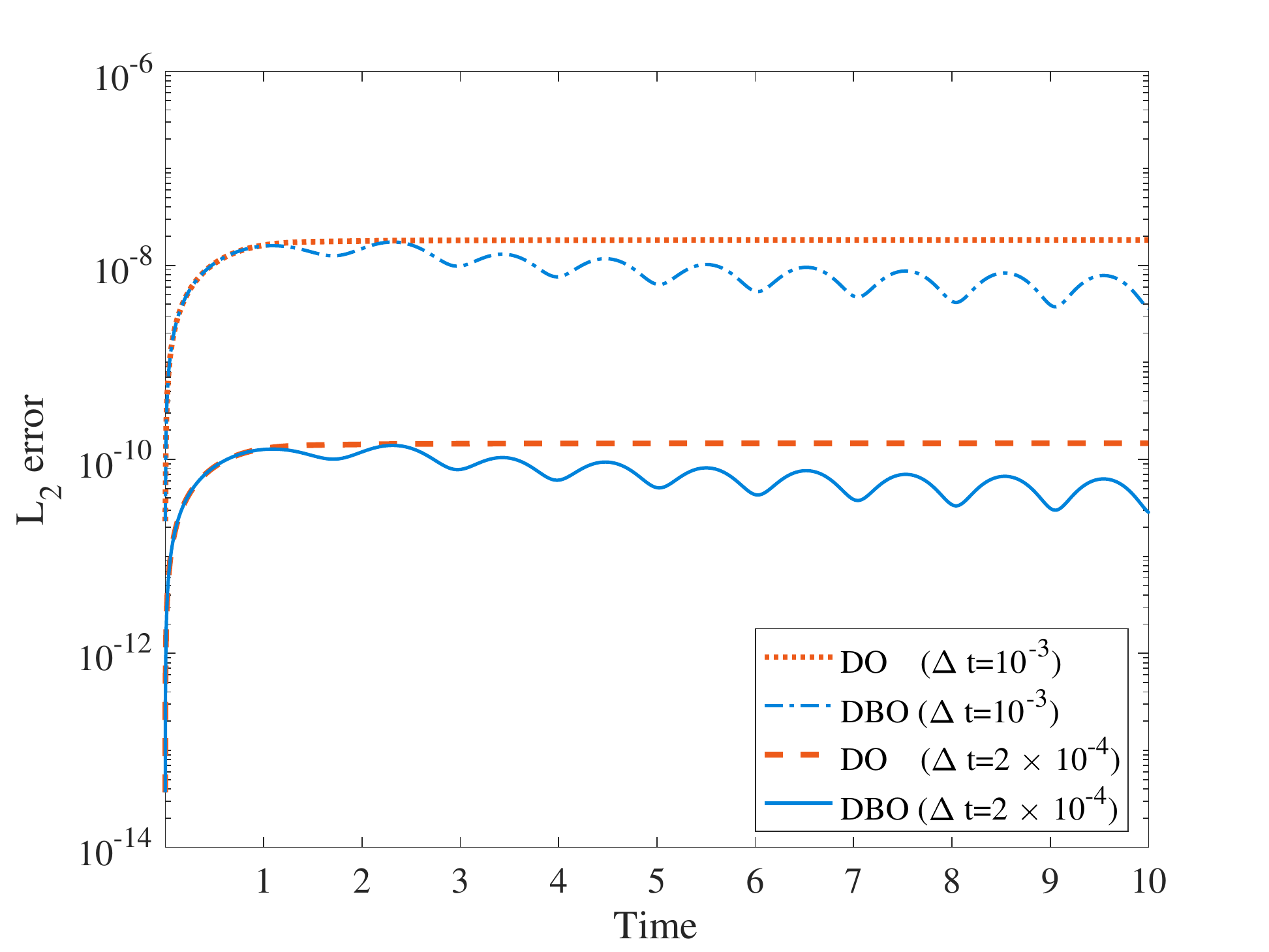}\label{fig:LinAdMean}}
     \hfill
     \subfloat[Variance error]{
     \includegraphics[width=0.48\textwidth]{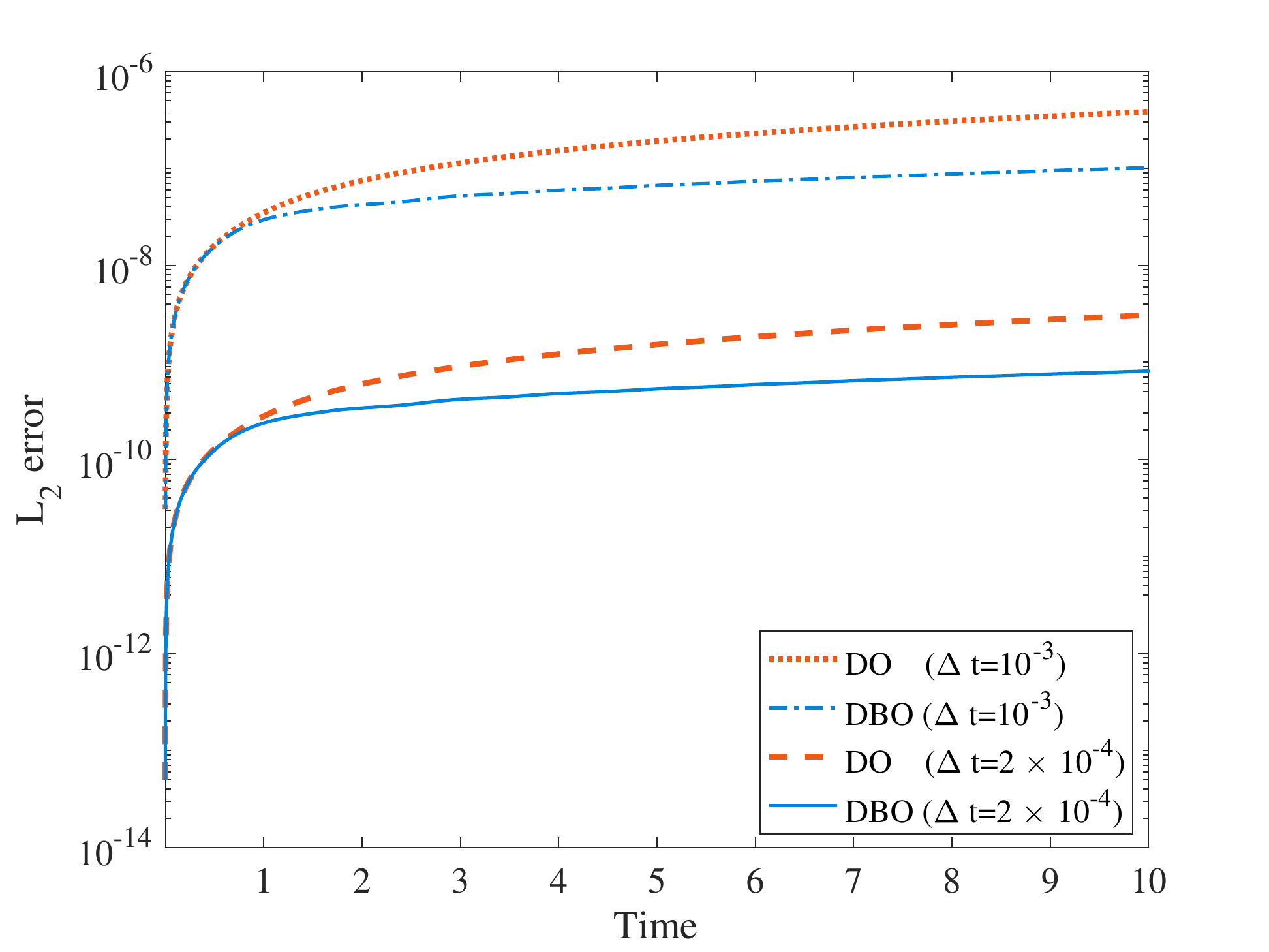}\label{fig:LinAdVar}}
     \caption{Stochastic linear advection equation: The $L_2$ errors for the mean and the variance are compared with the DO method.  The code used in this example is  available on GitHub at \url{https://github.com/ppatil1708/DBO.git}}
 \end{figure}

In Fig.(\ref{fig:LinAdMean}-\ref{fig:LinAdVar}), the $L_2$ error of the mean  and  variance  for both DO and DBO methods are shown, respectively. Since the solution of this problem can be exactly expressed with two DBO modes, the errors in the mean and variance come from the temporal, spatial and the PCM discretization of the random space. To the end, we present mean and variance errors for two values of $\Delta t =10^{-3}$ and $2 \times 10^{-4}$, in which the smaller $\Delta t$ shows smaller errors. We also refined the resolution for spatial and random discretizations, and we did not, however,  observe  noticeable change in the mean and variance errors.  This demonstrates that  the temporal discretization is the main source of error. For long time  integration, the resolution of solving  Eq.(\ref{subeq:DBOY})  must increase in time i.e., higher number of samples of $\xi$,  to maintain a desired level of accuracy as increasing time increases the wave number of $y_i(t;\omega)$ modes. However, in the DBO decomposition, the computational cost of increasing resolution in the random space is insignificant, as we solve the stochastic ODE of small order (here $r=2$) given by Eq.(\ref{subeq:DBOY}). This is in contrast to the PCM method, in which to maintain the desired level of accuracy the PCE order must increase with time, which results in solving  larger system of PDEs.  See reference \cite{wan2006long} for detailed error analysis of the stochastic linear advection equation using PCM. The BO method for this case would  diverge because of eigenvalue crossing. It is clear that both  DBO and  DO show similar errors as they are  equivalent. However, the DBO shows slightly smaller errors in both mean and the variance. 
 
\subsection{Stochastic Burgers' equation with manufactured solution}
We consider the stochastic Burgers' equation governed by:
\begin{subequations}\label{eqn:Burg}
\begin{align}
    \pfrac{u}{t} + u\pfrac{u}{x} &= \nu \pfrac{^2 u}{x^2} + f(x,t; \omega), &&x \in [0, 2\pi] \quad \mbox{and} \quad  t\in[0,t_f].\\
     u(x,0;\omega) &= g(x),    &&x \in [0, 2\pi].
\end{align}
\end{subequations}
We consider the following  manufactured solution expressed by  the KL decomposition with $r=2$ modes:
\begin{align*}
    &\bar{u}(x,t) = \sin(x-t),\\
    &u_1(x,t) = \frac{1}{\sqrt{\pi}} \cos(x-t), &&u_2(x,t) = \frac{1}{\sqrt{\pi}} \cos(2x-3t),\\
    &y_1(t; \omega) =  \sin(\pi \xi_1 (\omega) -t), &&y_2(t; \omega) =  \cos(\pi \xi_2 (\omega) -t),\\
    &\lambda_1(t) = (4.5 + \sin(t))^2, &&\lambda_2(t) = \epsilon^2 \cdot(1.5+ \cos(3t) )^2.
\end{align*}
We initialize the DBO systems with KL modes similar to the previous example. 
The  stochastic forcing $f(x,t;\omega)$  is calculated accordingly such that the above decomposition satisfies Eq.(\ref{eqn:Burg}). In the above equation 
 $\nu =0.05$ and $\xi_d \sim \mathcal{U}[-1,1]$. Here, $d$ is the dimension of the random space, which for this case is taken to be $d=2$. The parameter $\epsilon$ scales the smaller eigenvalue i.e., $\lambda_2(t)$, which in turn controls the condition number of the covariance matrix. The physical domain is considered to be periodic.  We  discretize the spatial domain using the Fourier spectral method with $N_s = 128$ modes. The random space is two-dimensional and is discretized with the ME-PCM (Multi-Element Probabilistic Collocation Method) \cite{wan2006multi} with 8 elements each containing 4 points in each random direction. Thus, the total points in every random direction is 32, which results in $N_r= 1024$. The third-order Runge-Kutta method is used for the time integration with $\Delta t= 10^{-3}$. Since at $t=0$ the stochasticity is zero, the numerical computation is started from $t_s=0.01$. The system is numerically evolved till $t_f = 3\pi$. 

The purpose of this case is to compare the performance of DO, BO and DBO methods for  cases with  ill-conditioned covariance matrices. We also compare the performance of DBO with pseudo-inverse DO (PI-DO) \cite{babaee2017robust}, where the authors proposed using pseudo inverse in the presence of singular or near-singular covariance matrices.   Two values of $\epsilon$ are considered and the evolution of the system for DO, PI-DO, BO and the DBO methods are studied. We use the $L_2$ error for evaluation of the mean and variance errors i.e., Eq.(\ref{eq:errormean}) and  Eq.(\ref{eq:errorvariance}) between the four methods. 

\begin{figure}
  \centering
  \setlength{\tempwidth}{0.43\textwidth}
    \settoheight{\tempheight}{\includegraphics[width=\tempwidth]{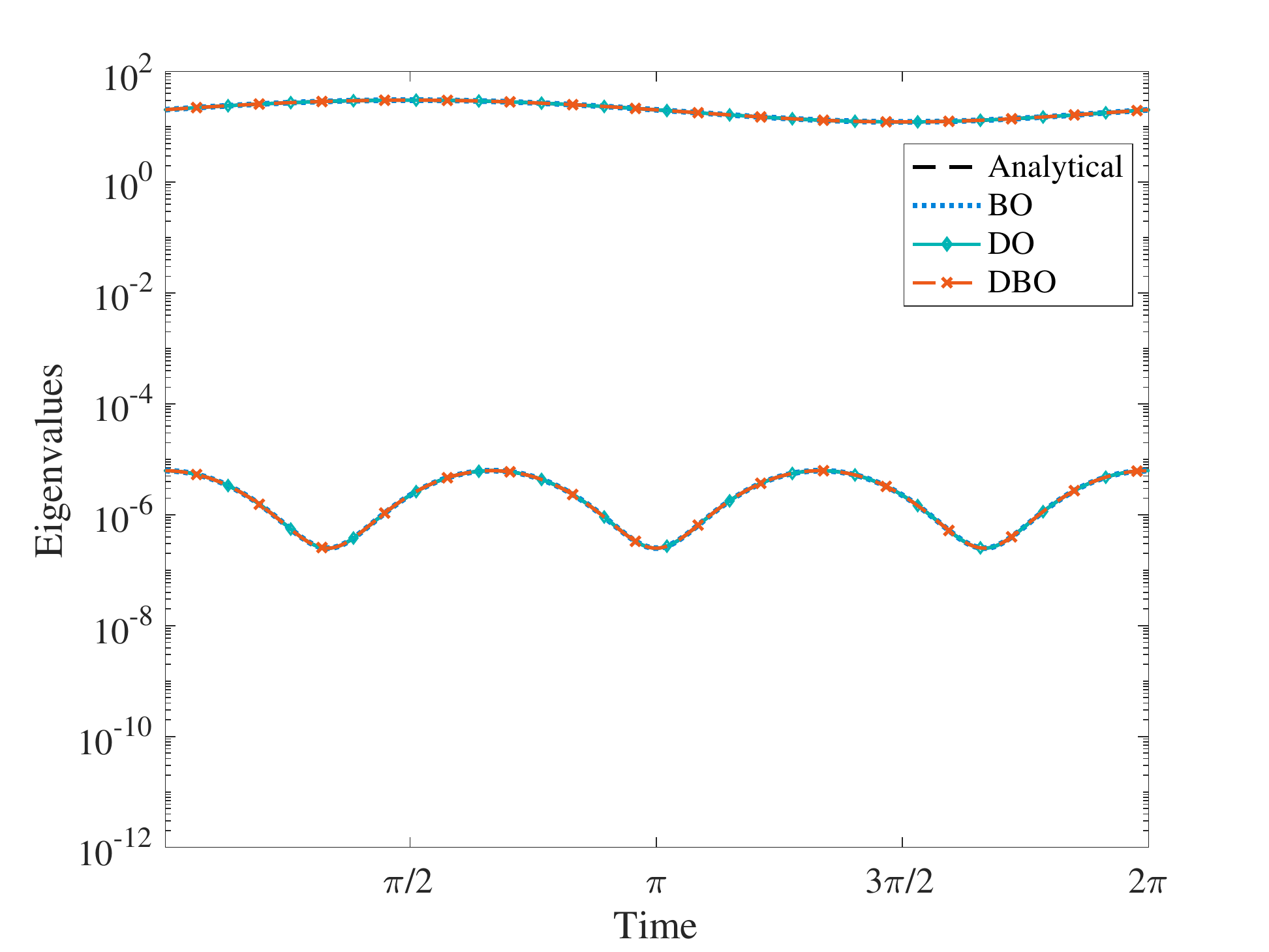}}%
    \columnname{$\epsilon= 10^{-3}$}
    \columnname{$\epsilon= 10^{-5}$}\\
    \subfloat[Eigenvalues] {\includegraphics[width=0.49\textwidth]{ManufacturedForcingdY/epsilon10_3/Eigenvalues.pdf}\label{fig:f4}}
    \subfloat[Eigenvalues] {\includegraphics[width=0.49\textwidth]{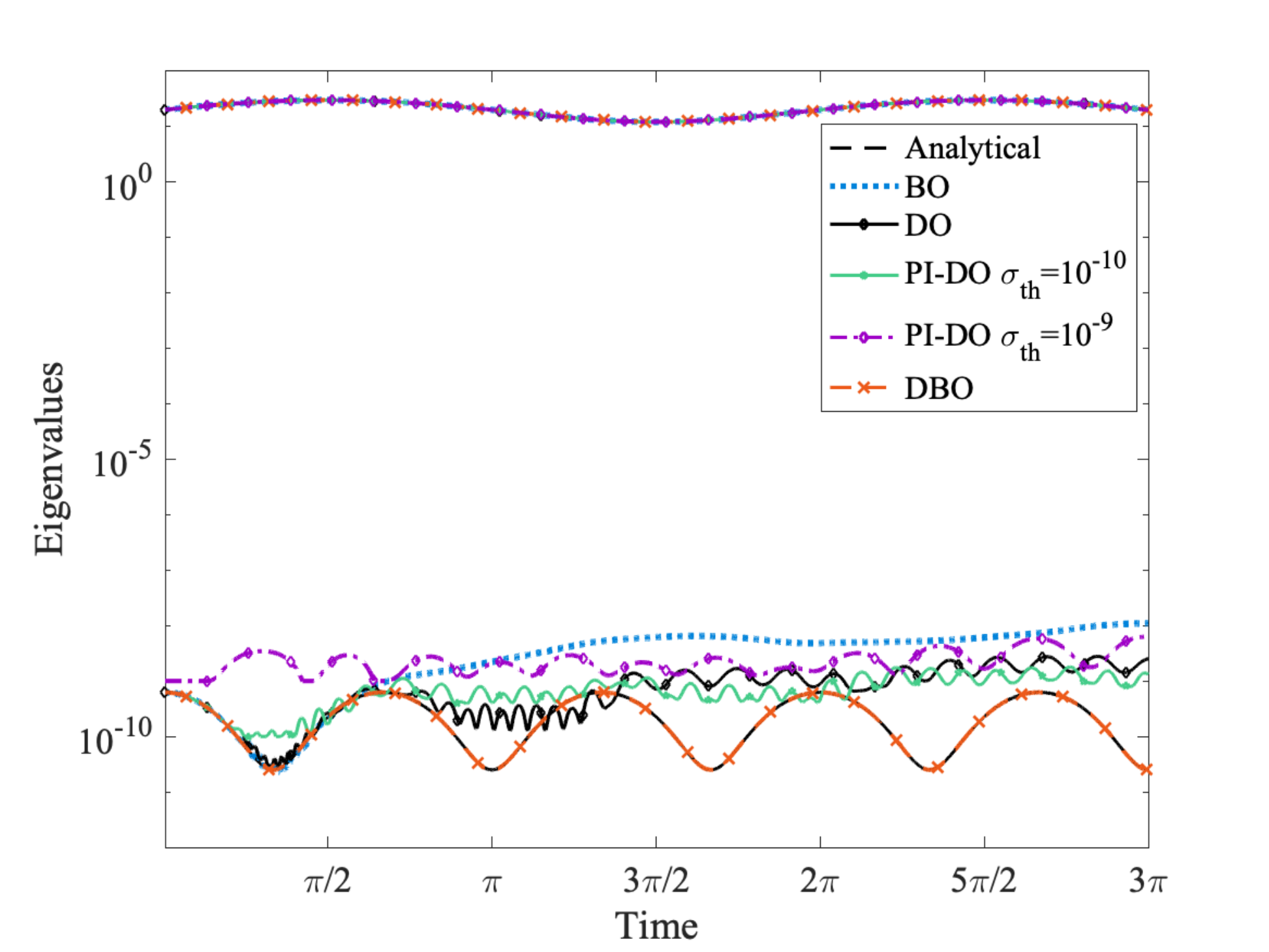}\label{fig:f5}}\\ 
    \subfloat[Mean error]   {\includegraphics[width=0.49\textwidth]{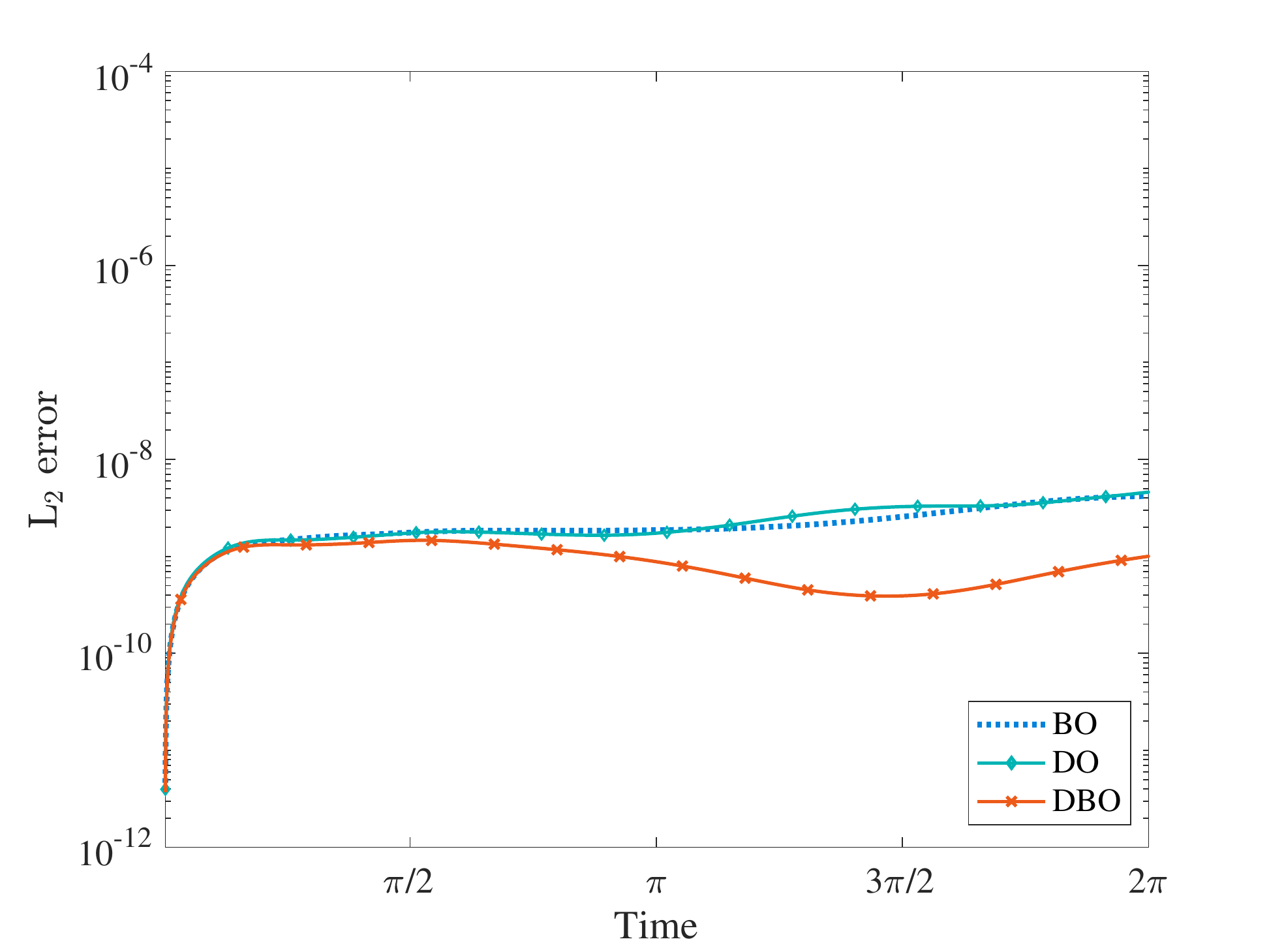}\label{fig:f6}}
    \subfloat[Mean error] {\includegraphics[width=0.49\textwidth]{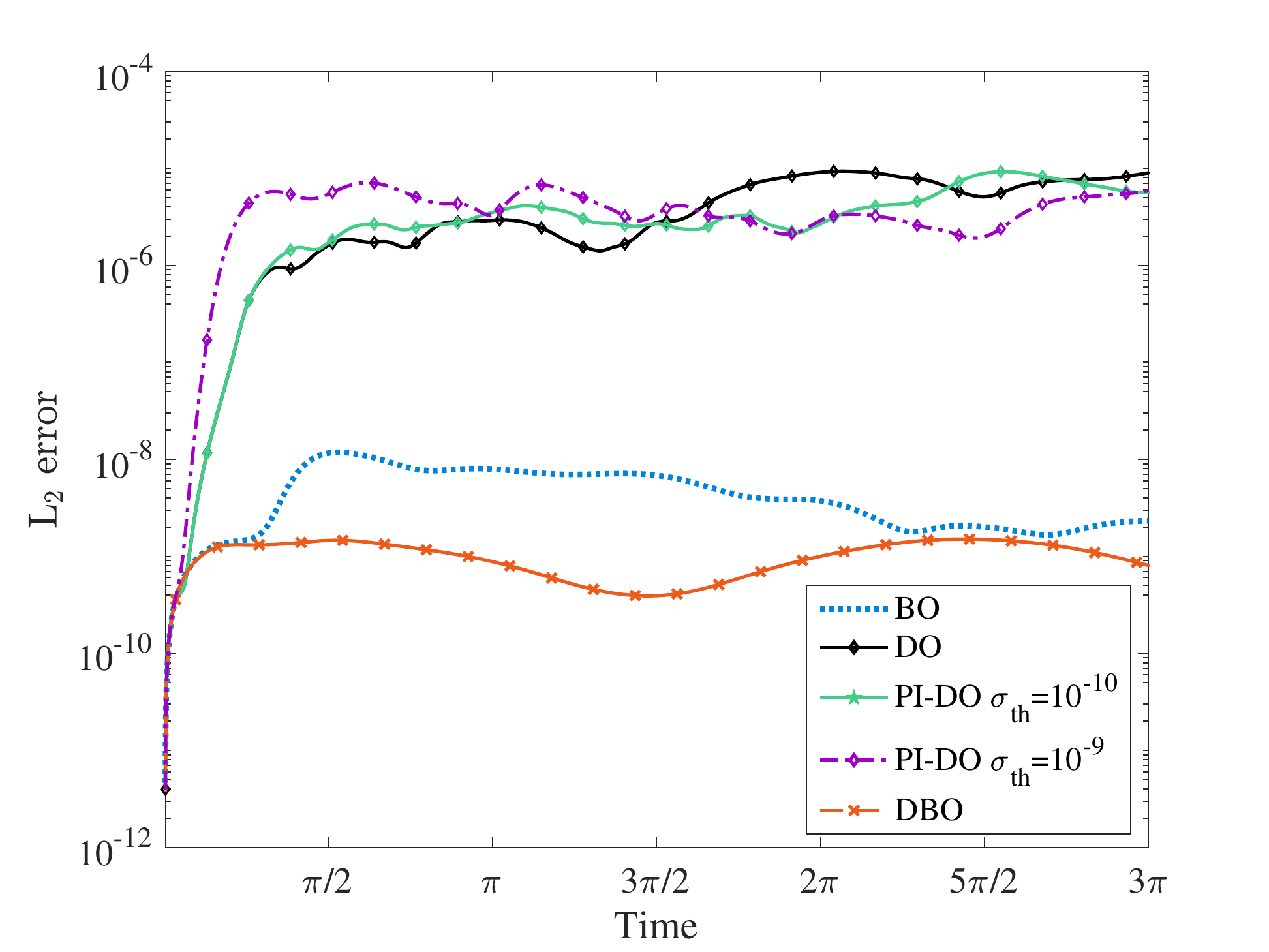}\label{fig:f7}}\\
    \subfloat[Variance error]{\includegraphics[width=0.49\textwidth]{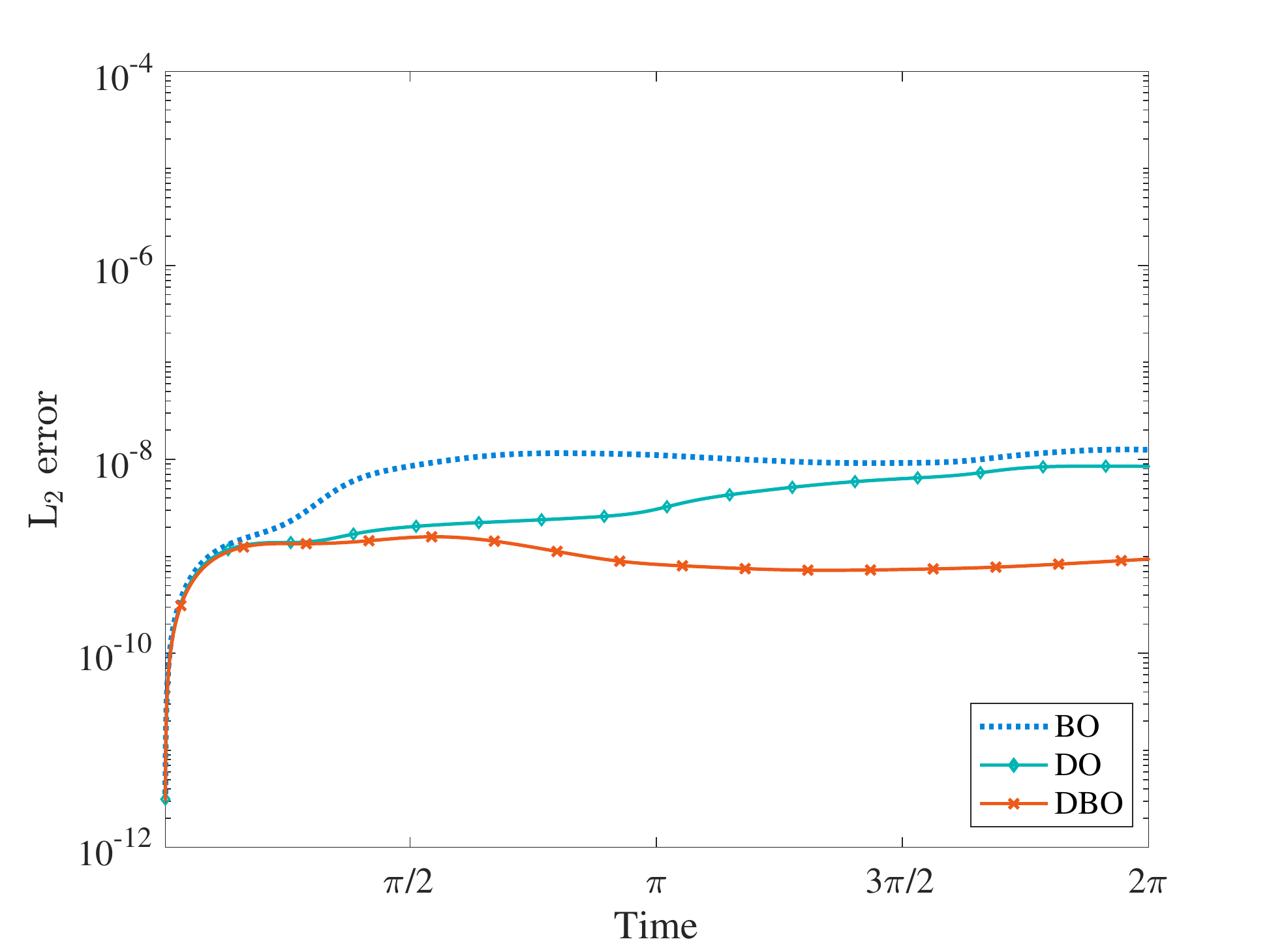}\label{fig:f8}}
  \subfloat[Variance error]{\includegraphics[width=0.49\textwidth]{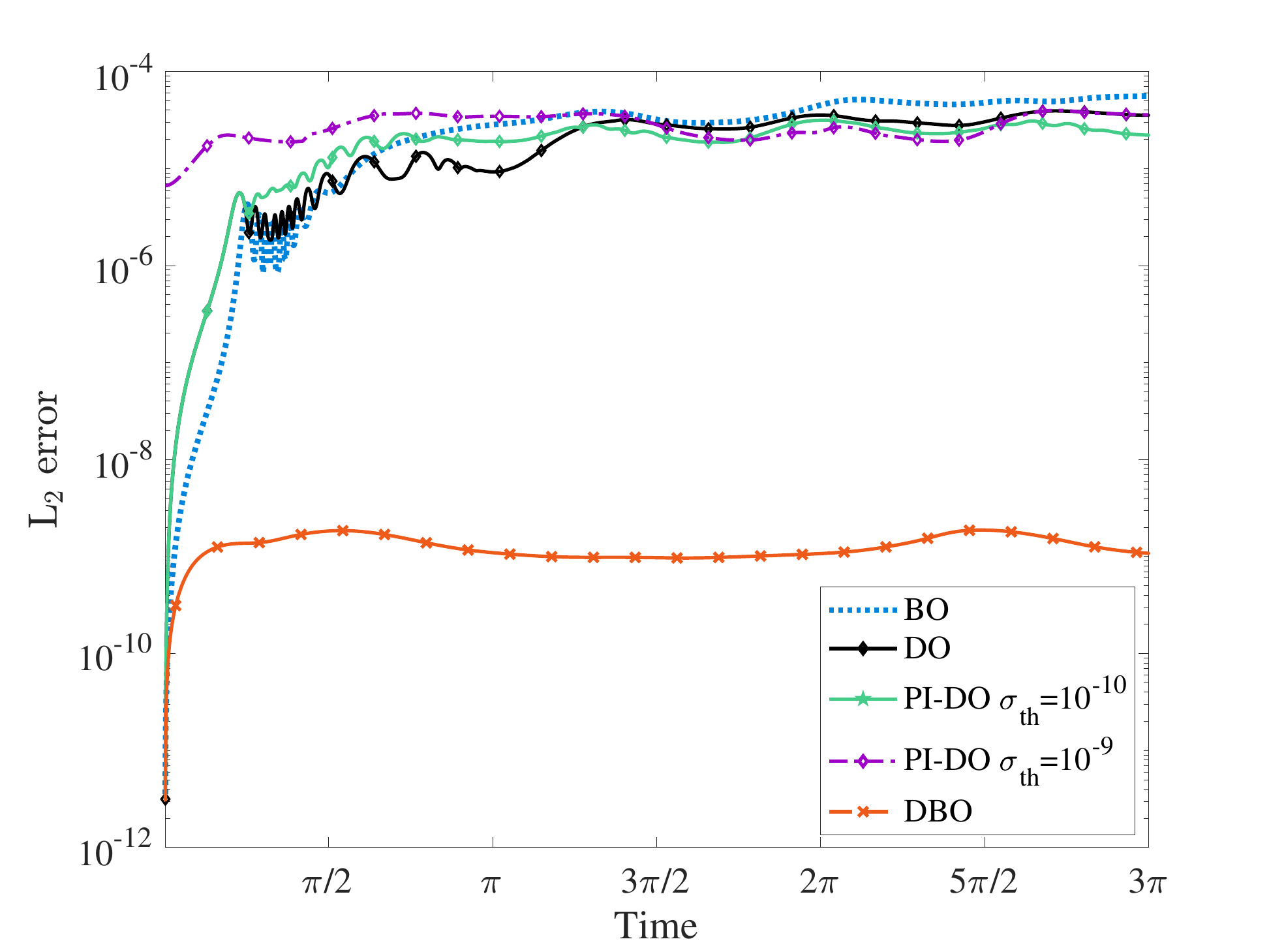}\label{fig:f9}}
  \caption{Burgers' equation with manufactured forcing: A comparison between two values of $\epsilon$, which controls the condition number of the system, is shown. The left column:(a),(c) and (e) correspond to the eigenvalues, mean error and variance error for the case with $\epsilon= 10^{-3}$, respectively. The right column:(b),(d) and (f) correspond to the eigenvalues, mean error and variance error for the case with $\epsilon= 10^{-3}$, respectively. It is observed that as the system becomes ill-conditioned for $\epsilon=10^{-5}$, the errors for the DO, PI-DO and the BO method increase whereas the DBO maintains the same accuracy for both the $\epsilon$ values. The code used in this example is available on GitHub at \url{https://github.com/ppatil1708/DBO.git}}
  \label{fig:MFerror}
\end{figure}

In Fig.(\ref{fig:MFerror}), the evolution of the eigenvalues, mean and variance error are shown for two values of $\epsilon= 10^{-3}$ and $\epsilon=10^{-5}$. Fig.(\ref{fig:f6}) and Fig.(\ref{fig:f7}) show a comparison between the mean errors for $\epsilon$ values $10^{-3}$ and $10^{-5}$, respectively. Similarly, Fig.(\ref{fig:f8}) and Fig.(\ref{fig:f9}) show the variance error for $\epsilon$ values $10^{-3}$ and $10^{-5}$ respectively. The PI-DO case is studied only for the case with $\epsilon = 10^{-5}$, since for the case with $\epsilon = 10^{-3}$ the covariance matrix does not become singular. Two threshold values are used for the inversion of the covariance matrix in the PI-DO method: $ \sigma_{th}= 10^{-9}$ and $\sigma_{th}=10^{-10} $. See reference \cite{babaee2017robust} for more details on the threshold values.  As shown in \cite{babaee2017robust}, the choice of the threshold value can play a significant role in the performance of PI-DO.  Based on the formulation of the eigenvalues, lower values of $\epsilon$ creates an ill-conditioned covariance matrix for DO, BO as well as an ill-conditioned ${\Sigma}$ matrix for DBO. However, in both DO and BO the condition number of the covariance matrix is $\kappa_{DO,BO}=\lambda_1(t)/\lambda_2(t)$, which scales with $1/\epsilon^2$, while the condition number of  $\Sigma$ in the DBO decomposition is $\kappa_{DBO}=\sqrt{\lambda_1(t)/\lambda_2(t)}$, which scales with $1/\epsilon$. Since DO, BO and DBO are equivalent, it is expected that they all perform similarly for the well-condition covariance matrix, i.e., $\epsilon=10^{-3}$. This can be seen in  Fig.(\ref{fig:f4}), Fig.(\ref{fig:f6}) and Fig.(\ref{fig:f8}), where all three methods exhibit the same  levels of error in mean and variance and the eigenvalues of the covariance matrix match well with the true eigenvalues.  However, for the case with $\epsilon=10^{-5}$,   it is expected that DBO performs better than BO and DO and this can be seen in Fig.(\ref{fig:f5}), Fig.(\ref{fig:f7}) and Fig.(\ref{fig:f9}). For this case neither DO, BO nor PI-DO can capture the smallest eigenvalue i.e., $\lambda_2(t)$ correctly. As a result they introduce error of the order of   $\sqrt{\lambda_2(t)} \sim \mathcal{O}(\epsilon)$, which can be observed in Fig.(\ref{fig:f7}) and Fig.(\ref{fig:f9}). As seen in Fig.(\ref{fig:f7}) and Fig.(\ref{fig:f9}), the threshold value of $\sigma_{th} = 10^{-9}$ for pseudo-inverse introduces higher order errors than that of the $\sigma_{th} = 10^{-10}$. The pseudo-inverse method introduces $\mathcal{O}(\sigma_{th})$ in the simulation whenever the lowest eigenvalue attains a value lower than the threshold $\sigma_{th}$.

\begin{figure}
    \centering
    \setlength{\tempwidth}{0.30\textwidth}
    \settoheight{\tempheight}{\includegraphics[width=\tempwidth, height=0.16\textheight]{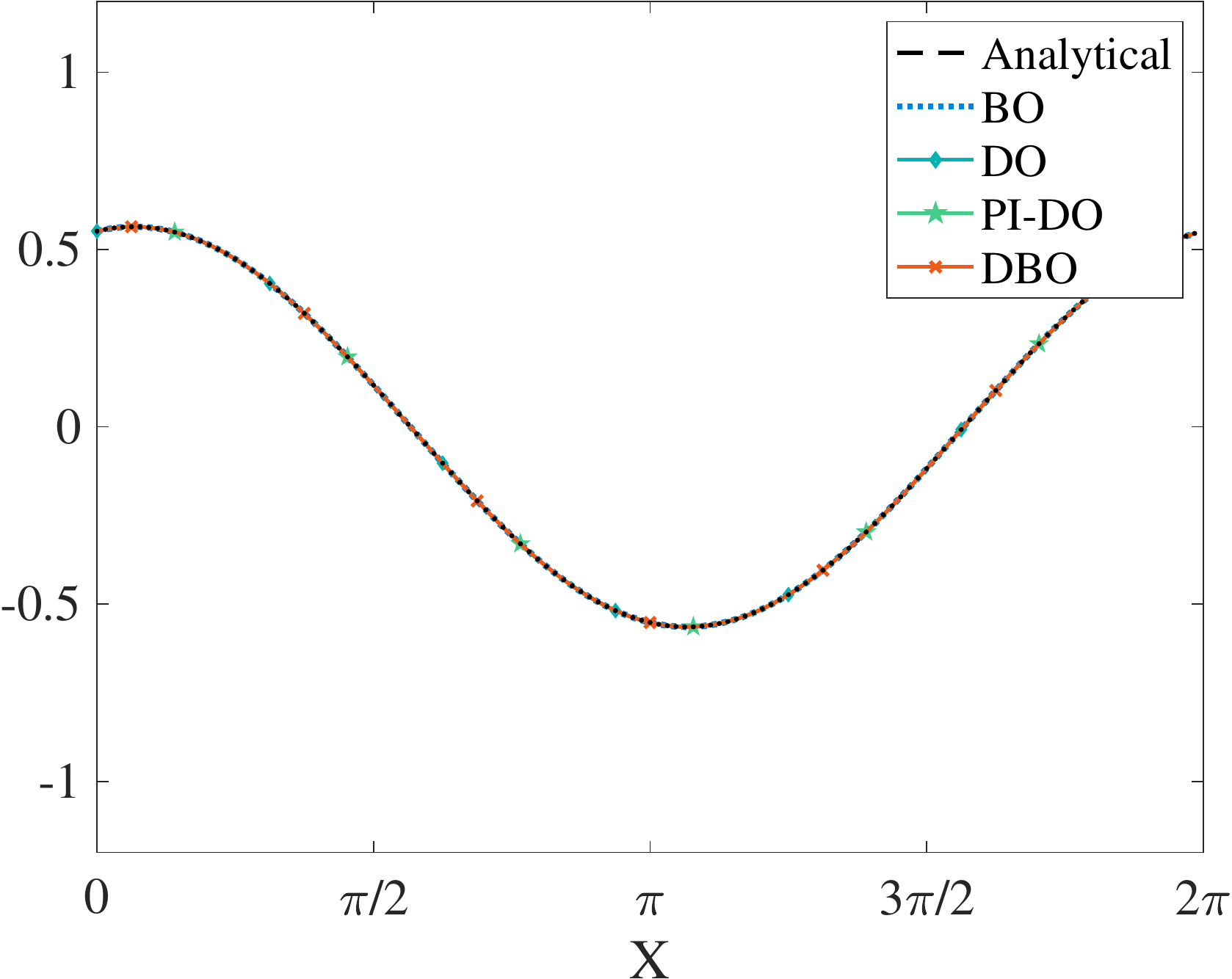}}%
    \columnname{\quad \qquad \small{$u_1(x,t)$}}\hfil
    \columnname{ \quad \small{$u_2(x,t)$}}\hfil
    \columnname{\small{ $y_1(t;\omega)$ vs. $y_2(t;\omega)$}}\\
    \rowname{\small{$t = 0.2$}}
    \subfloat[]{
    \includegraphics[width=\tempwidth, height=0.16\textheight]{ManufacturedForcingdY/epsilon10_5/PhyBasisMode1PI_200.pdf}\label{fig:fa10}}
    \subfloat[]{
    \includegraphics[width=\tempwidth, height=0.16\textheight]{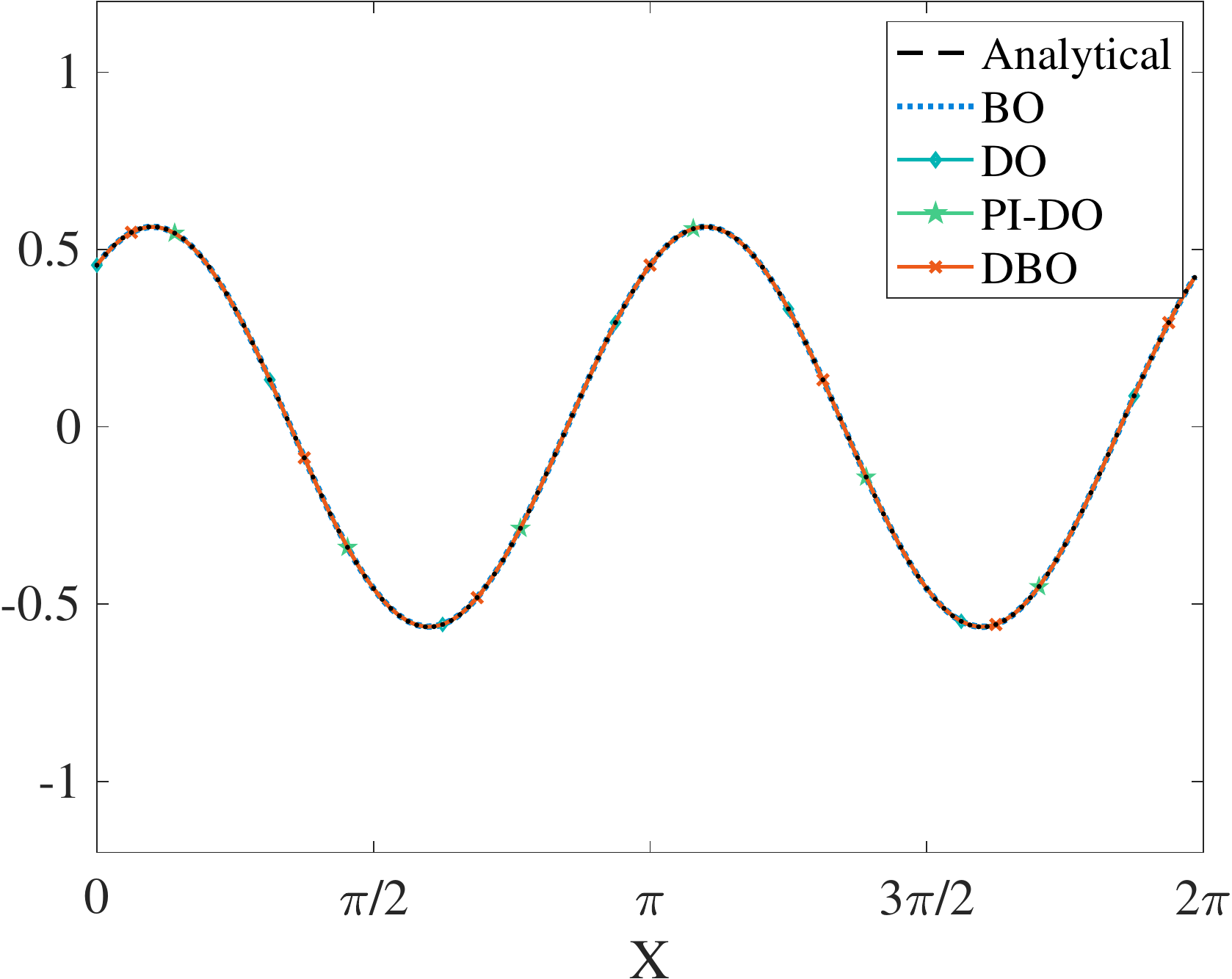}\label{fig:fa11}}
    \subfloat[]{
    \includegraphics[width=\tempwidth, height= 1.0125\tempheight]{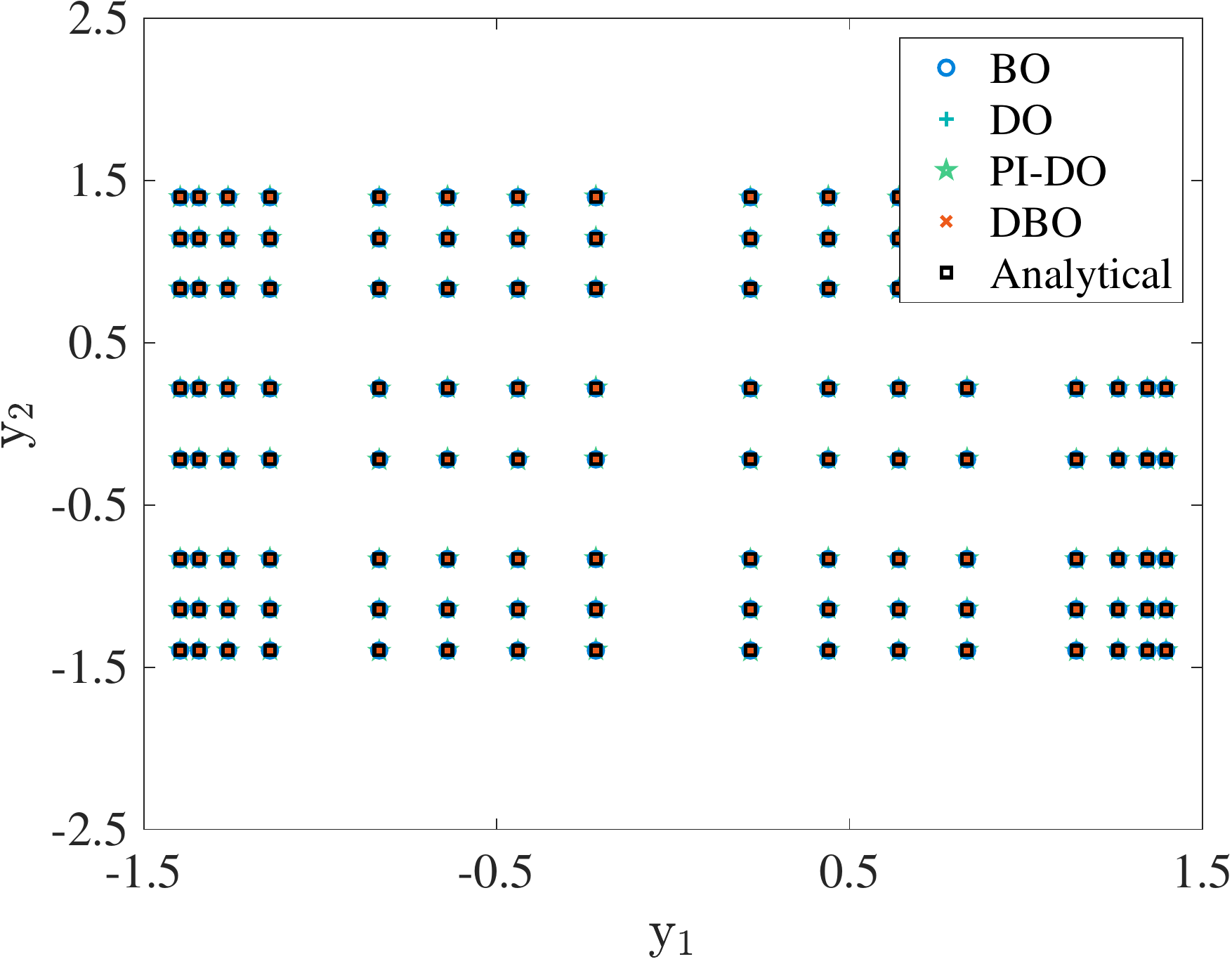}\label{fig:fa12}}\\ 
    \rowname{\small{$t = 1.2$ }}
    \subfloat[]{
    \includegraphics[width=\tempwidth, height=0.16\textheight]{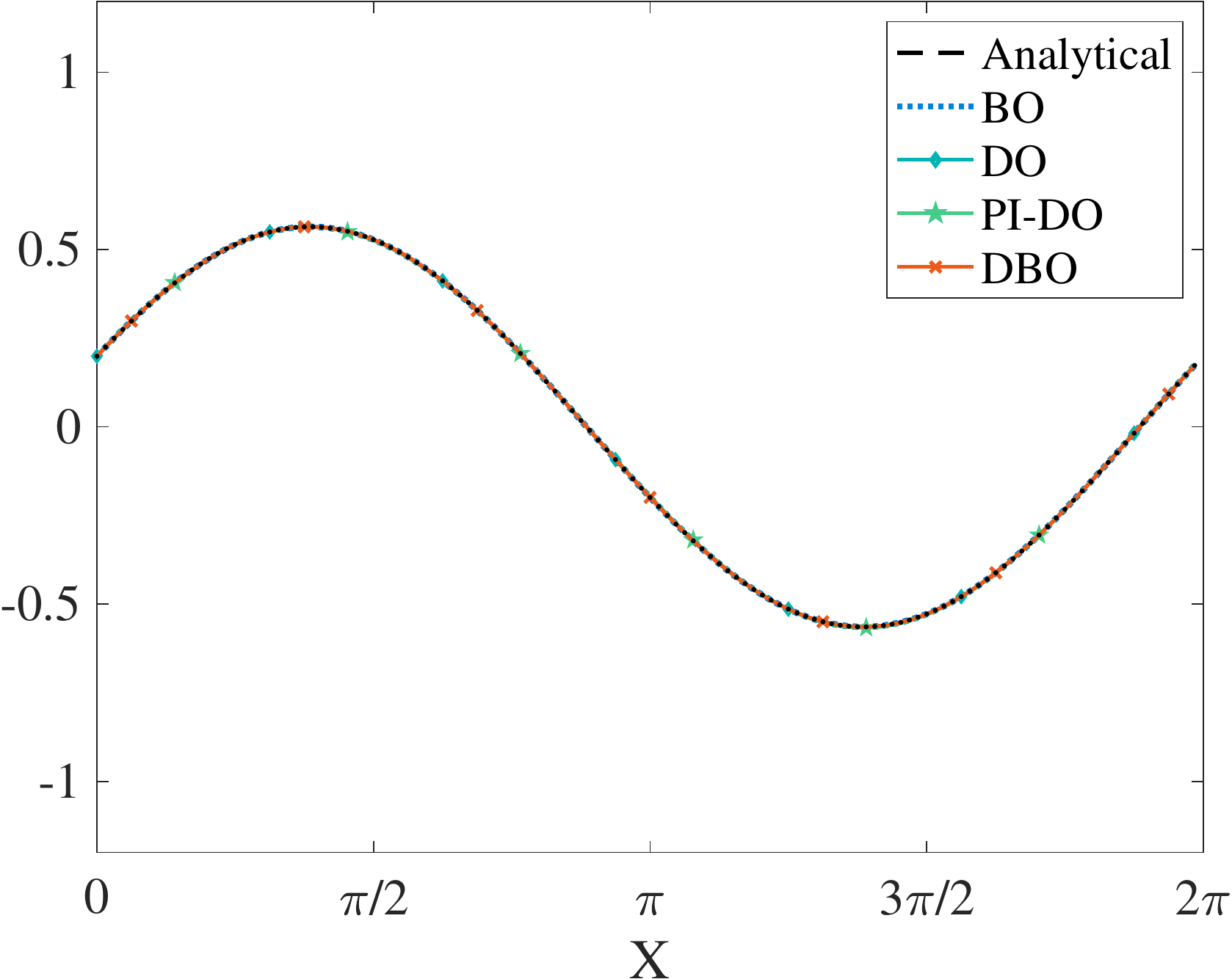}\label{fig:fa13}}
    \subfloat[]{
    \includegraphics[width=\tempwidth, height=0.16\textheight]{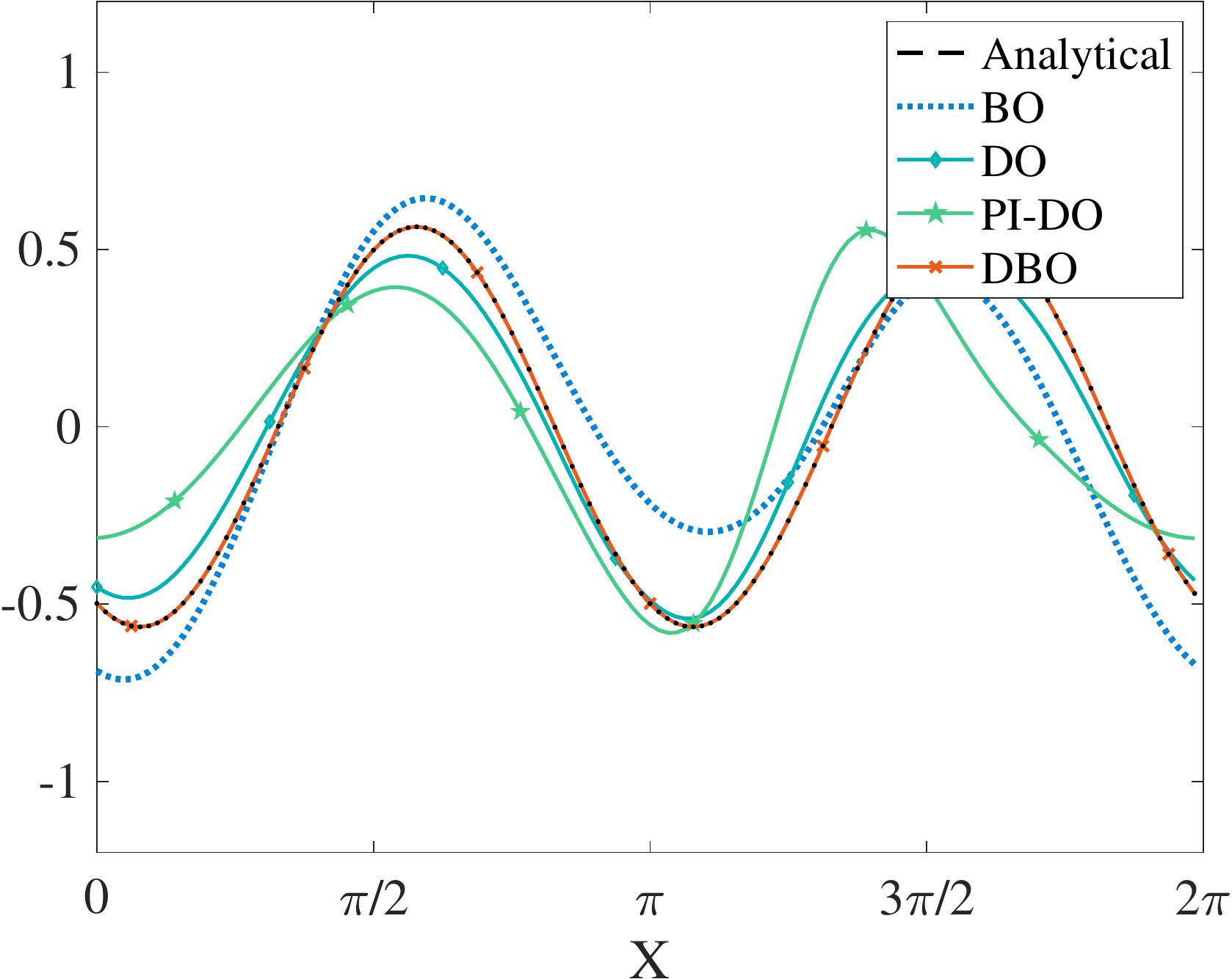}\label{fig:fa14}}
    \subfloat[]{
    \includegraphics[width=\tempwidth, height= 1.0125\tempheight]{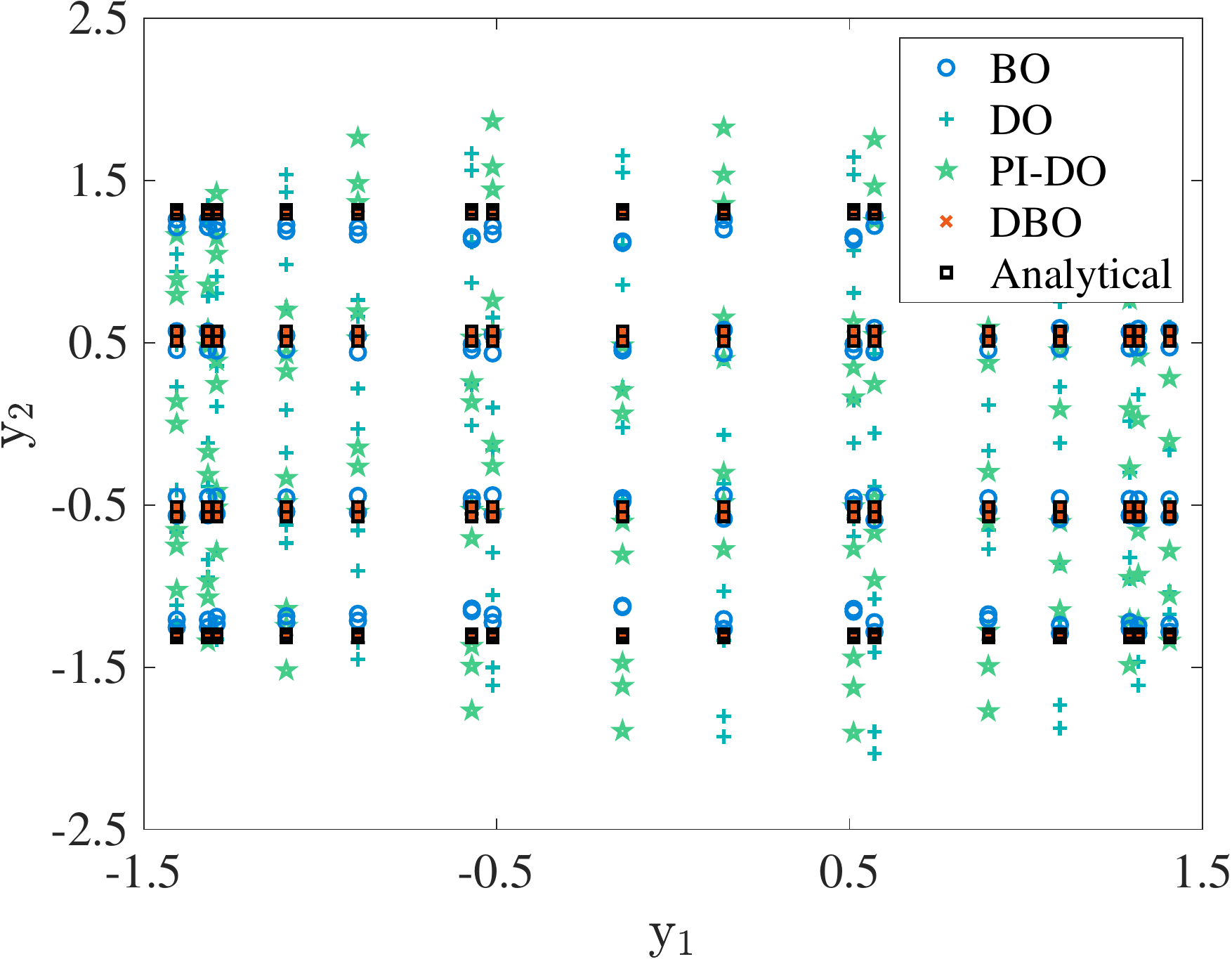}\label{fig:fa15}}\\
    \rowname{\small{$t = 3.2$ }}
    \subfloat[]{
    \includegraphics[width=\tempwidth, height=0.16\textheight]{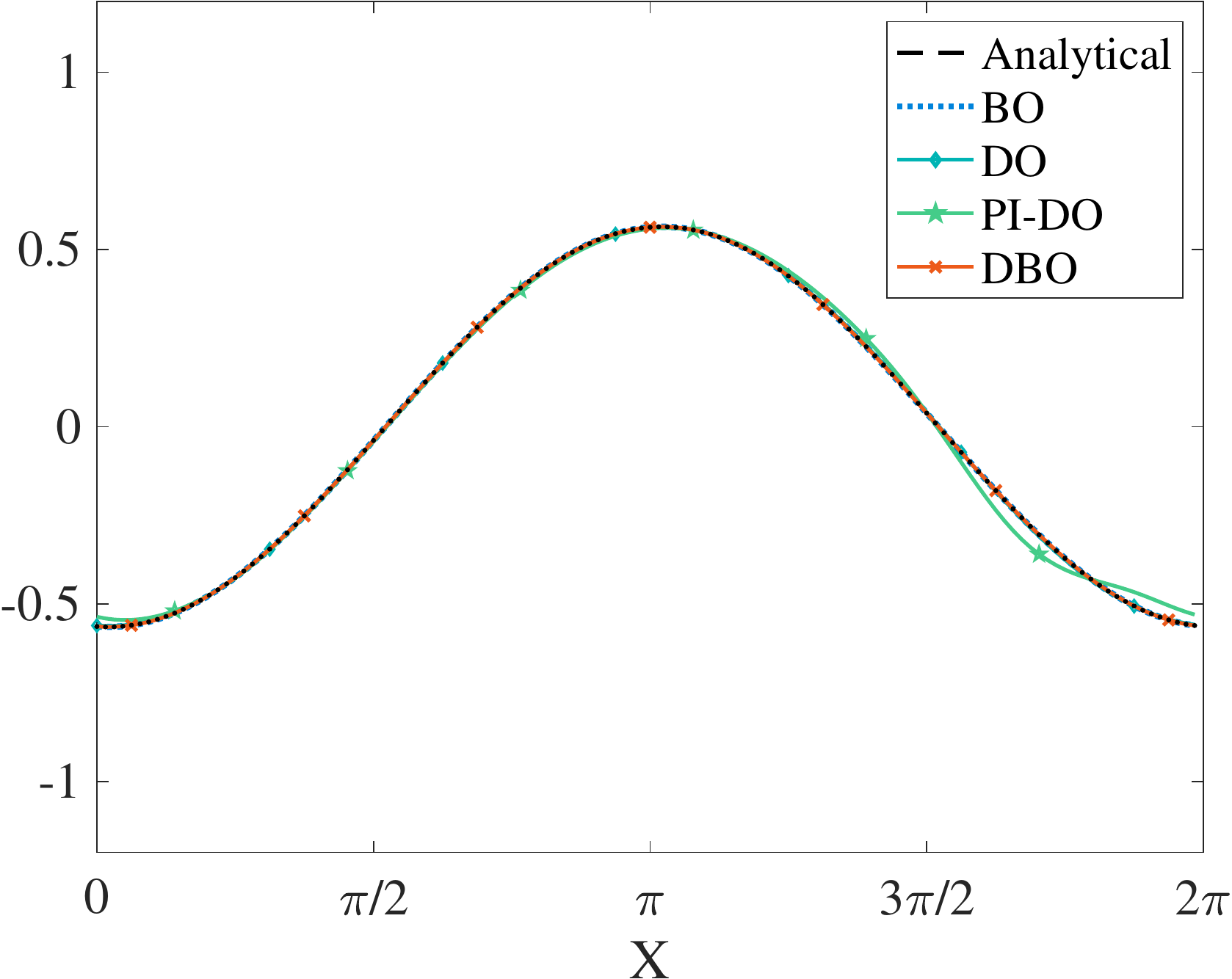}\label{fig:fa16}}
    \subfloat[]{
    \includegraphics[width=\tempwidth, height=0.16\textheight]{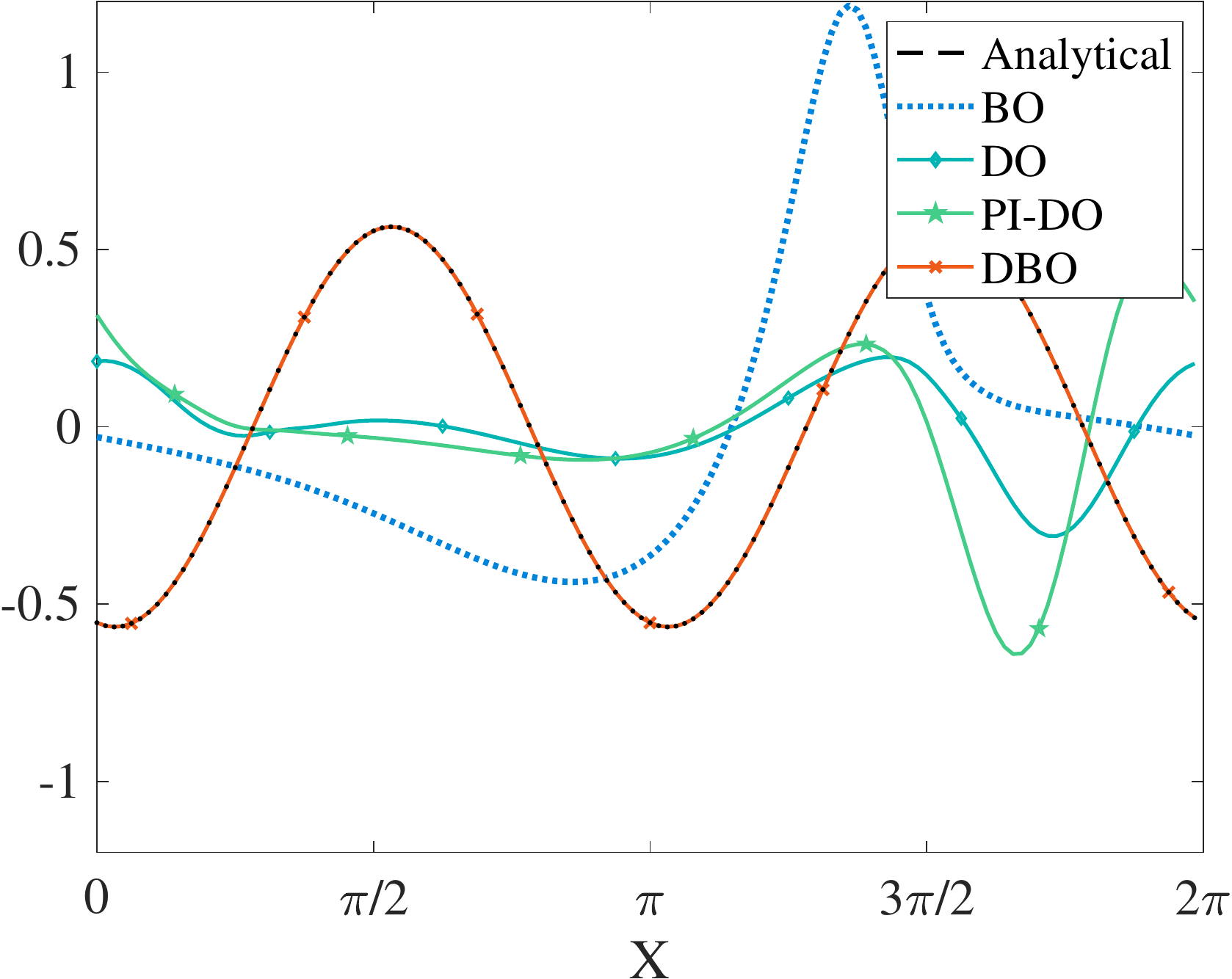}\label{fig:fa17}}
    \subfloat[]{
    \includegraphics[width=\tempwidth, height= 1.0125\tempheight]{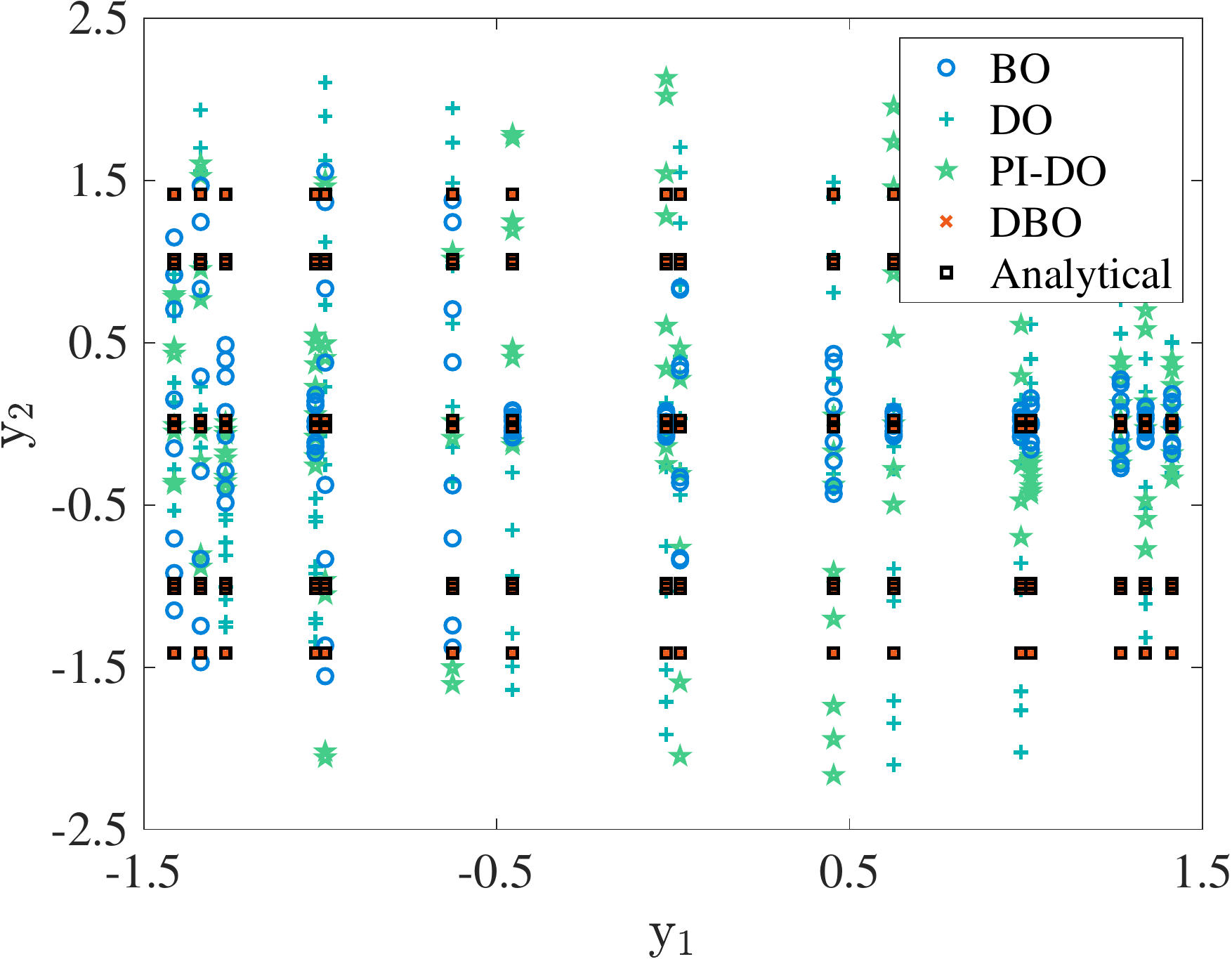}\label{fig:fa18}}\\ 
    \rowname{\small{$t = 5.2$ }}
    \subfloat[]{
    \includegraphics[width=\tempwidth, height=0.16\textheight]{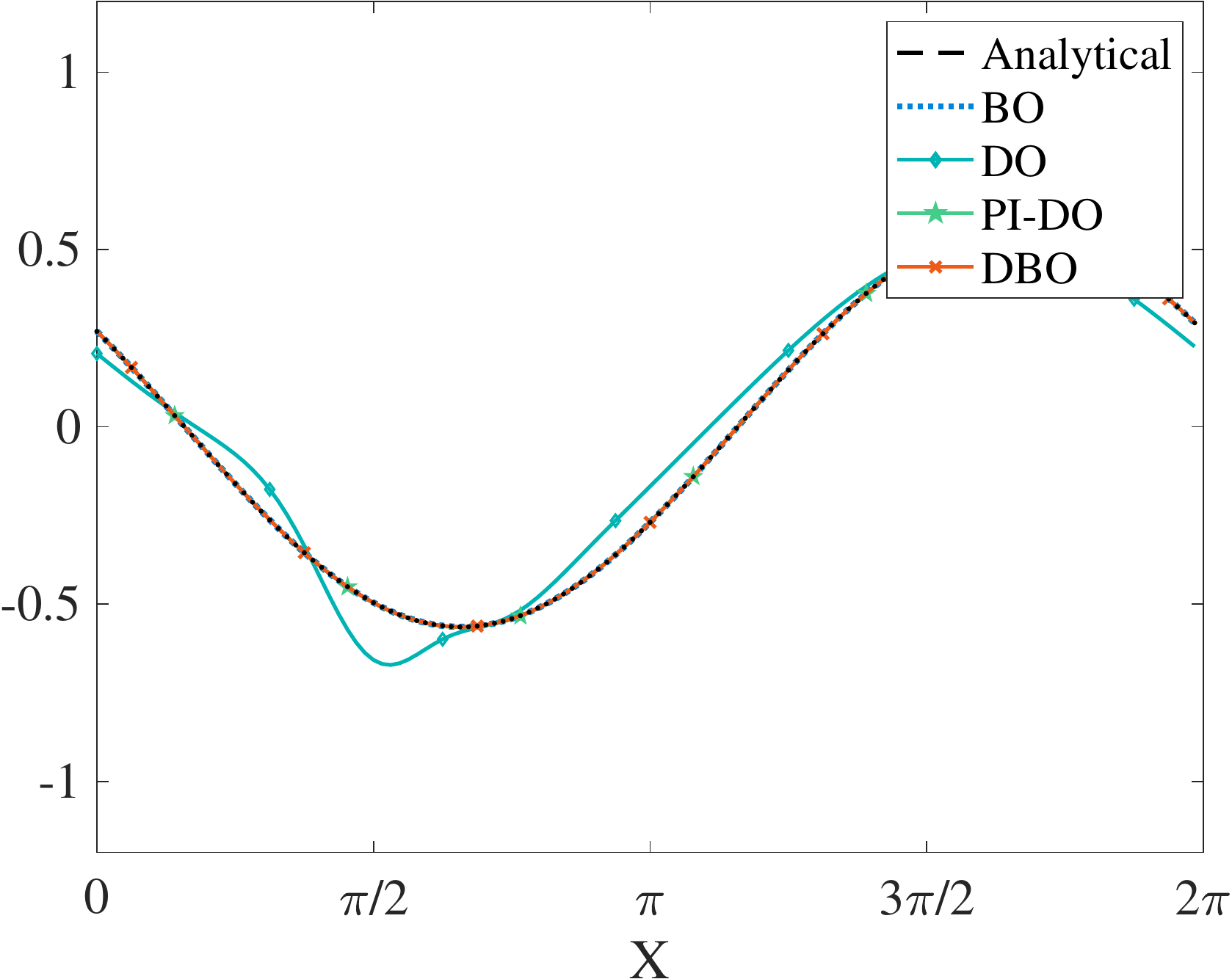}\label{fig:fa19}}
    \subfloat[]{
    \includegraphics[width=\tempwidth, height=0.16\textheight]{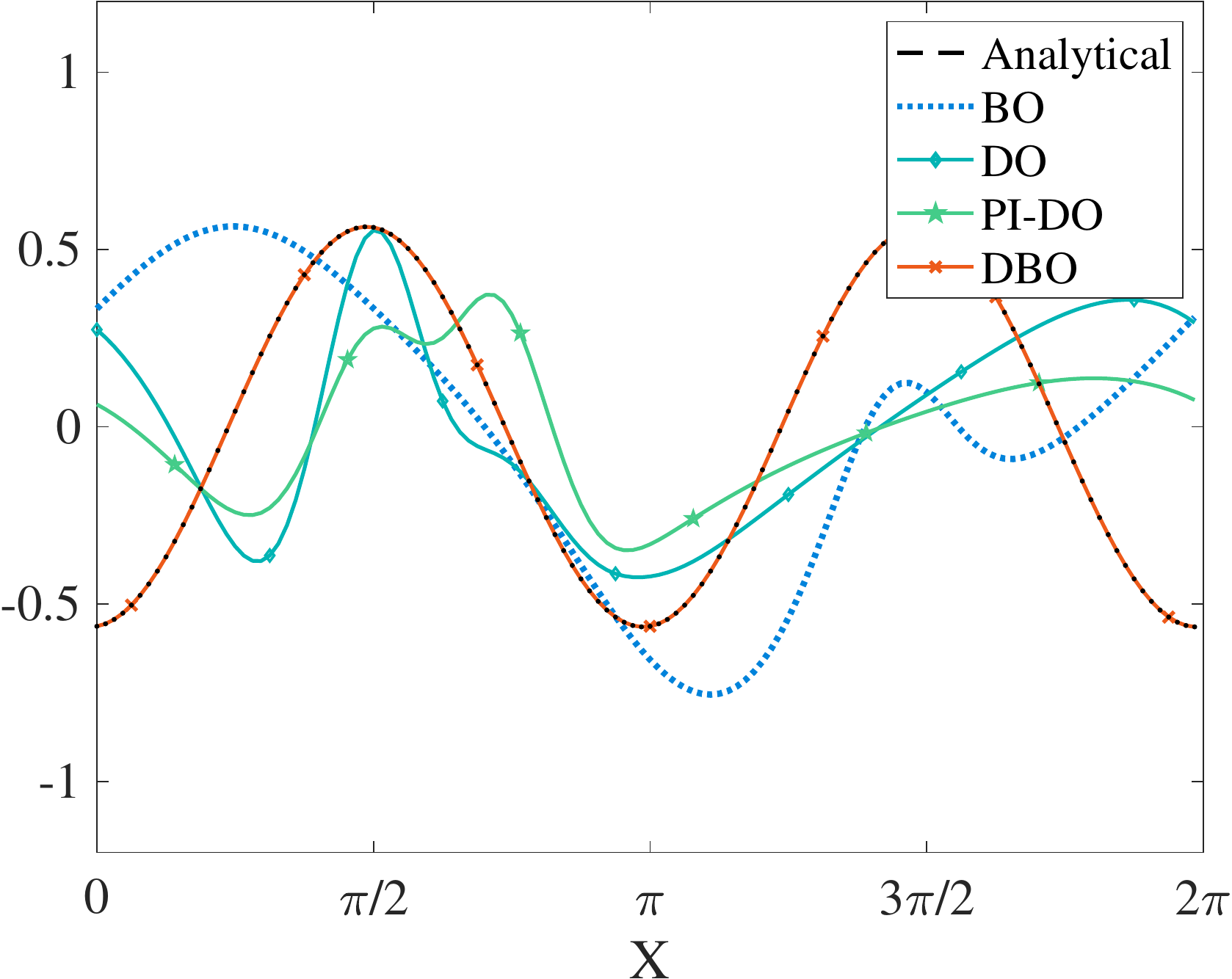}\label{fig:fa20}}
    \subfloat[]{
    \includegraphics[width=\tempwidth, height= 1.0125\tempheight]{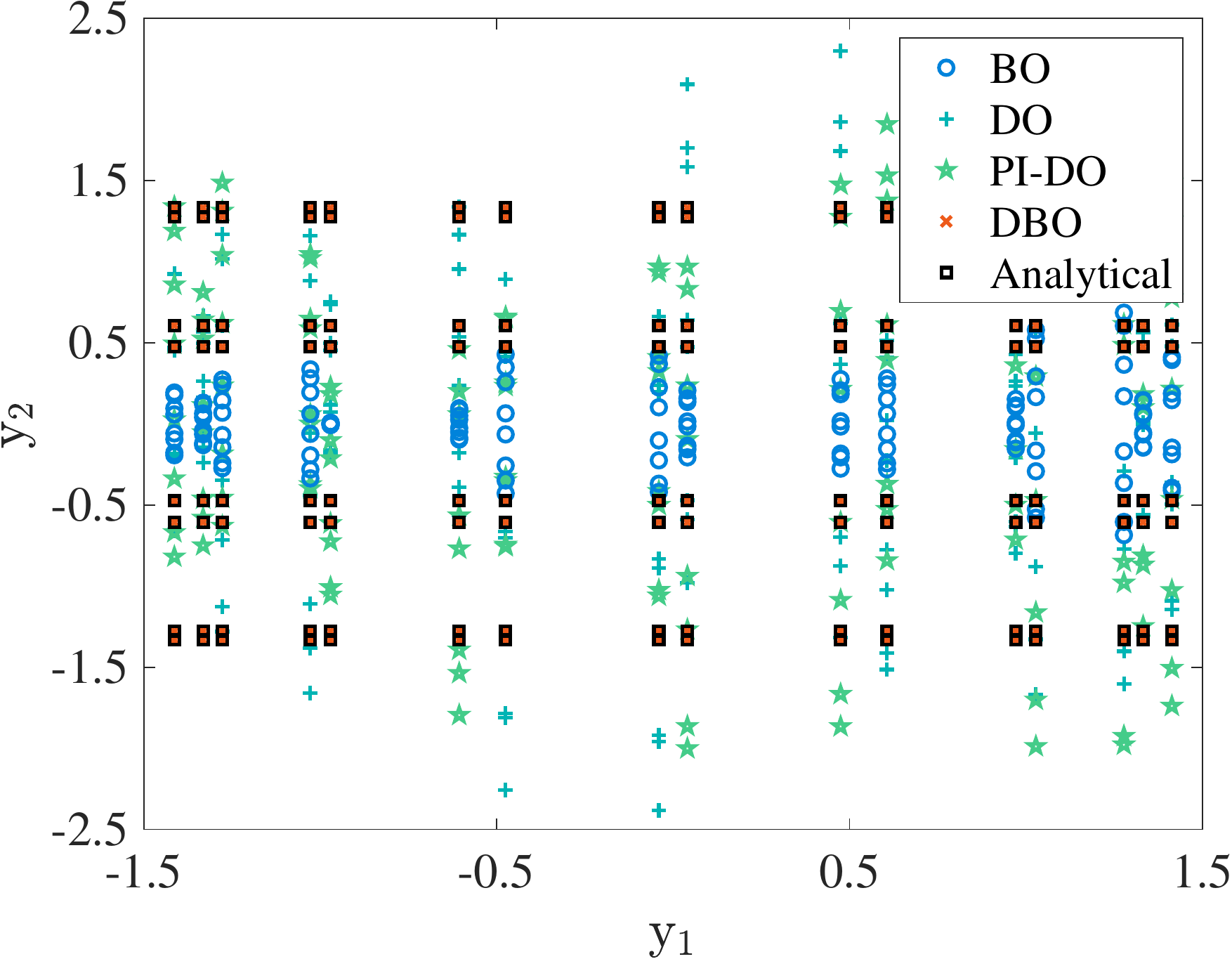}\label{fig:fa21}}\\ 
    \caption{Burgers' equation with manufactured forcing: The two physical modes and the phase space for the stochastic basis are shown at different times as the simulations progresses. The first row shows the modes and phase space at $t=0.1$. All the methods start from the same initial condition. In the second row, the modes and phase space are shown for $t=1.2$. The next rows show the system at $t=3.2$ and $5.2$, respectively. It is observed that the low variance mode is affected first and subsequently as the evolution continues the higher variance mode loses its accuracy as well. The code used in this example is available on GitHub at \url{https://github.com/ppatil1708/DBO.git}. }
    \label{fig:epsiloncompare}
\end{figure}

We have also investigated the effect of the condition number of the system on the spatial and stochastic modes.  In Fig.(\ref{fig:epsiloncompare}), the two  spatial modes and the phase space i.e.,  $y_1(t;\omega)$ vs. $y_2(t;\omega)$,  are shown for four different times: $t=0.2, 1.2, 3.2$ and  $5.2$. At $t=0.2$,  the  spatial modes and stochastic coefficients match well with those of the KL decomposition as  shown in Fig.(\ref{fig:fa10}-\ref{fig:fa12}). However, as time progresses to $t=1.2$  and $t=3.2$ the ability of the BO, DO, and PI-DO to retain the near-singular mode deteriorate as shown in Fig.(\ref{fig:fa14}-\ref{fig:fa15}) and Fig.(\ref{fig:fa17}-\ref{fig:fa18}). At time $t=5.2$, BO, DO, and PI-DO  completely fail to capture the lowest variance mode. Moreover, for both DO and PI-DO, the inability to accurately resolve the low-variance mode adversely affects first mode. See Fig.(\ref{fig:fa16}) and Fig.(\ref{fig:fa19}). 
\subsection{Burgers' equation with stochastic forcing}
In this section, we consider  Burgers' equation subject to random forcing where a large number of modes are needed to resolve the system accurately due to nonlinear interaction between the modes. We investigate the effect of low eigenvalues on the accuracy of the solution and the effect of long time integration on the solution for both DO and the DBO methods. The governing equation is given by: 
\begin{subequations}\label{eq:StocBurgF}
\begin{align}
    \pfrac{u}{t} + u\pfrac{u}{x} &= \nu \pfrac{^2 u}{x^2} + \frac{(1+\xi)}{2} \sin(2\pi t),   &&x \in [0, 2\pi] \quad \mbox{and} \quad  t\in[0,t_f],\\
    u(x,0;\omega) &= g(x) &&x \in [0, 2\pi],
\end{align}
where $ \nu =0.04$ and $\xi \sim \mathcal{U}[-1,1]$ is a one-dimensional uniform random variable and the initial condition is taken to be:
\begin{equation}
    g(x) = 0.5(\exp(\cos(x))-1.5) \sin(x+2\pi\cdot 0.37).
\end{equation}
\end{subequations}
We use the Fourier spectral method  for space discretization with  $N_s =128$ Fourier modes,  and PCM is used for the discretization of the one-dimensional random space $\xi$. We use $N_r=64$ Legendre-Gauss collocation points. The third-order Runge-Kutta scheme is used for evolving the discrete systems in time with $\Delta t = 10^-3$. At $t=0$ the system is deterministic, hence the covariance matrix is singular. Therefore, neither DO nor DBO decompositions can be initialized at $t=0$. To this end, we evolve the stochastic systems up to $t_s=2$ using PCM and the KL decomposition of the solution at this time is taken as the initial condition. This is in accordance to methodology presented in \cite{choi2013convergence}. 

\begin{figure}
    \centering
    \includegraphics[width=0.48\textwidth]{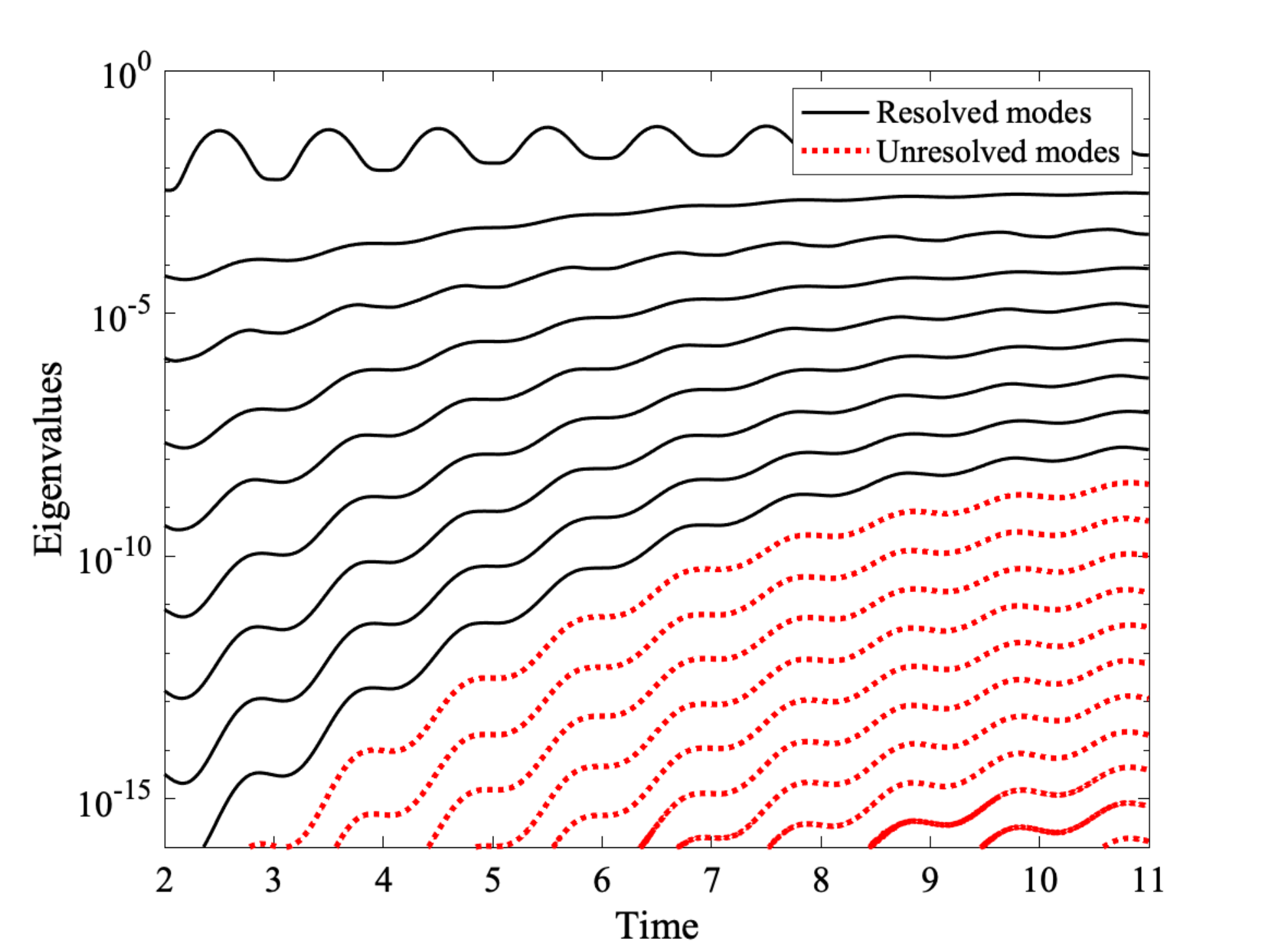}
    \caption{Burgers' equation with stochastic forcing: Growth in the eigenvalues as the system evolves. The modes shown in red dotted lines are the unresolved modes i.e., modes which are not included in the simulations. These eigenvalues are obtained by performing \KL decomposition on the instantaneous samples. }
    \label{fig:EigStocForcing}
\end{figure}

\begin{figure}
            \centering
            \subfloat[Mean error for r=7]{
            \includegraphics[width=0.48\textwidth]{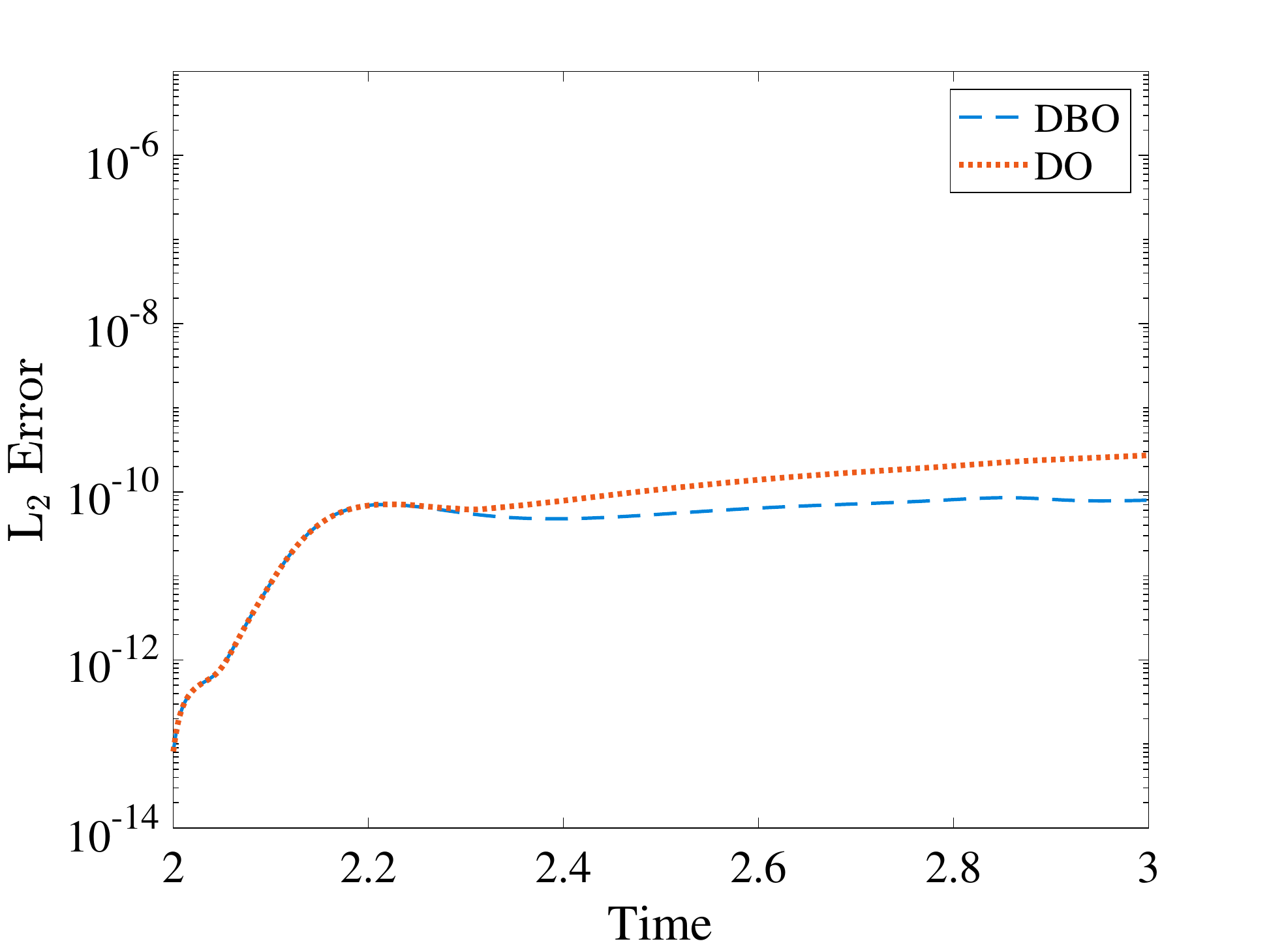}\label{fig:f22}}
            \hfill
            \subfloat[Mean error for r=9]{
            \includegraphics[width=0.48\textwidth]{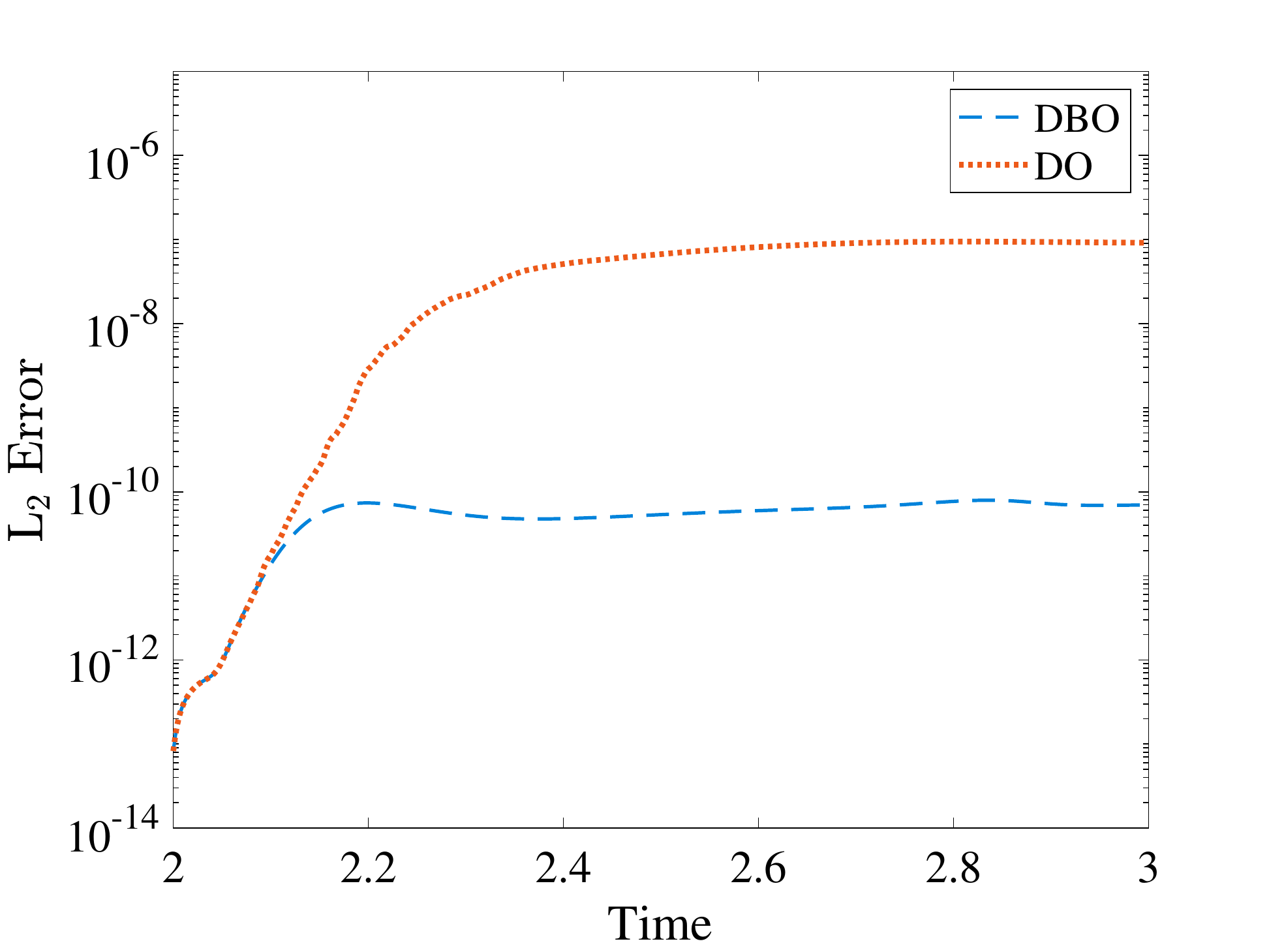}\label{fig:f23}}
            \hfill
            \subfloat[Variance error for r=7]{
            \includegraphics[width=0.48\textwidth]{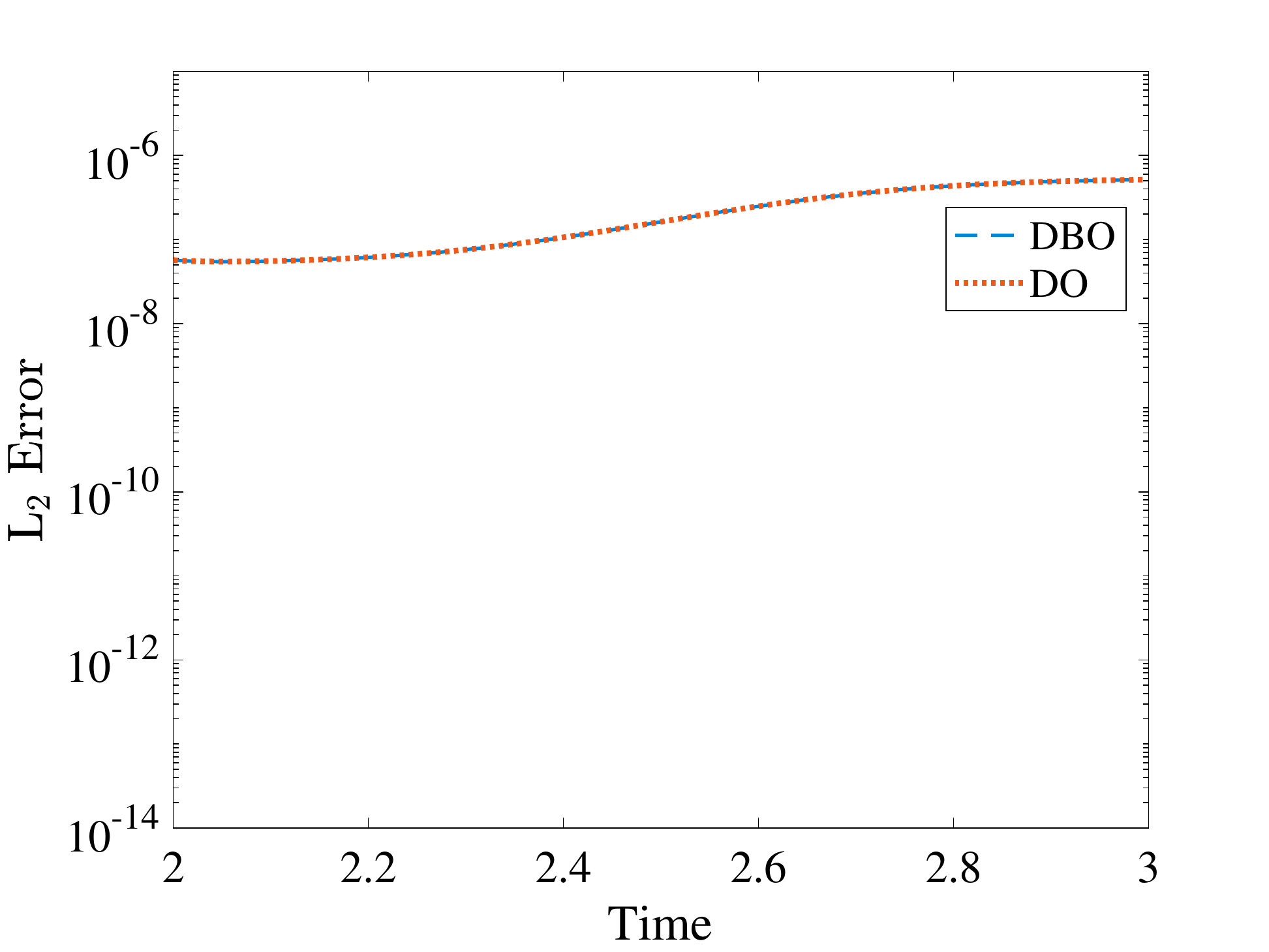}\label{fig:f24}}
            \hfill
            \subfloat[Variance error for r=9]{
            \includegraphics[width=0.48\textwidth]{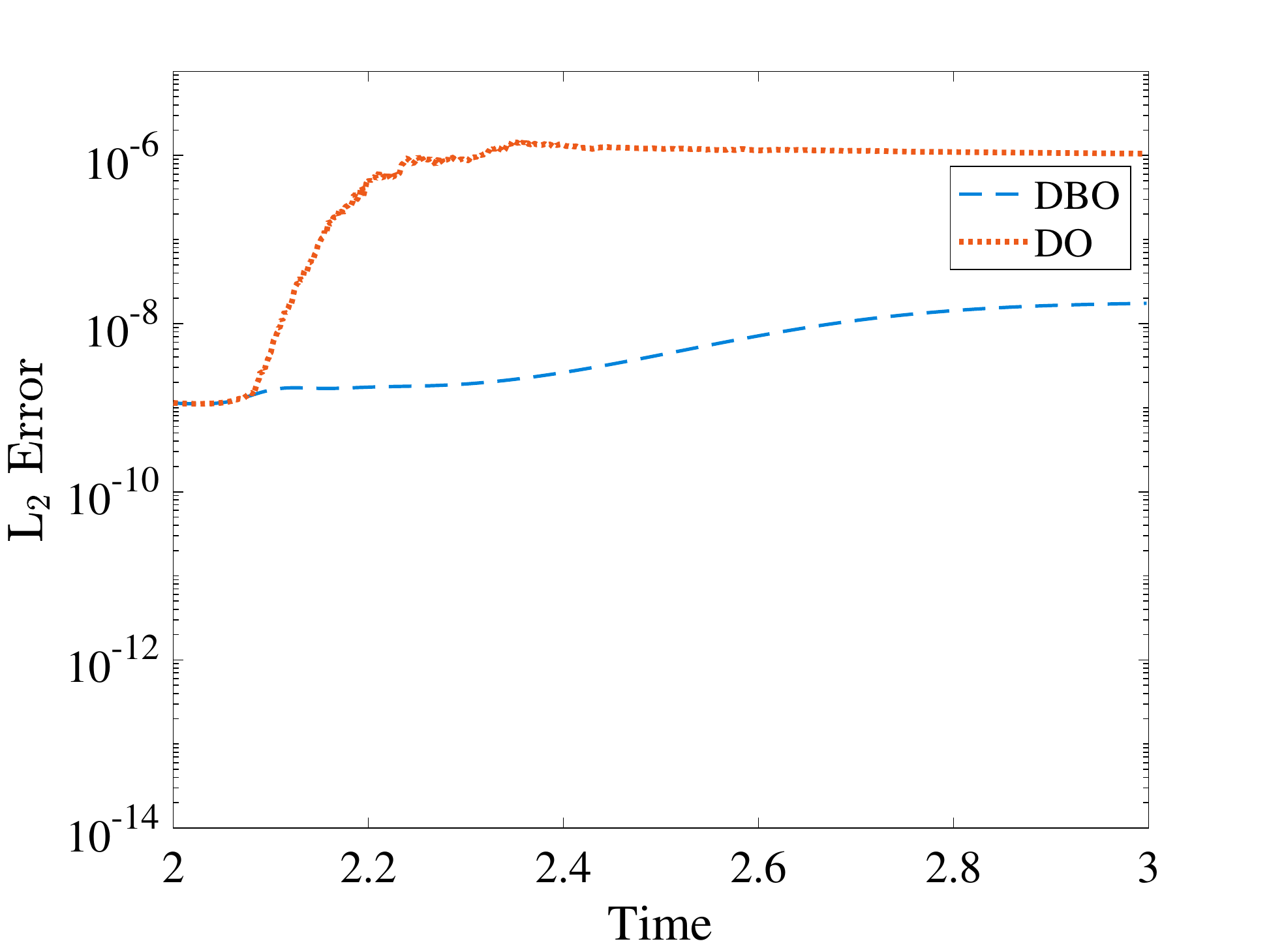}\label{fig:f25}}
            \hfill
            \caption{Burgers' equation with stochastic forcing (effect of low variance modes on the accuracy of the solution): It is observed that effectively resolving the modes with lower variance improves the numerical accuracy of the solution.  The DO method fails to resolve the lower eigenvalues and hence the error for DO is higher than that of the DBO method. The code used in this example is available on GitHub at \url{https://github.com/ppatil1708/DBO.git}}
            \label{fig:StocForcingError}
        \end{figure}

This case is used to study two properties of an ill-conditioned system on the overall accuracy of the mean and variance: (i) effect of low eigenvalues resulting in an ill-conditioned covariance matrix, (ii) effect of unresolved modes on long term integration. To study the effect of low eigenvalues we consider two reduction sizes of $r=7$ and $r=9$ and the system is evolved till $t_f=3$. Fig.(\ref{fig:EigStocForcing}) shows the eigenvalues for this case as extracted from the PCM solution. It is observed that modes 8 and 9 (shown in red) have eigenvalues which are the order of $10^{-15}$, rendering the covariance matrix $C$ highly ill-conditioned. The mean error for reduction sizes $r=7$ and $r=9$  can be seen in Fig.(\ref{fig:f22}-\ref{fig:f23}), respectively. The variance error is plotted in Fig.(\ref{fig:f24}-\ref{fig:f25}). 
It can be seen that the lower modes affect the accuracy of the solution for DO. The error affects the solution of the higher modes and we observe an increased error for the DO method in case of reduction order $r=9$.  The DBO method, on the other hand, resolves the lower mode accurately without affecting the accuracy of the higher modes. In fact adding additional modes, improves the accuracy of the DBO solution as seen from the variance error plots in Fig.(\ref{fig:f25}).  
        \begin{figure}
            \centering
            \subfloat[Mean error]{
            \includegraphics[width=0.48\textwidth]{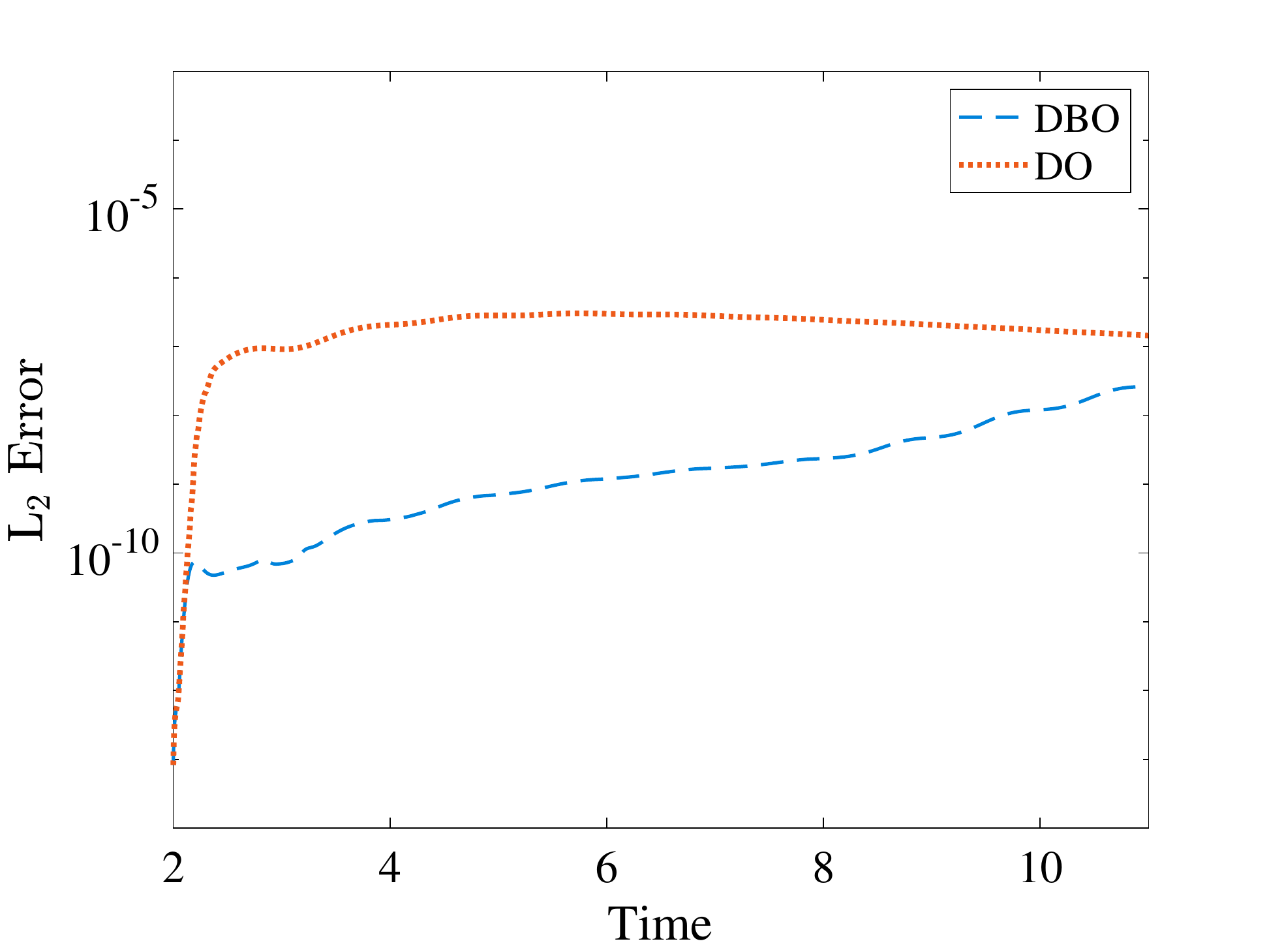}}
            \hfill
            \subfloat[Variance error]{
            \includegraphics[width=0.48\textwidth]{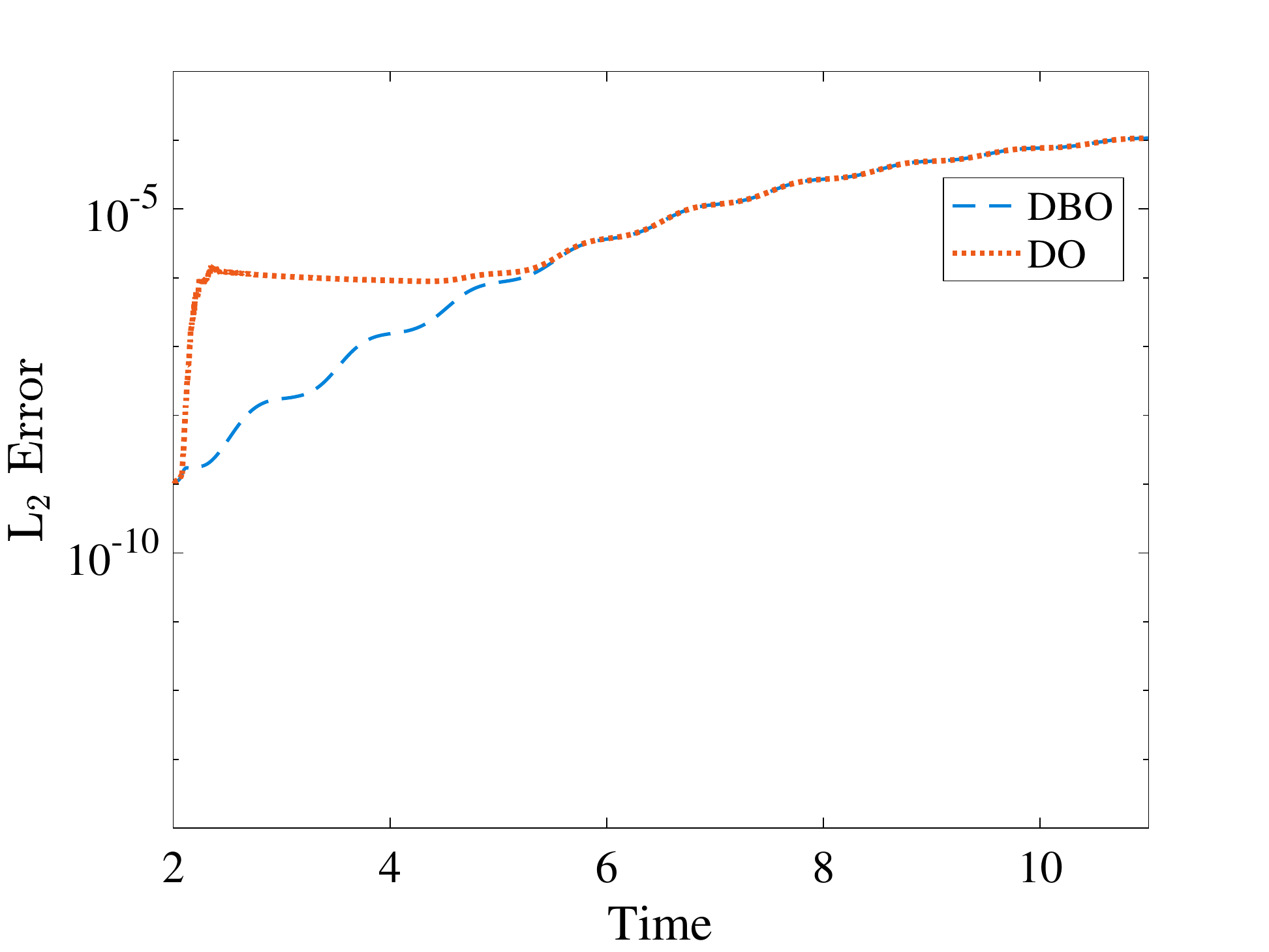}}
            \hfill
            \caption{Burgers' equation with stochastic forcing (long time integration effects): The 9 dominant modes are used to resolve the system. The mean error and variance error for DBO and DO as compared with PCM are shown in (a) and (b). It is observed that DBO performs better for short time (i.e., till 4 time units). After 4 time units  the lower unresolved modes gain variance and the effect of these unresolved modes dominate the error which is equal for both DO and DBO methods. The code used in this example is available on GitHub at \url{https://github.com/ppatil1708/DBO.git}.}
            \label{fig:LongtimeStocForcing}
        \end{figure}
        
The solutions for the long time integration case for the stochastic Burgers' equation is shown in Fig.(\ref{fig:LongtimeStocForcing}). Between $t=2$ and $t=3$, we observe that the DO has higher error as the lower modes affect the accuracy of the higher modes. This result is same as seen from the previous case Fig.(\ref{fig:StocForcingError}). As the lower modes start gaining energy, the error from the unresolved modes dominates the error of the effect of lower modes and hence, we observe that the error for both the DO and the DBO methods is the same as time progresses. 

\subsection{Stochastic incompressible Navier-Stokes: Flow over a bump} \label{Sec:SNS}
\begin{figure}
    \subfloat[Schematic of the flow ]{\includegraphics[width=0.355\textwidth]{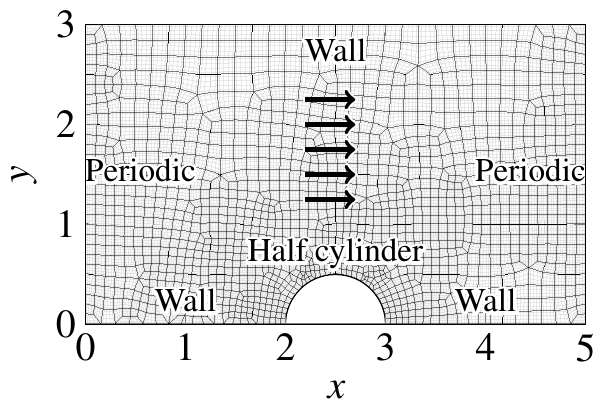}\label{fig:schematic}}
    \subfloat[First POD mode: ${\Phi}_{y_1}(x,y)$]{\includegraphics[width=0.317\textwidth]{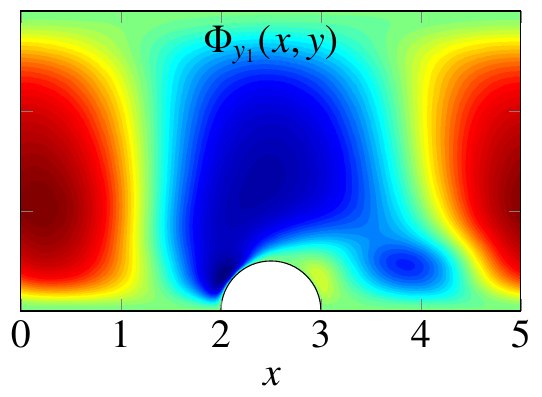}\label{fig:PODMode1}}
    \subfloat[First POD mode: ${\Phi}_{y_2}(x,y)$]{\includegraphics[width=0.317\textwidth]{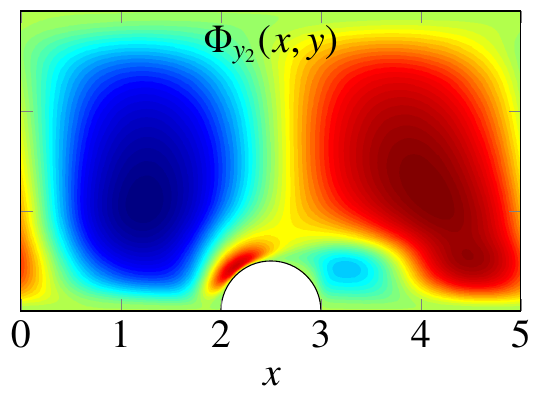}\label{fig:PODMode2}}
    \caption{Flow over a bump in a channel flow: (a) The schematic of the problem and the mesh for the spectral/hp element. (b) and (c)  The $y$-velocity component of the two dominant POD modes.}
\end{figure}

In this example, we apply the DO and DBO decompositions to solve stochastic incompressible   Navier-Stokes equations. The governing equations are given by:
\begin{subequations}
\begin{equation}
\frac{\partial \mathbf{u}}{\partial t} + \left(\mathbf{u} \cdot \nabla \right) \mathbf{u} = -\frac{1}{\rho}\nabla p + \nu \nabla^2 \mathbf{u}   + \mathbf{f},
\end{equation}
\begin{equation}
\nabla \cdot \mathbf{u} = 0. 
\end{equation}
\end{subequations}
where: $\bm{u}=(u_x, u_y)$ is the velocity vector field, $\mathbf{f}=(f_x,f_y)=(1,0)$ is the forcing and $p$ is the pressure field.
We solve the flow over a bump in a channel as shown in Fig.(\ref{fig:schematic}), where flow is from left to  right. Periodic boundary condition is imposed in the streamwise direction and no-slip boundary condition is imposed at the bottom and top walls. We consider $\nu=0.04$ and $\rho=1$ and the Reynolds number is based on the channel height and time-averaged centerline horizontal velocity which is roughly equal to $Re=1500$. For these parameters the flow in not chaotic, but it is time dependent due to constant shedding of separated region behind the bump. The stochasticity is introduced in the flow via random initial conditions given by the following equation:
\begin{equation} 
 \mathbf{u}(x,y,0;\omega) = \mathbf{u}_0(x,y)+\sum_{i=1}^d \sigma  \xi_i(\omega) \bm{\Phi}_i(x,y),
\end{equation}
where $\mathbf{u}_0(x,y)$ is the solution of a deterministic simulation at $t=50$. The deterministic solution at this time has reached the statistically steady state.  In the above initial condition  $\mathbf{\Phi}_i=(\Phi_{x_i}, \Phi_{y_i})$ are the proper orthogonal decomposition (POD) modes obtained from the deterministic simulation of the flow over a bump at $Re=1500$. We consider $d=2$ and the $\Phi_y$ component of the  two corresponding POD modes are shown in  Fig.(\ref{fig:PODMode1}-\ref{fig:PODMode2}). 
For the spatial discretization of the mean flow and the spatial basis,  we  use  spectral/hp element method with quadrilateral elements for $N_e= 1451$ and polynomial order 5. The spectral element mesh is shown in  Fig.(\ref{fig:schematic}). A first-order time-splitting scheme is used  for the evolution of mean and the spatial basis, in which the nonlinear terms are treated explicitly and the diffusion terms are treated implicitly. The  time-integration step of $\Delta t=10^{-4}$ is used. The random space is two-dimensional and discretization of the stochastic coefficients in the random space is performed using ME-PCM  with 4 elements in each random direction and  4 quadrature points in each element. Therefore, the total number of quadrature points in every direction of the random space is 16 and  hence, the total  number of quadrature points in the two dimensional random space is $N_r=16^2=256$. We solved both DO and DBO systems with identical discretization schemes as described above till $t_f=5$, which amounts to 20 flow through periods.

\begin{figure}
      \centering
     \subfloat[Reduction size $r=2$]{\includegraphics[width=0.49\textwidth]{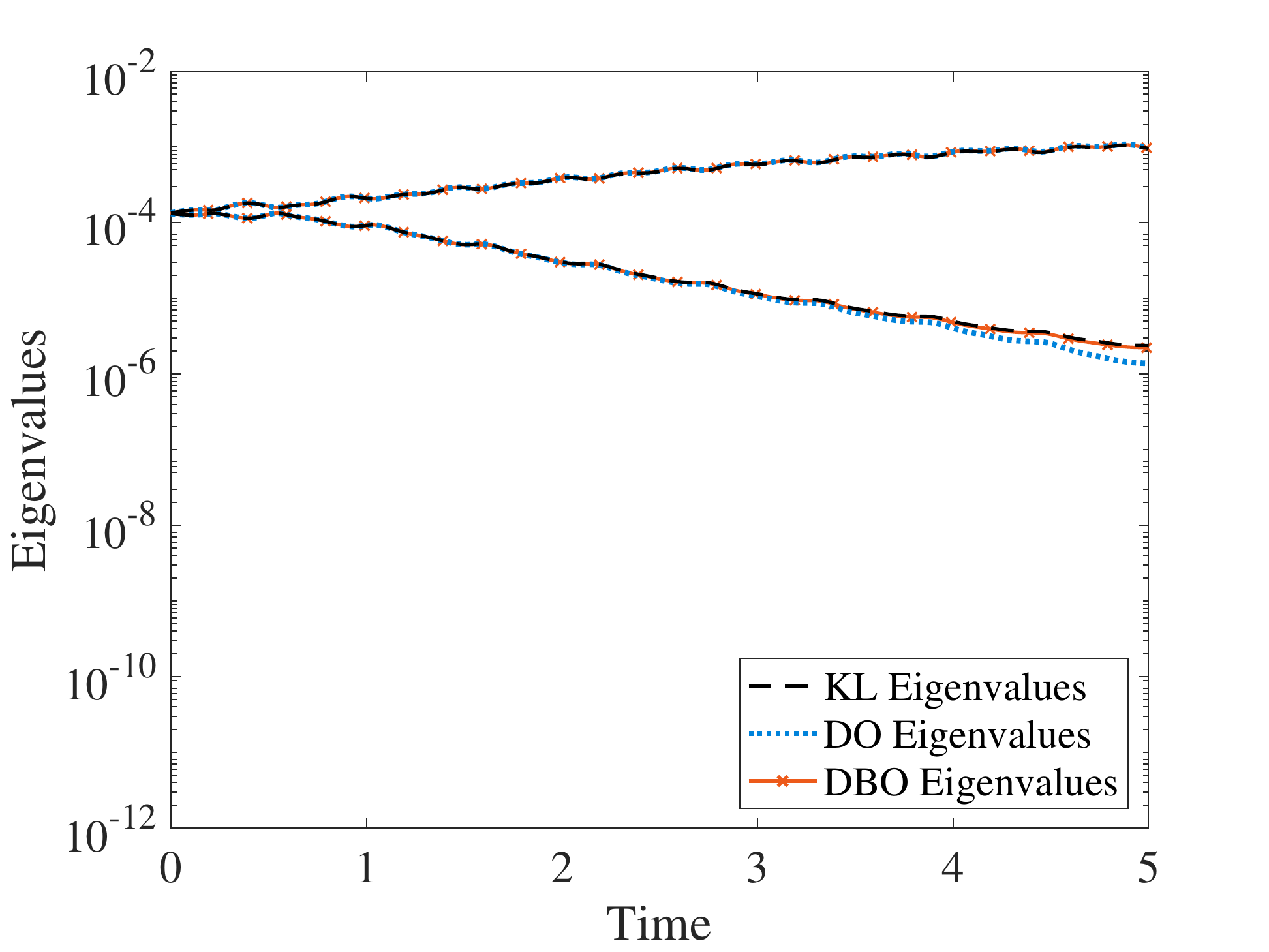}\label{fig:f26}}
      \subfloat[Reduction size $r=3$]{\includegraphics[width=0.49\textwidth]{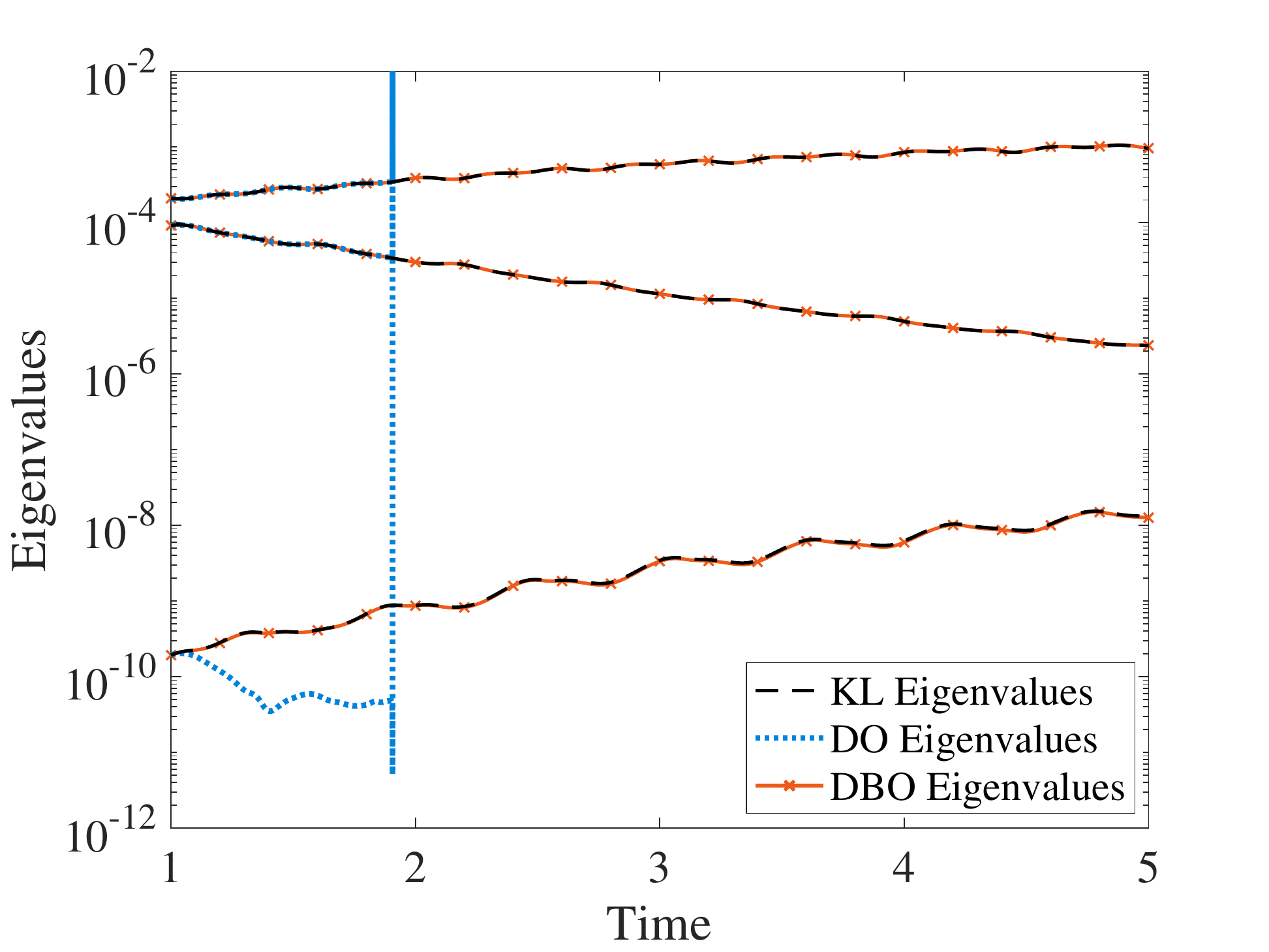}\label{fig:f28}\label{fig:f27}}
      \hfill
      \caption{Flow over a bump in a channel: A comparison between eigenvalues for two reduction orders $r={2,3}$ between KL, DO and DBO. For $r=3$, it is observed that the DO method is not able to resolve lower modes when the condition number for inverting the covariance matrix is high and it eventually diverges, whereas the DBO does not have the aforementioned issue due to a better condition number for $\Sigma$ inversion hence can resolve low variance modes with better accuracy. }
\end{figure}

To compare the performance of DO and DBO we performed  simulations for two reduction sizes: $r=2$ and $r=3$. For the reference solution, we  performed 256 non-intrusive direct numerical simulation (DNS) at the same ME-PCM quadrature points. We then performed KL decomposition of the 256 sample at each time step.
The eigenvalues of the covariance matrix of  DO, DBO for the case of $r=2$ and the two largest  KL eigenvalues  are shown in Fig.(\ref{fig:f26}).  It is clear that both methods perform well and match the two most energetic KL modes, although the eigenvalues of DBO are more accurate than that of the DO. 

In the case of $r=3$, the eigenvalue associated with the third mode has very small values. In fact at $t=0$ the third eigenvalue is zero. This eigenvalue gradually grows due to nonlinearity of Navier-Stokes equations.  To avoid an exact singularity, the DO and DBO simulations for $r=3$  are initialized at $t=1$ from the solution of the corresponding KL decomposition.  The system is ill-conditioned for $r=3$ due to the low variance of the third mode. At $t=1$, the third eigenvalue is roughly equal to $10^{-10}$  as shown in Fig.(\ref{fig:f27}). The third eigenvalue of the DO decomposition deviates from the truth due to the near singularity and it eventually leads to the divergence of the DO system, while  DBO performs accurately and all three eigenvalues match those of the KL.

\begin{figure}
    \centering
    \setlength{\tempwidth}{0.22\textwidth}
    \settoheight{\tempheight}{\includegraphics[width=\tempwidth, height=0.833in]{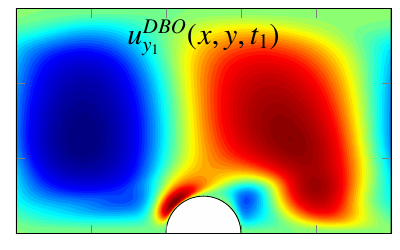}}%
    \columnname{{$\bar{u}_y(x,y)$}}
    \columnname{{$\bar{u}_{y_1}(x,y)$}}
    \columnname{{$\bar{u}_{y_2}(x,y)$}}
    \columnname{{$\bar{u}_{y_3}(x,y)$}} \\
    \rowname{\scriptsize{DBO  ($t = 1$)}}
    \includegraphics[width=0.23\textwidth, height=0.833in]{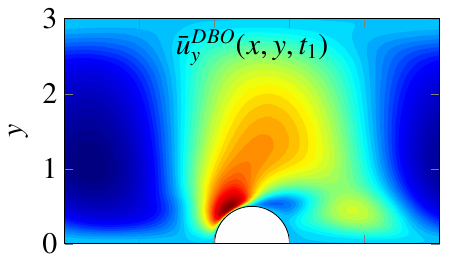}
    \includegraphics[width=0.22\textwidth]{TikzFiles/Evolve2.pdf}
    \includegraphics[width=0.22\textwidth]{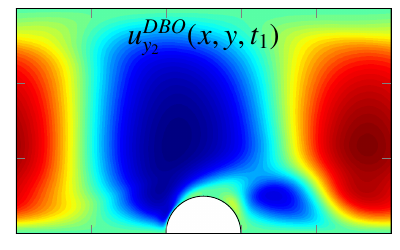}
    \includegraphics[width=0.22\textwidth]{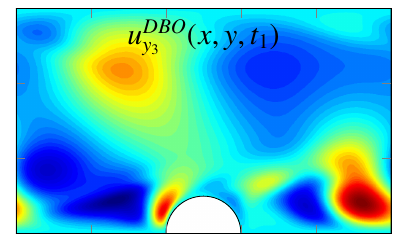}\\
    \rowname{\scriptsize{KL ($t = 1$)}}
    \includegraphics[width=0.23\textwidth, height=0.833in]{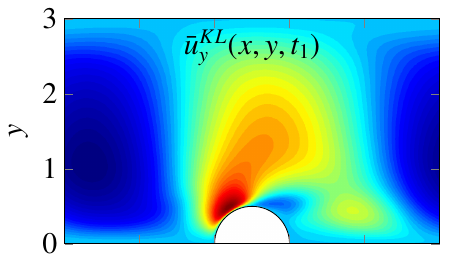}
    \includegraphics[width=0.22\textwidth]{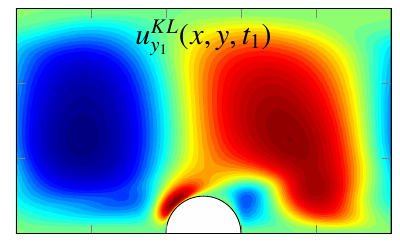}
    \includegraphics[width=0.22\textwidth]{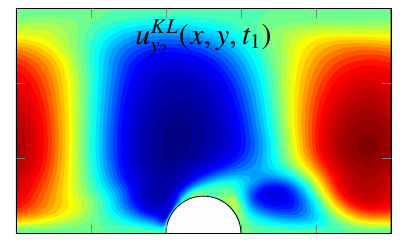}
    \includegraphics[width=0.22\textwidth]{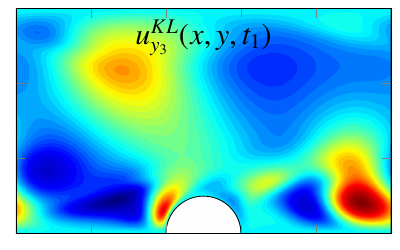}\\
    \rowname{\scriptsize{DBO  ($t = 2$)}}
    \includegraphics[width=0.23\textwidth, height=0.833in]{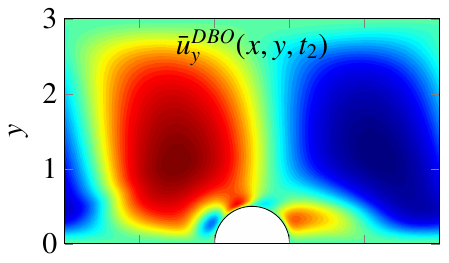}
    \includegraphics[width=0.22\textwidth]{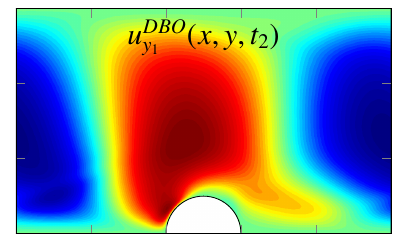}
    \includegraphics[width=0.22\textwidth]{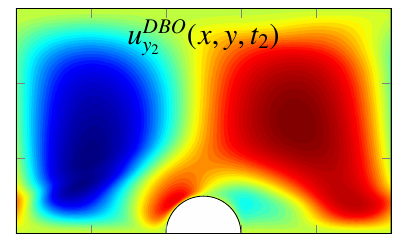}
    \includegraphics[width=0.22\textwidth]{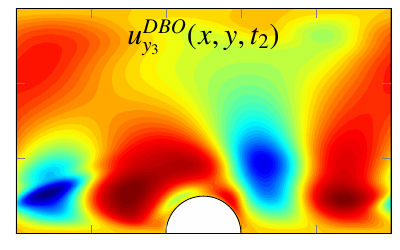}\\
    \rowname{\scriptsize{KL  ($t = 2$)}}
    \includegraphics[width=0.23\textwidth, height=0.833in]{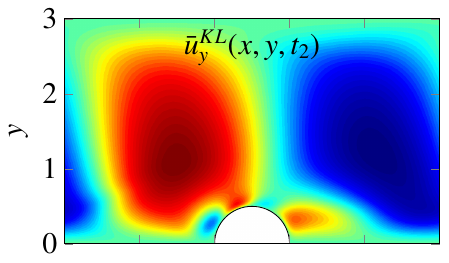}
    \includegraphics[width=0.22\textwidth]{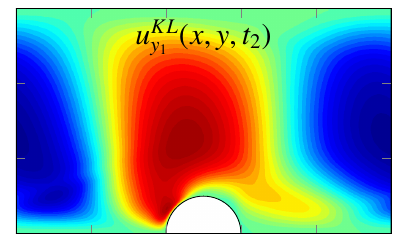}
    \includegraphics[width=0.22\textwidth]{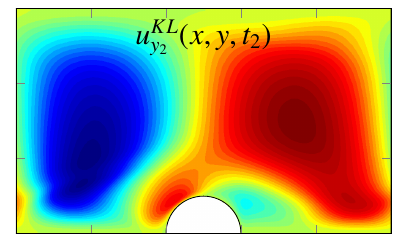}
    \includegraphics[width=0.22\textwidth]{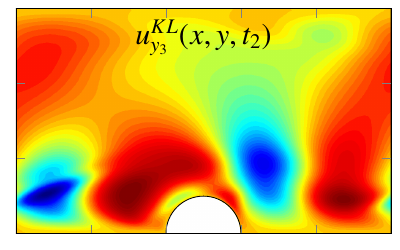}\\
    \rowname{\scriptsize{DBO  ($t = 3$)}}
    \includegraphics[width=0.23\textwidth, height=0.833in]{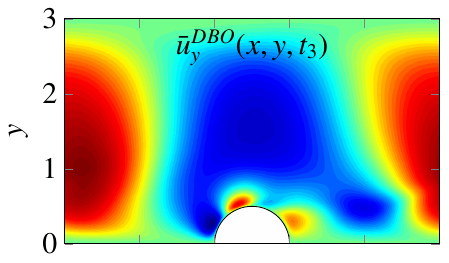}
    \includegraphics[width=0.22\textwidth]{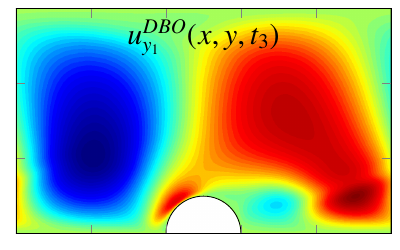}
    \includegraphics[width=0.22\textwidth]{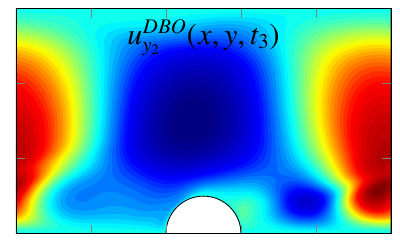}
    \includegraphics[width=0.22\textwidth]{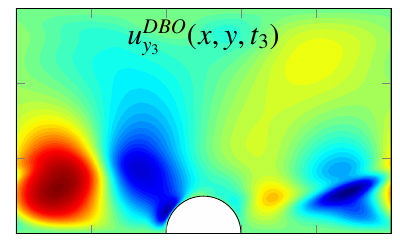}\\
    \setlength{\tempwidth}{0.22\textwidth}
    \settoheight{\tempheight}{\includegraphics[width=\tempwidth, height=0.965in]{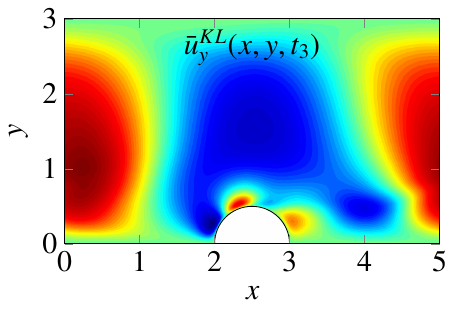}}%
    \rowname{\scriptsize{KL  ($t = 3$)}}
    \includegraphics[width=0.23\textwidth, height=0.965in]{TikzFiles/Evolve9KL.pdf}
    \includegraphics[width=0.22\textwidth]{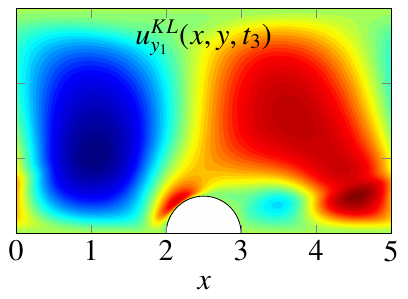}
    \includegraphics[width=0.22\textwidth]{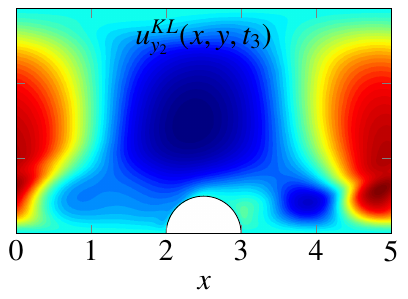}
    \includegraphics[width=0.22\textwidth]{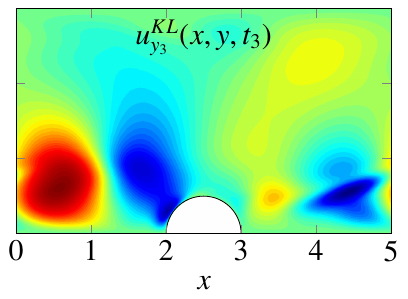}
    \caption{Flow over a bump in a channel flow: The spatial modes of DBO and KL for the stochastic flow in a channel with bump are visualized for comparison in the figure above. Column 1: The $\bar{u}_y(x,t)$ for different time instants. Column 2, 3 \& 4: The three dominant spatial modes for the DBO and KL simulation. Rows 1 and 2 correspond to the DBO and KL spatial modes for $t=1$ respectively. Rows 3 and 4 correspond to the DBO and KL spatial modes at $t=2$ respectively. Finally, rows 5 and 6 correspond to the DBO and KL spatial modes at $t=3$ respectively. }
    \label{fig:f30}
\end{figure}

Fig.(\ref{fig:f30}) shows evolution of the $u_y$ of the mean and three dominant spatial modes of the DBO and KL system at $t=1,2$ and $3$. By visual comparison we can observe that the KL and DBO modes are similar at every time step. Mode 1 and 2 of the system are the POD modes we have used as an initialization for the stochastic random conditions, convected through the channel by the mean velocity, $\bar{u}_x(x,y,t)$ of the flow. It is necessary to consider the lower eigenvalues into the flow field as we observe that overtime the lower eigenvalues can gain energy and alter the system dynamics. 


\section{Summary}\label{sec:Sum}
\begin{figure}
    \centering
    \subfloat[Chaos test]{\includegraphics[width=0.49\textwidth]{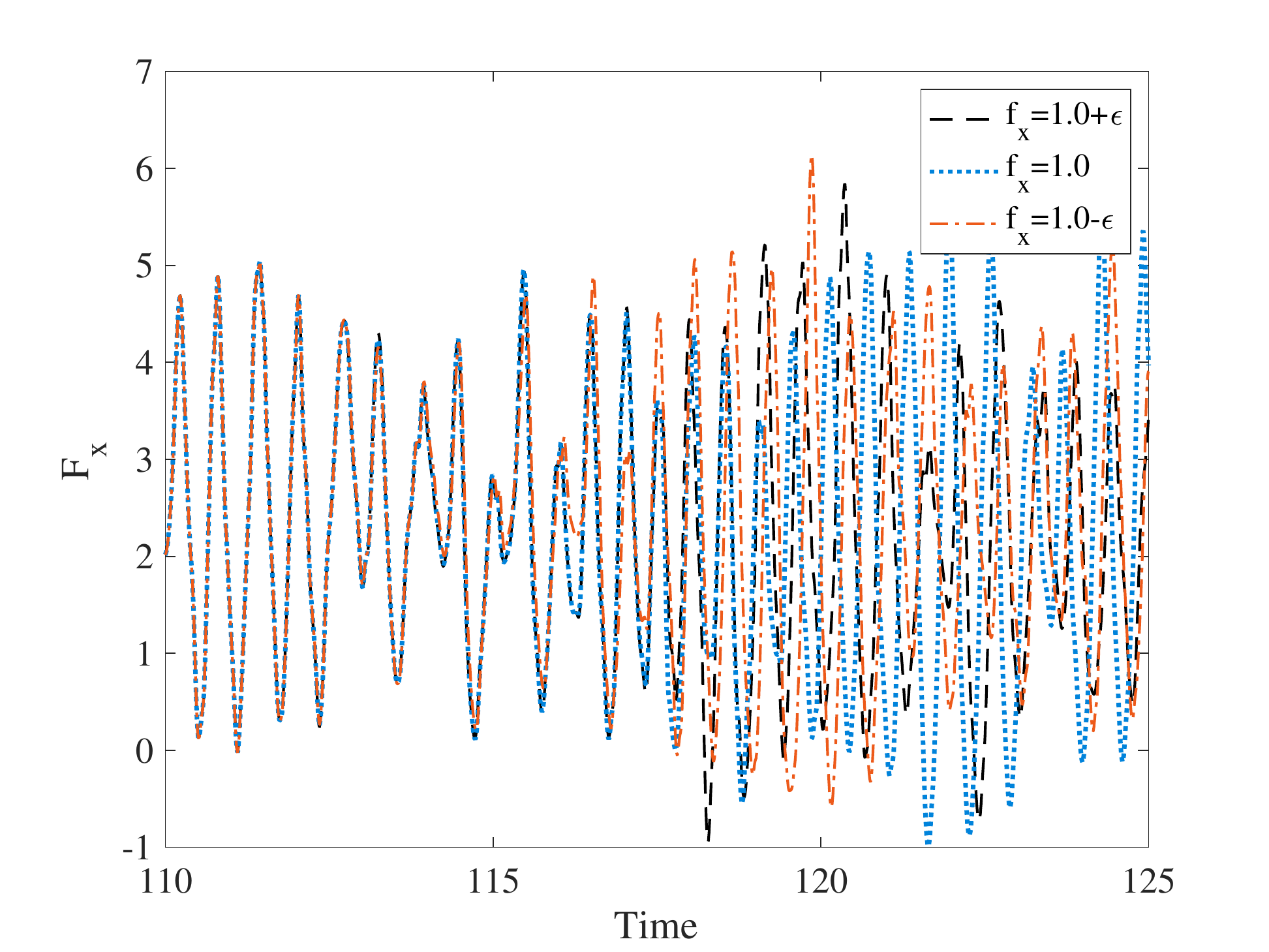}\label{fig:f31}}
    \subfloat[2 Modes eigenvalue comparison]{\includegraphics[width=0.49\textwidth]{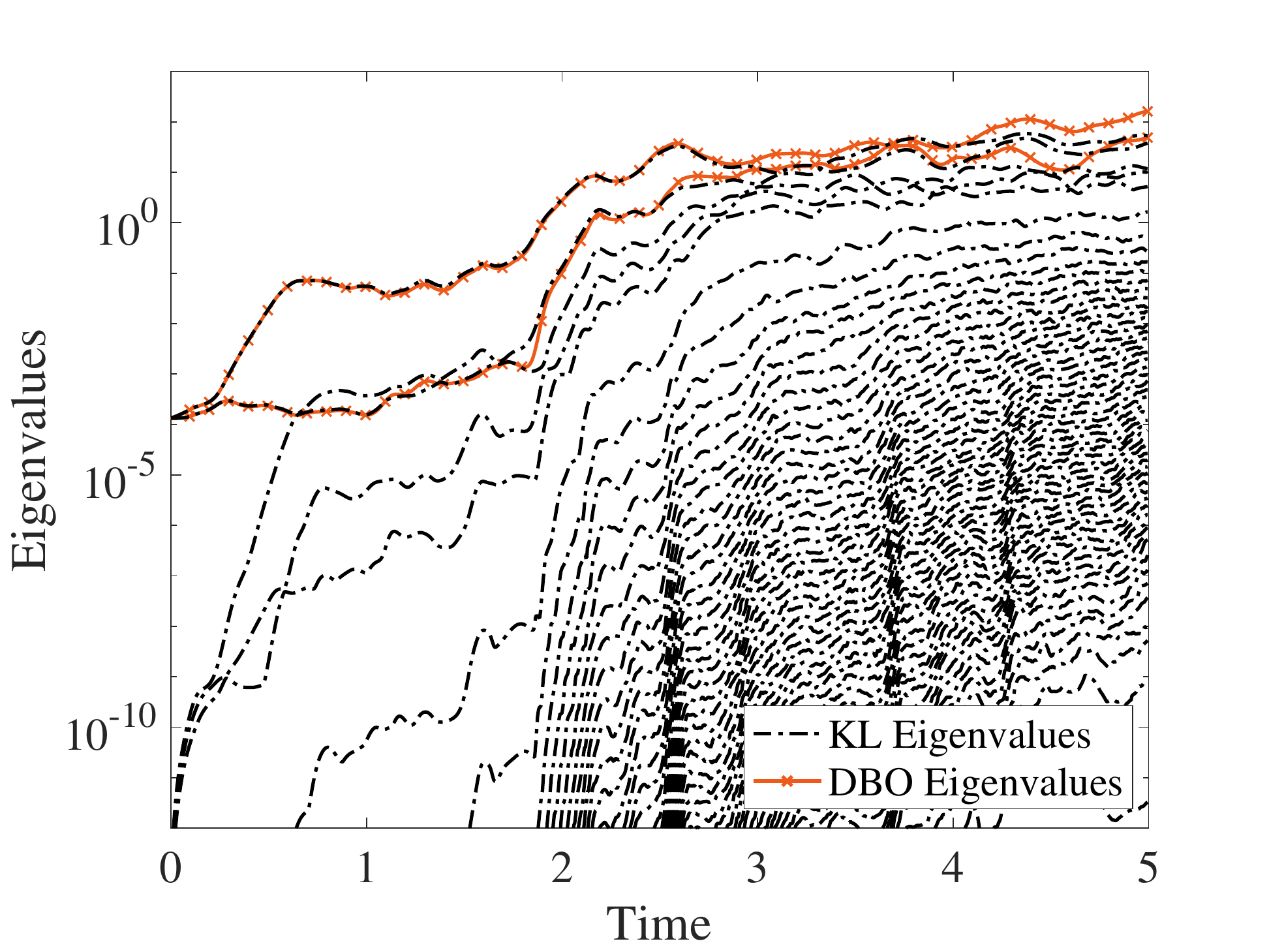}\label{fig:f32}}
    \caption{Dynamically bi-orthonormal decomposition for flow over a bump in a channel in chaotic regime: (a) The growth of the  small perturbations in the forcing measured by the horizontal viscous shear force on the walls. The signals are observed to completely diverge after $t=116$. (b) The growth in the eigenvalues of the DBO system with $r=2$ and the eigenvalues  of the \KL decomposition.}
    \label{fig:fchaos}
\end{figure}
In this paper, we present a new real-time reduced order modeling methodology called the \emph{dynamically bi-orthonormal} (DBO) decomposition  for solving stochastic partial differential equations. The presented method  approximates  a random field by decomposing it to a set of time-dependent orthonormal spatial basis,  a set of time-dependent orthonormal stochastic basis and a low-rank factorization of the covariance matrix. 
 We  derived closed form evolution equations for above components of the decomposition as well as the time-dependent mean field.
 
 We  show that the presented method is equivalent to the dynamically orthogonal and bi-orthonormal decompositions via an invertible  matrix transformation. We derive evolution equation for these  transformation matrices. Although DBO is equivalent to both DO and BO decompositions, it exhibits superior numerical performance especially in highly ill-conditioned systems. In both BO and DO decompositions, the condition number of covaraince matrix, whether diagonal (BO) or full (DO), is $\lambda_{max}(t)/\lambda_{min}(t)$, where $\lambda_{min}(t)$ and  $\lambda_{max}(t)$  are the smallest and largest eigenvalues of the covariance matrix, respectively. However, in the DBO decomposition, a factorization of the covariance matrix ($\Sigma(t)$) is inverted, and $\Sigma(t)$   has the condition number of  $\sqrt{\lambda_{max}(t)/\lambda_{min}(t)}$. The improvement in the condition number of the DBO systems compared with BO or DO is important for adaptive reduced order modeling as the newly added or removed mode has very small eigenvalues. The DBO decomposition  tolerates significantly smaller eigenvalues compared to BO and DO without degrading the accuracy. Moreover, in comparison with BO, DBO does not become singular in the case of eigenvalue crossing, and in comparison with DO, the DBO stochastic coefficients are orthonormal, resulting in better-conditioned representation of the stochastic subspace compared to that of DO.  
 
We demonstrated the DBO decomposition for several benchmark SPDEs: (i)  linear advection equation (ii) Burgers' equation with manufactured solution, (iii) and Burgers' equation with random initial condition. We also applied DBO to stochastic incompressible   Navier-Stokes equation. We  compared  the performance of DBO against BO and DO. We  conclude that for well-conditioned cases, the numerical accuracy of all three decompositions are similar. However, for ill-conditioned systems, where BO and DO either diverge or show poor numerical performance, the DBO decomposition performs well.

We conclude by showing a limitation of the presented method. In particular we revisit the demonstration case for stochastic Navier-Stokes equation as presented in Section \ref{Sec:SNS}. 
We consider the same problem setup as the previous case of Reynolds number 1500 except that the kinematic viscosity is chosen to be $\nu=0.015$ which changes the Reynolds number to $Re=5000$. For this Reynolds number the flow  is chaotic. To ensure that the flow is chaotic,  we solved three deterministic cases by perturbing the horizontal forcing with three  values $f_x = 1-\epsilon, 1$ and $1+\epsilon$ with $\epsilon = 10^{-3}$. The resulting shear viscous force in the $x$-direction on the top and bottom walls   is plotted in Fig.(\ref{fig:f31}). It is clear that difference between the three solutions due to the perturbation grows and after $t>116$ becomes $\mathcal{O}(1)$ -- verifying that the flow is chaotic.  We consider DBO reduction with $r=2$. The eigenvalues of the covariance of the DBO system and those of the KL decomposition are plotted in Fig.(\ref{fig:f32}). We observe that for the chaotic regime a fast decay of the eigenvalues is not observed, since the randomness in the initial condition quickly propagates on large number of independent dimensions in the phase space of the dynamical system due to strong non-linear interaction between the modes and fast growth of small perturbations.  As a result, the effect of unresolved modes must be accounted for.

\section{Acknowledgements}
The author has been supported by NASA Transformational Tools and Technologies (TTT) Project,   grant number 80NSSC18M0150  and an award by American Chemical Society, Petroleum Research Fund, grant number 59081-DN19. This research was supported in part by the University of Pittsburgh Center for Research Computing through the resources provided. All 1D codes used in this manuscript are available on GitHub at \url{https://github.com/ppatil1708/DBO.git}.

\begin{appendices}
\section{Derivation of the DBO evolution equations}
\label{appendix:AppA}
\subsection{Proof of Theorem(\ref{MainTh})}
For the sake of brevity in notation, we denote $\ol{u}(x,t) $ as $\ol{u}$, $u_i(x,t)$ as $u_i$, $y_i(t;\omega)$ as $y_i$ and $\Sigma_{ij}(t)$ as $\Sigma_{ij}$. The complete stochastic field given by $u(x,t;\omega)$, will be denoted as $u$. 
To obtain the evolution equations of each of the DBO components, we first substitute the DBO decomposition, given by Eq.(\ref{eq:DBODecomposition}), into a general form of SPDE as given by Eq.(\ref{eq:SPDE}). This follows:
\begin{equation}\label{eq:FullEq}
    \pfrac{\ol{u}}{t} + \pfrac{u_i}{t} \Sigma_{ij} y_j + u_i \frac{d\Sigma_{ij}}{dt} y_j + u_i \Sigma_{ij} \frac{dy_j}{dt} = \mathscr{F}(u)
\end{equation}
We take expectation of the above equation:
\begin{equation}
    \pfrac{\ol{u}}{t} = \mathbb{E}[\mathscr{F}(u)],
\end{equation}
where we have used  $\mathbb{E}[y_i] = 0$ and  $d \mathbb{E}[y_i]/dt = 0$. The above equation denotes the evolution of the mean field, which is given by the first equation in the theorem i.e, Eq.(\ref{subeq:DBOMean}). We proceed further by obtaining a mean subtracted form of the original SPDE, by subtracting the above mean evolution equation from Eq.(\ref{eq:FullEq}). This follows: 
\begin{equation}\label{eq:Meansubtracted}
    \pfrac{u_i}{t} \Sigma_{ij} y_j + u_i \frac{d\Sigma_{ij}}{dt} y_j + u_i \Sigma_{ij} \frac{dy_j}{dt} = \mathscr{\tilde{F}}(u),
\end{equation}
where $\mathscr{\tilde{F}}(u)= \mathscr{F}(u)-\mathbb{E}[\mathscr{F}(u)]$. We then project the mean-subtracted equation onto the stochastic modes $y_k$,
\begin{equation*}
    \pfrac{u_i}{t} \Sigma_{ij} \mathbb{E}[y_j y_k] + u_i \frac{d\Sigma_{ij}}{dt} \mathbb{E}[y_j y_k] + u_i \Sigma_{ij} \mathbb{E}[\frac{dy_j}{dt} y_k] =  \mathbb{E}[\mathscr{\tilde{F}}(u) y_k].
\end{equation*}
The  stochastic modes are orthonormal i.e., $\mathbb{E}[y_j y_k] = \delta_{jk}$ and dynamically orthogonal i.e., $\mathbb{E}[\dfrac{dy_j}{dt} y_k]=0$. Using these two conditions and changing index $k$ to $j$, the above equation simplifies to:
\begin{equation}\label{eq:UEq}
    \pfrac{u_i}{t} \Sigma_{ij} + u_i \frac{d\Sigma_{ij}}{dt} = \mathbb{E}[\mathscr{\tilde{F}}(u) y_j].
\end{equation}
We now project the above equation onto the spatial modes $u_k$,
\begin{equation*}
    \left<  u_k, \pfrac{u_i}{t} \right> \Sigma_{ij} + \left< u_k, u_i \right> \frac{d\Sigma_{ij}}{dt} = \left< u_k,\mathbb{E}[\mathscr{\tilde{F}}(u) y_j] \right>.
\end{equation*}
By enforcing  the orthonormality property i.e., $\left< u_k, u_i \right> = \delta_{ki}$ and the dynamical orthogonality property i.e., $\left< \pfrac{u_k}{t}, u_i \right>=0$ of the spatial basis, we obtain the evolution equation of the $\Sigma_{ij}$ corresponding to Eq.(\ref{subeq:DBOSigma}):
\begin{equation}\label{eqn:apn_Sigma}
    \frac{d\Sigma_{ij}}{dt} = \left< u_i,\mathbb{E}[\mathscr{\tilde{F}}(u) y_j] \right>.
\end{equation}
To obtain the evolution equations for the spatial modes, we substitute  Eq.(\ref{eqn:apn_Sigma}) into Eq.(\ref{eq:UEq}) and we then multiply both sides by $\Sigma_{ij}^{-1}$. This results in:
\begin{equation*}
    \pfrac{u_i}{t} = \left[ \mathbb{E}[\mathscr{\tilde{F}}(u) y_j] - u_k \left< u_k,\mathbb{E}[\mathscr{\tilde{F}}(u) y_j] \right> \right] \Sigma_{ij}^{-1} .
\end{equation*}
Similarly, to obtain the evolution equation for the stochastic modes, we project Eq.(\ref{eq:Meansubtracted}) onto the spatial modes $u_k$. This results in:
\begin{equation*}
    \left<u_k, \pfrac{u_i}{t} \right>\Sigma_{ij} y_j + \left<u_k, u_i \right> \frac{d\Sigma_{ij}}{dt} y_j + \left<u_k, u_i \right> \Sigma_{ij} \frac{dy_j}{dt} = \left< u_k, \mathscr{\tilde{F}}(u) \right>.
\end{equation*}
Once again we utilize the orthonormality and dynamical orthogonality  of the spatial modes and substitute Eq.(\ref{eqn:apn_Sigma}) into the above equation. We finally swap the indices $j$ and $i$ to get the form in Eq.(\ref{subeq:DBOY}). The resulting equation is:
\begin{equation*}
    \frac{dy_i}{dt} = \left[ \left< u_j,\mathscr{\tilde{F}}(u)  \right> - \left< u_j,\mathbb{E}[\mathscr{\tilde{F}}(u) y_k] \right> y_k \right] \Sigma_{ji} ^{-1}.
\end{equation*}
Since the boundary conditions are deterministic, the boundary conditions for the mean and the spatial modes are given by:
\begin{align*}
    \mathscr{B}[\ol{u}(x,t)] &=h(x,t),   & x \in \partial D,\\
    \mathscr{B}[u_i(x,t)]    &=0, & x \in \partial D.
\end{align*}
The initial conditions for the mean are given by applying the mean value to the stochastic field at $t=0$:
\begin{equation*}
    \ol{u}_0(x, t_0) = \mathbb{E}[u_0(x;\omega)].
\end{equation*}
This completes the proof. 
\section{Equivalence of DO and DBO methods}
\label{appendix:AppB}
\subsection{Proof of Lemma (\ref{lemma1})}
\begin{enumerate}[label=(\roman*)]
    \item The transformation matrix $R_u$ can be obtained by projecting the equivalence relation $U_{DO} = U_{DBO}R_u$ onto $U_{DBO}$. This results in: 
    \begin{align}\label{eq:Rudef}
        U_{DO} &= U_{DBO}  {R}_u, \nonumber \\
        R_u &=   \left< {U}_{DBO},  {U}_{DO} \right>, 
    \end{align}
    where we have used the orthonormality property of $U_{DBO}$ basis: $\left< {U}_{DBO},  {U}_{DBO} \right>= I$, where $I$ is the identity matrix. 
    Similarly projecting the equivalence relation onto $U_{DO}$ and using the orthonormality property of the $U_{DO}$ basis:$\left< {U}_{DO},  {U}_{DO} \right>= I$ we obtain,
    \begin{align*}
          \left< {U}_{DO},  {U}_{DBO}  \right>  {R}_u &= I, \\
         {R}_u^{-1} &=\left< {U}_{DO},  {U}_{DBO}  \right>.
     \end{align*}
         It follows from the definition of inner product of quasimatrices i.e., Eq.(\ref{eq:quasiinner}), that the transpose of the inner product can be written as $\left< V(x,t), U(x,t) \right> = \left< U(x,t), V(x,t) \right>^T$. The above equation can be re-written as the transpose of inner product of quasimatrices in the following form:
     \begin{align*}
            {R}_u^{-1} &=\left< {U}_{DBO},  {U}_{DO}  \right>^T.
     \end{align*}
     Now, using the result from Eq.(\ref{eq:Rudef}), the above equation can be written as,
     \begin{align*}
         {R}_u^{-1} &=  {R}_u^T, \\
         {R}_u^T {R}_u &= I.
    \end{align*}
   This equation shows that $R_u^T$ is an inverse of $R_u$, which is a property of orthogonal matrices. Therefore, ${R}_u$ is an orthogonal matrix. 
    \item Since the two decompositions are equivalent, we have 
    \begin{align*}
         {U}_{DBO} {\Sigma}_{DBO} {Y}_{DBO}^T &=  {U}_{DO}  {Y}_{DO}^T.
         \end{align*}
         Using the transformation definition $U_{DO} = U_{DBO}R_u$ and $Y_{DO} = Y_{DBO} W_y$, the DO decomposition can be expressed as:
         \begin{align*}
             {U}_{DBO} {\Sigma}_{DBO}  {Y}_{DBO}^T &=  {U}_{DBO} {R}_u  {W}_y^T  {Y}_{DBO}^T.
         \end{align*}
        Projecting the above equation on the $U_{DBO}$ basis and using the orthonormality property of the DBO basis i.e., $\left< {U}_{DBO},  {U}_{DBO} \right>= I$, we get: 
         \begin{align*}
             \left<  {U}_{DBO},  {U}_{DBO} \right>  {\Sigma}_{DBO} {Y}_{DBO}^T &= \left<  {U}_{DBO},  {U}_{DBO} \right>  {R}_u  {W}_y^T {Y}_{DBO}^T , \\
             {\Sigma}_{DBO} {Y}_{DBO}^T &= {R}_u  {W}_y^T {Y}_{DBO}^T. 
         \end{align*}
         We now project the above equation on the stochastic DBO basis, i.e. $Y_{DBO}$: 
         \begin{align*}
             {\Sigma}_{DBO} \mathbb{E}[{Y}_{DBO}^T  Y_{DBO}] &= {R}_u  {W}_y^T \mathbb{E}[{Y}_{DBO}^T Y_{DBO}].
         \end{align*}
         The stochastic basis of DBO are orthonormal i.e., $\mathbb{E}[Y_{DBO}^T Y_{DBO}]= I$. We apply this property to the above equation and simplify it further, which results in: 
        \begin{align*}
         {\Sigma}_{DBO} &=  {R}_u  {W}_y^T.
          \end{align*}
          Multiplying the above equation by $R_u^T$ from left and using $R_u^T=R_u^{-1}$ and  transposing the resulting equation yields:
          \begin{align*}
         {W}_y &=  {\Sigma}_{DBO}^T  {R}_u.
    \end{align*}
    \item We now prove that the $R_u$ matrix does not evolve in time. The evolution equation for ${U}_{DO}$ in a quasimatrix form can be written as:
\begin{equation*}
    \frac{\partial  {U}_{DO}}{\partial t} = \left[ \mathbb{E}[\mathscr{\tilde{F}}  {Y}_{DO}]-  {U}_{DO} \mathbb{E}\left[\left<  {U}_{DO},  \mathscr{\tilde{F}} \right> {Y}_{DO} \right] \right]   {C}_{DO}^{-1}. 
\end{equation*}
Substituting the transformation $ {U}_{DO} =  {U}_{DBO}  {R}_u$ and $ {Y}_{DO} =  {Y}_{DBO}  {W}_y$ in the above equation results in: 
\begin{equation*}
    \pfrac{ {U}_{DBO}}{t}  {R}_u +  {U}_{DBO} \frac{d {R}_u}{dt} = \left[ \mathbb{E}[\mathscr{\tilde{F}} {Y}_{DBO}] -  {U}_{DBO} {R}_u  {R}_u^T \mathbb{E}[ \left<  {U}_{DBO}, \mathscr{\tilde{F}}\right>  {Y}_{DBO} ]\right]  {W}_y  {C}_{DO}^{-1}.
\end{equation*}
Projecting the above equation on the ${U}_{DBO}$ bases, using the dynamically orthogonal condition i.e., $\left<\dot{U}_{DBO}, U_{DBO} \right> = 0$, orthonormality property of DBO spatial modes i.e., $\left< U_{DBO}, U_{DBO} \right> =I$ and orthogonal matrix property i.e., $R_u R_u^T= I$ on the previous equation results in: 
\begin{equation*}
    \frac{d {R}_u}{dt} = \left[\left< {U}_{DBO}, \mathbb{E}[\mathscr{\tilde{F}} {Y}_{DBO}] \right> - \mathbb{E}[ \left<  {U}_{DBO}, \mathscr{\tilde{F}}\right>  {Y}_{DBO} ] \right] {W}_y  {C}_{DO}^{-1}.
    \end{equation*}
    The expectation operator and the spatial inner product operations commute, which results in:
    \begin{equation*}
    \frac{d {R}_u}{dt} = 0.
\end{equation*}
\end{enumerate}
This completes the proof. 

\subsection{Proof of Theorem (\ref{DBOtoDO})}
In this section, we prove that the DO and DBO decompositions of SPDE in Eq.(\ref{eq:SPDEall}) remain equivalent for all time. We begin with the evolution equations for the stochastic and spatial DO bases in the quasimatrix form: 
\begin{subequations}
    \begin{align}
    \frac{\partial {U}_{DO}}{\partial t} &= \left[ \mathbb{E}[\mathscr{\tilde{F}}  {Y}_{DO}]-  {U}_{DO} \mathbb{E}\left[\left<  {U}_{DO},  \mathscr{\tilde{F}} \right> {Y}_{DO} \right] \right]   {C}_{DO}^{-1},  \label{eq:UDOvec}\\
    \frac{d {Y}_{DO}}{dt} &= \left< \mathscr{\tilde{F}}, {U}_{DO} \right>. \label{eq:YDOvec}
\end{align}
\end{subequations}
We substitute the transformation $ {U}_{DO} =  {U}_{DBO}  {R}_u$ and $ {Y}_{DO} =  {Y}_{DBO}  {W}_y$ in the evolution equation for spatial DO modes i.e., Eq.(\ref{eq:UDOvec}). The equation thus becomes: 
\begin{align*}
     \pfrac{ {U}_{DBO}}{t}  {R}_u +  {U}_{DBO}\frac{ {dR}_u}{dt} &= \left[ \mathbb{E}[\mathscr{\tilde{F}} {Y}_{DBO}] -  {U}_{DBO} {R}_u  {R}_u^T \mathbb{E}[ \left<  {U}_{DBO}, \mathscr{\tilde{F}}\right>  {Y}_{DBO} ]\right]  {W}_y  {C}_{DO}^{-1}.
\end{align*}
Using the results of (i) and (iii) from Lemma (\ref{lemma1}), the above equation can be simplified as:
\begin{equation}\label{eq:eq1001}
    \pfrac{ {U}_{DBO}}{t}  {R}_u = \left[ \mathbb{E}[\mathscr{\tilde{F}} {Y}_{DBO}] -  {U}_{DBO} \mathbb{E}[ \left<  {U}_{DBO}, \mathscr{\tilde{F}}\right>  {Y}_{DBO} ]\right]  {W}_y  {C}_{DO}^{-1}.
\end{equation}
The covariance matrix for DO is defined by the following equation:
\begin{equation}\label{eq:eq1007}
    C_{DO} = \mathbb{E}[Y_{DO}^T Y_{DO}]. 
 \end{equation}
 We can simplify the above equation by using the transformation $Y_{DO}= Y_{DBO} W_y$ and using the orthonormality of the DBO stochastic modes: 
 \begin{align*}
     C_{DO} &= \mathbb{E}[Y_{DO}^T Y_{DO}], \\
     C_{DO} &= W_y^T \mathbb{E}[Y_{DBO}^T Y_{DBO}] W_y,\\
     C_{DO} &= W_y^T W_y.
 \end{align*}
 
Thus, $C_{DO}^{-1}$ can be written as $C_{DO}^{-1} = W_y^{-1} W_y^{-T}$. We now simplify the $ {W}_y {C}_{DO}^{-1}$ which appears in Eq.(\ref{eq:eq1001}) and using inverse of $W_y^T$ from the property (ii) from Lemma (\ref{lemma1}).
\begin{align*}
     {W}_y  {C}_{DO}^{-1} &=  {W}_y  {W}_y^{-1}  {W}_y^{-T}, \\ 
                                 &=  {\Sigma}_{DBO}^{-1}  {R}_u.
\end{align*}
Multiplying Eq.(\ref{eq:eq1001}) by $ {R}_u^T$ from right and using the value of $ {W}_y  {C}_{DO}^{-1}$ from the above equations and using the property of orthogonal matrix $R_u$ i.e., $R_u^T R_u =I$, the evolution equation simplifies to: 
\begin{equation*}
    \pfrac{ {U}_{DBO}}{t} = \left[ \mathbb{E}[\mathscr{\tilde{F}} {Y}_{DBO}] -  {U}_{DBO} \mathbb{E}[ \left<  {U}_{DBO} \mathscr{\tilde{F}}\right>  {Y}_{DBO} ]\right]  {\Sigma}_{DBO}^{-1}.
\end{equation*}
The above equation is the evolution equation of the DBO spatial modes in quasimatrix form. Similarly, substituting the transformations ${U}_{DO} =  {U}_{DBO}  {R}_u$ and $ {Y}_{DO} =  {Y}_{DBO}  {W}_y$ in the evolution equation for $ {Y}_{DO}$, i.e. Eq.(\ref{eq:YDOvec}), results in: 
\begin{equation}\label{eq:eq1002}
   \frac{d {Y}_{DBO}}{dt}  {W}_y +  {Y}_{DBO} \frac{d {W}_y}{dt}   =   \left<  \mathscr{\tilde{F}}, {U}_{DBO} \right>{R}_u.
\end{equation}
From  parts (ii) and (iii) of Lemma (\ref{lemma1}), we have: $\dfrac{d {W}_y}{dt} = \dfrac{d {\Sigma}_{DBO}^T}{dt} {R}_u$. Using this relation in Eq.(\ref{eq:eq1002}): 
\begin{align*}
   \frac{d {Y}_{DBO}}{dt} {W}_y + {Y}_{DBO} \dfrac{d {\Sigma}_{DBO}^T}{dt} {R}_u  &=   \left< \mathscr{\tilde{F}}, {U}_{DBO} \right> R_u,\\
    \frac{d {Y}_{DBO}}{dt} {W}_y &=  \left[  \left< \mathscr{\tilde{F}}, {U}_{DBO} \right> -  {Y}_{DBO} \mathbb{E}\left[{Y}^T_{DBO} \left<  \mathscr{\tilde{F}},{U}_{DBO} \right>   \right]    \right]R_u,
\end{align*}
where evolution of $\Sigma_{DBO}^T$ given by: $\frac{d\Sigma^T_{DBO}}{dt} = \mathbb{E}\left[{Y}^T_{DBO} \left<  \mathscr{\tilde{F}},{U}_{DBO} \right>\right]   $ is substituted in the above equation.
Multiplying both sides of the equation by $W_y^{-1}$ from the right and using the result of part (ii) of Lemma (\ref{lemma1}), we get: 
\begin{equation*}
     \frac{d {Y}_{DBO}}{dt} =  \left[  \left<\mathscr{\tilde{F}},  {U}_{DBO}  \right> -  {Y}_{DBO} \mathbb{E}\left[ {Y}_{DBO}^T \left< \mathscr{\tilde{F}},  {U}_{DBO}  \right> \right] \right]  {\Sigma}_{DBO}^{-T}.
\end{equation*}
The above equation is the evolution equation of the DBO stochastic modes in the quasimatrix form. Thus, we see that the equivalence between the stochastic basis is maintained $\forall t > 0$. This completes the proof. 
\section{Equivalence of DBO and BO methods}
\label{appendix:AppC}
\subsection{Proof of Lemma (\ref{lemma2})}
\begin{enumerate}[label=(\roman*)]
    \item We begin with the transformation equation for the stochastic modes given by:
    \begin{equation*}
        Y_{DBO} = Y_{BO} R_y.
    \end{equation*}
    We project the above equation onto the DBO stochastic modes and use the orthonormality property of the DBO modes i.e., $\mathbb{E}[Y_{DBO}^T Y_{DBO}]=I$:
    \begin{align*}
         Y_{DBO} &= Y_{BO} R_y, \nonumber \\ 
         \mathbb{E}[Y_{DBO}^T Y_{DBO}] &= \mathbb{E}[Y_{DBO}^T Y_{BO}] R_y, \nonumber 
    \end{align*}
    By using the orthonormality of the DBO stochastic coefficients, the above equation is simplified to: 
    \begin{equation}\label{eq:YDBOBO}
     R_y^{-1} = \mathbb{E}[Y_{DBO}^T Y_{BO}]  
     \end{equation}
    We also project the transformation equation on the BO stochastic modes and use the orthonormality condition of the BO modes i.e., $\mathbb{E}[Y_{BO}^T Y_{BO}]= I$:
    \begin{align*}
        Y_{DBO}  &= Y_{BO} R_y, \\ 
        \mathbb{E}[Y_{BO}^T Y_{DBO}] &= \mathbb{E}[Y_{BO}^T Y_{BO}] R_y, 
    \end{align*}
        which results in 
    \begin{equation*}
         R_y =  \mathbb{E}[Y_{BO}^T Y_{DBO}].
     \end{equation*}
    Taking transpose of the above equation and using Eq.(\ref{eq:YDBOBO}):
    \begin{align*}
        \mathbb{E}[Y_{DBO}^T Y_{BO}] &= R_y^T, \\ 
        R_y^{-1} &= R_y^T,\\
        R_y R_y^T &= I. 
    \end{align*}
    This equation shows that $R_y^T$ is an inverse of $R_y$; which is a property of orthogonal matrices. Therefore, $R_y$ is an orthogonal matrix. 
    \item Now, since the two decompositions are equivalent, we have
    \begin{equation*}
         U_{DBO} \Sigma_{DBO} Y_{DBO}^T = U_{BO} Y_{BO}^T. 
    \end{equation*}
    Using the transformation equations i.e., $Y_{DBO} = Y_{BO} R_y$ and $U_{DBO} =U_{BO} W_u$ in the above equation: 
    \begin{equation*}
        U_{BO} W_u \Sigma_{DBO} R_y^T Y_{BO}^T = U_{BO} Y_{BO}^T.
    \end{equation*}
    We now project the above equation on the $U_{BO}$ bases and use the BO condition, i.e., $\left< U_{BO}, U_{BO} \right>= \Lambda$, which results in: 
    \begin{align*}
        \left< U_{BO}, U_{BO} \right> W_u \Sigma_{DBO} R_y^T Y_{BO}^T &=\left< U_{BO}, U_{BO} \right> Y_{BO}^T, \\ 
        \Lambda W_u \Sigma_{DBO} R_y^T Y_{BO}^T &= \Lambda Y_{BO}^T,\\ 
        W_u \Sigma_{DBO} R_y^T Y_{BO}^T &= Y_{BO}^T.
    \end{align*}
    We now project the above equation onto the stochastic BO bases and use the orthonormality property of the bases:
    \begin{align}\label{eq:SigmaWR}
        W_u \Sigma_{DBO} R_y^T \mathbb{E}[Y_{BO}^T  Y_{BO}] &= \mathbb{E}[Y_{BO}^T Y_{BO}], \nonumber\\
        W_u \Sigma_{DBO} R_y^T &= I, \nonumber\\ 
        \Sigma_{DBO} &= W_u^{-1} R_y.
    \end{align}
    \item We now derive the evolution equation for $W_u$. We begin by using the transformation relation for the spatial modes given by:
    \begin{equation*}
        U_{DBO} =U_{BO} W_u.
    \end{equation*}
    We project the above equation onto the DBO spatial modes and use  the orthonormality condition of the DBO modes, i.e., $\left< U_{DBO}, U_{DBO} \right> =I $:
    \begin{align}\label{eq:UDBOBO}
        U_{DBO} &= U_{BO} W_u, \nonumber\\ 
        \left< U_{DBO}, U_{DBO}\right> &= \left< U_{DBO}, U_{BO} \right> W_u, \nonumber \\ 
        \left< U_{DBO}, U_{BO} \right> &= W_u^{-1}. 
    \end{align}
    We also project the transformation equation onto the BO spatial modes and use the BO condition i.e., $\left< U_{BO}, U_{BO} \right>= \Lambda $: 
    \begin{align*}
        U_{DBO} &= U_{BO} W_u, \\
        \left< U_{BO}, U_{DBO} \right> &= \left< U_{BO}, U_{BO} \right> W_u, \\ 
        \left< U_{BO}, U_{DBO} \right> &= \Lambda W_u.
    \end{align*}
    Taking transpose of the above equation and using Eq.(\ref{eq:UDBOBO}):
    \begin{align}\label{eq:WWT}
        \left< U_{DBO}, U_{BO} \right> &= W_u^T \Lambda, \nonumber \\ 
        W_u^{-1} &= W_u^T \Lambda, \nonumber \\ 
        W_u W_u^T &= \Lambda^{-1}. 
    \end{align}
    We now consider the evolution equation of $U_{DBO}$ given by: 
    \begin{equation*}
        \pfrac{ {U}_{DBO}}{t} = \left[ \mathbb{E}[\mathscr{\tilde{F}} {Y}_{DBO}] -  {U}_{DBO}  \left<  {U}_{DBO}, \mathbb{E}[\mathscr{\tilde{F}}{Y}_{DBO} ]\right>  \right]  {\Sigma}_{DBO}^{-1}.
    \end{equation*}
    Substituting the transformation $Y_{DBO} = Y_{BO} R_y$ and $U_{DBO} = U_{BO} W_u$ in the above equation, we obtain: 
    \begin{align*}
        \pfrac{U_{BO}}{t} W_u + U_{BO} \frac{d W_u}{dt} &= \left[ \mathbb{E}[\mathscr{\tilde{F}} {Y}_{BO}]R_y -  {U}_{BO}W_u W_u^T \left<  {U}_{BO}, \mathbb{E}[\mathscr{\tilde{F}}{Y}_{BO} ]\right> R_y  \right]  {\Sigma}_{DBO}^{-1}.
    \end{align*}
    We now use Eq.(\ref{eq:SigmaWR}) to obtain $\Sigma_{DBO}^{-1}$ versus $W_u$ and $R_y$ and use Eq.(\ref{eq:WWT}) to simplify the above equation further: 
    \begin{equation}\label{eq:eq101}
        \pfrac{U_{BO}}{t} W_u + U_{BO} \frac{d W_u}{dt} = \left[ \mathbb{E}[\mathscr{\tilde{F}} {Y}_{BO}] -  {U}_{BO}\Lambda^{-1} \left<  {U}_{BO}, \mathbb{E}[\mathscr{\tilde{F}}{Y}_{BO} ]\right>   \right]  W_u.
    \end{equation}
    The evolution equation for $U_{BO}$ is given by:
    \begin{align*}
        \pfrac{U_{BO}}{t} &= {U}_{BO} {M}  + \mathbb{E}[\mathscr{\tilde{F}} {Y}_{BO}],\\
        M &= \mathbb{E}[Y_{BO}^T \frac{d Y_{BO}}{dt}].
    \end{align*}
    We further simplify the equation by substituting the evolution equation for $U_{BO}$ in Eq.(\ref{eq:eq101}):
    \begin{align*}
         \left[ {U}_{BO} {M}  + \mathbb{E}[\mathscr{\tilde{F}} {Y}_{BO}] \right] W_u + U_{BO} \frac{d W_u}{dt} &= \left[ \mathbb{E}[\mathscr{\tilde{F}} {Y}_{BO}] -  {U}_{BO}\Lambda^{-1} \left<  {U}_{BO}, \mathbb{E}[\mathscr{\tilde{F}}{Y}_{BO} ]\right>   \right]  W_u, \\
           U_{BO} \frac{d W_u}{dt} &= -{U}_{BO} {M} W_u -  {U}_{BO}\Lambda^{-1} \left<  {U}_{BO}, \mathbb{E}[\mathscr{\tilde{F}}{Y}_{BO} ]\right> W_u.
    \end{align*}
    We project the above equation onto the $U_{BO}$ bases and use the BO orthogonality condition of the spatial modes: 
    \begin{equation*}
        \Lambda \frac{d W_u}{dt} = - \Lambda \left[M  +  \Lambda^{-1} \left<  {U}_{BO}, \mathbb{E}[\mathscr{\tilde{F}}{Y}_{BO} ]\right> \right] W_u.
    \end{equation*}
   Denoting $G = \left<  {U}_{BO}, \mathbb{E}[\mathscr{\tilde{F}}{Y}_{BO} ]\right> $ according to the notation in reference \cite{choi2014equivalence}, the evolution equation for $W_u$ becomes: 
    \begin{equation*}
        \frac{d W_u}{dt} = - \left[M + \Lambda^{-1} G \right] W_u.
    \end{equation*}
    \item We now derive the evolution equation for $R_y$. We begin with the evolution equation for $Y_{DBO}$ given by: 
    \begin{equation*}
        \frac{d {Y}_{DBO}}{dt} =  \left[  \left<\mathscr{\tilde{F}},  {U}_{DBO}  \right> -  {Y}_{DBO} \mathbb{E}\left[ {Y}_{DBO}^T \left< \mathscr{\tilde{F}},  {U}_{DBO}  \right> \right] \right]  {\Sigma}_{DBO}^{-T}.
    \end{equation*}
    The transformation equations i.e., $Y_{DBO} =Y_{BO} R_y$ and $U_{DBO} = U_{BO} W_u$ are substituted in the above equation. We also use Eq.(\ref{eq:SigmaWR}) to replace ${\Sigma}_{DBO}^{-T}$ versus $W_u$ and $R_y$. The equation, thus, becomes: 
    \begin{equation*}
        \frac{d {Y}_{BO}}{dt} R_y + Y_{BO} \frac{d R_y}{dt} = \left[  \left<\mathscr{\tilde{F}},  {U}_{BO}  \right>W_u -  {Y}_{BO}R_y R_y^T\mathbb{E}\left[ {Y}_{BO}^T \left< \mathscr{\tilde{F}},  {U}_{BO}  \right> W_u \right] \right]  W_u^T R_y.
    \end{equation*}
    Using property of orthogonal of matrix $R_y$ and Eq.(\ref{eq:WWT}) to simplify the above equation simplifies to: 
    \begin{equation}\label{eq:eq102}
        \frac{d {Y}_{BO}}{dt} R_y + Y_{BO} \frac{d R_y}{dt} = \left[  \left<\mathscr{\tilde{F}},  {U}_{BO}  \right> -  {Y}_{BO} \mathbb{E}\left[ {Y}_{BO}^T \left< \mathscr{\tilde{F}},  {U}_{BO}  \right>  \right] \right]  \Lambda^{-1} R_y.
    \end{equation}
    On the other hand, the evolution equation for $Y_{BO}$ is given by: 
    \begin{align*}
        \frac{ d{Y}_{BO}}{dt} &= \left[ \left< \mathscr{\tilde{F}},{U}_{BO} \right> -  {Y}_{BO}S^T\right]{\Lambda}^{-1},\\
        {S} &= \left<  {U}_{BO}, \pfrac{{U}_{BO}}{t} \right>.
    \end{align*}
    Substituting the evolution equation for $Y_{BO}$ in Eq.(\ref{eq:eq102}), results in:
    \begin{align*}
        \left[ \left< \mathscr{\tilde{F}},{U}_{BO} \right> -  {Y}_{BO}S^T\right]{\Lambda}^{-1}R_y + Y_{BO} \frac{d R_y}{dt} &= \left[  \left<\mathscr{\tilde{F}},  {U}_{BO}  \right> -  {Y}_{BO} \mathbb{E}\left[ {Y}_{BO}^T \left< \mathscr{\tilde{F}},  {U}_{BO}  \right>  \right] \right]  \Lambda^{-1} R_y,\\
        -  {Y}_{BO}S^T  {\Lambda}^{-1}R_y + Y_{BO} \frac{d R_y}{dt} &= -  {Y}_{BO} \mathbb{E}\left[ {Y}_{BO}^T \left< \mathscr{\tilde{F}},  {U}_{BO}  \right> \right]\Lambda^{-1} R_y.
    \end{align*}
    Projecting the above equation onto $Y_{BO}$ and using the orthonormality property of the stochastic BO modes, results in:
    \begin{equation*}
        -S^T \Lambda^{-1} R_y + \frac{d R_y}{dt} = -\mathbb{E}\left[ {Y}_{BO}^T \left< \mathscr{\tilde{F}},  {U}_{BO}  \right> \right]\Lambda^{-1} R_y.
    \end{equation*}
    The term $\mathbb{E}\left[ {Y}_{BO}^T \left< \mathscr{\tilde{F}},  {U}_{BO}  \right> \right]$ can be expressed as $G^T$ according to the notation in reference \cite{choi2014equivalence}. Thus, the evolution equation for $R_y$ can be written as:
    \begin{equation*}
        \frac{dR_y}{dt} = (S^T -G^T)\Lambda^{-1} R_y.
    \end{equation*}
\end{enumerate}
This completes the proof. 
\subsection{Proof of Theorem (\ref{DBOtoBO})}
In this theorem, we prove that the equivalence relation is valid for all $t>0$. The DBO evolution equations in the quasimatrix form are given by:
\begin{subequations}
    \begin{align}
    \pfrac{ {U}_{DBO}}{t} = \left[ \mathbb{E}[\mathscr{\tilde{F}} {Y}_{DBO}] -  {U}_{DBO} \mathbb{E}[ \left<  {U}_{DBO} \mathscr{\tilde{F}}\right>  {Y}_{DBO} ]\right]  {\Sigma}_{DBO}^{-1}, \label{eq:UDBOvec}\\ 
    \frac{d {Y}_{DBO}}{dt} =  \left[  \left<\mathscr{\tilde{F}},  {U}_{DBO}  \right> -  {Y}_{DBO} \mathbb{E}\left[ {Y}_{DBO}^T \left< \mathscr{\tilde{F}},  {U}_{DBO}  \right> \right] \right]  {\Sigma}_{DBO}^{-T}. \label{eq:YDBOvec}
    \end{align}
\end{subequations}
We plug $U_{DBO} = U_{BO} W_u$ and $Y_{DBO} = Y_{BO} R_y$ into the evolution equation for $U_{DBO}$ i.e., Eq.(\ref{eq:UDBOvec}). The equation thus becomes:
\begin{equation*}
    \pfrac{U_{BO}}{t} W_u + U_{BO} \frac{d W_u}{dt} = \left[ \mathbb{E}[\mathscr{\tilde{F}} {Y}_{BO}]R_y -  {U}_{BO}W_u W_u^T \left<  {U}_{BO}, \mathbb{E}[\mathscr{\tilde{F}}{Y}_{BO} ]\right> R_y  \right]  {\Sigma}_{DBO}^{-1}.
\end{equation*}
Using Eq.(\ref{eq:WWT}) and Eq.(\ref{eq:SigmaWR}) in the above equation, results in:
\begin{equation*}
    \pfrac{U_{BO}}{t} W_u + U_{BO} \frac{d W_u}{dt} = \left[ \mathbb{E}[\mathscr{\tilde{F}} {Y}_{BO}] -  {U}_{BO}\Lambda^{-1} \left<  {U}_{BO}, \mathbb{E}[\mathscr{\tilde{F}}{Y}_{BO} ]\right>   \right]  W_u.
\end{equation*}
Using property (iii) of Lemma (\ref{lemma2}) and definition of $G$, i.e., $G = \left<  {U}_{BO}, \mathbb{E}[\mathscr{\tilde{F}}{Y}_{BO} ]\right> $:
\begin{align*}
    \pfrac{U_{BO}}{t} W_u - U_{BO}\left[M+\Lambda^{-1} G \right] W_u &=  \left[ \mathbb{E}[\mathscr{\tilde{F}} {Y}_{BO}] -  {U}_{BO}\Lambda^{-1} G  \right]  W_u, \\ \pfrac{U_{BO}}{t} W_u &= U_{BO} M  W_u + \mathbb{E}[\mathscr{\tilde{F}} {Y}_{BO}] W_u, \\
    \pfrac{U_{BO}}{t} &= U_{BO} M + \mathbb{E}[\mathscr{\tilde{F}} {Y}_{BO}].
\end{align*}
The above equation is the evolution equation for $U_{BO}$. Thus, we see that the equivalence between the spatial bases is maintained $\forall t>0$. 
Similarly, we substitute the transformations $U_{DBO} = U_{BO} W_u$ and $Y_{DBO} = Y_{BO} R_y$ into the evolution equation for $Y_{DBO}$ i.e., Eq.(\ref{eq:YDBOvec}). The equation thus becomes:
\begin{equation*}
    \frac{d {Y}_{BO}}{dt} R_y + Y_{BO} \frac{d R_y}{dt} = \left[  \left<\mathscr{\tilde{F}},  {U}_{BO}  \right>W_u -  {Y}_{BO}R_y R_y^T\mathbb{E}\left[ {Y}_{BO}^T \left< \mathscr{\tilde{F}},  {U}_{BO}  \right> W_u \right] \right] \Sigma_{DBO}^{-T}.
\end{equation*}
Using Eq.(\ref{eq:WWT}) and Eq.(\ref{eq:SigmaWR}) to simplify the above equation:
\begin{equation*}
    \frac{d {Y}_{BO}}{dt} R_y + Y_{BO} \frac{d R_y}{dt} = \left[  \left<\mathscr{\tilde{F}},  {U}_{BO}  \right> -  {Y}_{BO} \mathbb{E}\left[ {Y}_{BO}^T \left< \mathscr{\tilde{F}},  {U}_{BO}  \right>  \right] \right]  \Lambda^{-1} R_y.
\end{equation*}
Using property (iv) of Lemma (\ref{lemma2}) and definition of $G^T$, i.e., $G^T = \mathbb{E}\left[ {Y}_{BO}^T \left< \mathscr{\tilde{F}},  {U}_{BO}  \right>  \right] $:
\begin{align*}
    \frac{d {Y}_{BO}}{dt} R_y + Y_{BO} \left[ S^T-G^T\right]\Lambda^{-1} R_y &= \left[  \left<\mathscr{\tilde{F}},  {U}_{BO}  \right> -  {Y}_{BO} G^T \right]  \Lambda^{-1} R_y,\\
    \frac{d {Y}_{BO}}{dt} &= \left[  \left<\mathscr{\tilde{F}},  {U}_{BO}  \right> - Y_{BO}S^T \right] \Lambda^{-1}.
\end{align*}
The above equation is the evolution equation of the BO  stochastic bases in the quasimatrix form. Thus, we see that the equivalence between the stochastic basis is maintained $\forall t>0$. This completes the proof.

\end{appendices}

\bibliographystyle{ieeetr}
\bibliography{main}

\begin{thebibliography}{10}

\bibitem{sapsis2009dynamically}
T.~P. Sapsis and P.~F. Lermusiaux, ``Dynamically orthogonal field equations for
  continuous stochastic dynamical systems,'' {\em Physica D: Nonlinear
  Phenomena}, vol.~238, no.~23-24, pp.~2347--2360, 2009.

\bibitem{cheng2013dynamicallyI}
M.~Cheng, T.~Y. Hou, and Z.~Zhang, ``A dynamically bi-orthogonal method for
  time-dependent stochastic partial differential equations {I: D}erivation and
  algorithms,'' {\em Journal of Computational Physics}, vol.~242, pp.~843--868,
  2013.

\bibitem{MM_DOEworks_11}
F.~Alexander, M.~Anitescu, J.~Bell, D.~Brown, M.~Ferris, M.~Luskin,
  S.~Mehrotra, B.~Moser, A.~Pinar, A.~Tartakovsky, {\em et~al.}, ``A
  multifaceted mathematical approach for complex systems,'' {\em Report of the
  DOE Workshop on Mathematics for the Analysis, Simulation, and Optimization of
  Complex Systems}, 2011.

\bibitem{giles2008multilevel}
M.~B. Giles, ``Multilevel monte carlo path simulation,'' {\em Operations
  Research}, vol.~56, no.~3, pp.~607--617, 2008.

\bibitem{barth2011multi}
A.~Barth, C.~Schwab, and N.~Zollinger, ``Multi-level monte carlo finite element
  method for elliptic pdes with stochastic coefficients,'' {\em Numerische
  Mathematik}, vol.~119, no.~1, pp.~123--161, 2011.

\bibitem{kuo2012quasi}
F.~Y. Kuo, C.~Schwab, and I.~H. Sloan, ``Quasi-monte carlo finite element
  methods for a class of elliptic partial differential equations with random
  coefficients,'' {\em SIAM Journal on Numerical Analysis}, vol.~50, no.~6,
  pp.~3351--3374, 2012.

\bibitem{ghanem2003stochastic}
R.~G. Ghanem and P.~D. Spanos, {\em Stochastic finite elements: a spectral
  approach}.
\newblock Courier Corporation, 2003.

\bibitem{wan2006multi}
X.~Wan and G.~E. Karniadakis, ``Multi-element generalized polynomial chaos for
  arbitrary probability measures,'' {\em SIAM Journal on Scientific Computing},
  vol.~28, no.~3, pp.~901--928, 2006.

\bibitem{xiu2005high}
D.~Xiu and J.~S. Hesthaven, ``High-order collocation methods for differential
  equations with random inputs,'' {\em SIAM Journal on Scientific Computing},
  vol.~27, no.~3, pp.~1118--1139, 2005.

\bibitem{xiu2002wiener}
D.~Xiu and G.~E. Karniadakis, ``The wiener--askey polynomial chaos for
  stochastic differential equations,'' {\em SIAM journal on scientific
  computing}, vol.~24, no.~2, pp.~619--644, 2002.

\bibitem{foo2010multi}
J.~Foo and G.~E. Karniadakis, ``Multi-element probabilistic collocation method
  in high dimensions,'' {\em Journal of Computational Physics}, vol.~229,
  no.~5, pp.~1536--1557, 2010.

\bibitem{foo2008multi}
J.~Foo, X.~Wan, and G.~E. Karniadakis, ``The multi-element probabilistic
  collocation method (me-pcm): Error analysis and applications,'' {\em Journal
  of Computational Physics}, vol.~227, no.~22, pp.~9572--9595, 2008.

\bibitem{babuvska2007stochastic}
I.~Babu{\v{s}}ka, F.~Nobile, and R.~Tempone, ``A stochastic collocation method
  for elliptic partial differential equations with random input data,'' {\em
  Siam J. Numer. Anal}, vol.~45, no.~3, pp.~1005--1034, 2007.

\bibitem{ganapathysubramanian2007sparse}
B.~Ganapathysubramanian and N.~Zabaras, ``Sparse grid collocation schemes for
  stochastic natural convection problems,'' {\em Journal of Computational
  Physics}, vol.~225, no.~1, pp.~652--685, 2007.

\bibitem{yang2012adaptive}
X.~Yang, M.~Choi, G.~Lin, and G.~E. Karniadakis, ``Adaptive anova decomposition
  of stochastic incompressible and compressible flows,'' {\em Journal of
  Computational Physics}, vol.~231, no.~4, pp.~1587--1614, 2012.

\bibitem{babaee2014effect}
H.~Babaee, X.~Wan, and S.~Acharya, ``Effect of uncertainty in blowing ratio on
  film cooling effectiveness,'' {\em Journal of Heat Transfer}, vol.~136,
  no.~3, p.~031701, 2014.

\bibitem{zhang2018stochastic}
D.~Zhang, H.~Babaee, and G.~E. Karniadakis, ``Stochastic domain decomposition
  via moment minimization,'' {\em SIAM Journal on Scientific Computing},
  vol.~40, no.~4, pp.~A2152--A2173, 2018.

\bibitem{wan2006long}
X.~Wan and G.~E. Karniadakis, ``Long-term behavior of polynomial chaos in
  stochastic flow simulations,'' {\em Computer methods in applied mechanics and
  engineering}, vol.~195, no.~41-43, pp.~5582--5596, 2006.

\bibitem{branicki2013fundamental}
M.~Branicki and A.~J. Majda, ``Fundamental limitations of polynomial chaos for
  uncertainty quantification in systems with intermittent instabilities,'' {\em
  Communications in Mathematical Sciences}, vol.~11, no.~1, pp.~55--103, 2013.

\bibitem{sirovich1987turbulenceI}
L.~Sirovich, ``Turbulence and the dynamics of coherent structures. {I.
  C}oherent structures,'' {\em Quarterly of applied mathematics}, vol.~45,
  no.~3, pp.~561--571, 1987.

\bibitem{sirovich1987turbulenceII}
L.~Sirovich, ``Turbulence and the dynamics of coherent structures. {II.
  S}ymmetries and transformations,'' {\em Quarterly of Applied mathematics},
  vol.~45, no.~3, pp.~573--582, 1987.

\bibitem{sirovich1987turbulenceIII}
L.~Sirovich, ``Turbulence and the dynamics of coherent structures. {III.
  D}ynamics and scaling,'' {\em Quarterly of Applied mathematics}, vol.~45,
  no.~3, pp.~583--590, 1987.

\bibitem{schmid2010dynamic}
P.~J. Schmid, ``Dynamic mode decomposition of numerical and experimental
  data,'' {\em Journal of fluid mechanics}, vol.~656, pp.~5--28, 2010.

\bibitem{Alvergue:2015aa}
L.~Alvergue, H.~Babaee, G.~Gu, and S.~Acharya, ``Feedback stabilization of a
  reduced-order model of a jet in crossflow,'' {\em AIAA Journal}, vol.~53,
  pp.~2472--2481, 2016/02/17 2015.

\bibitem{rowley2010reduced}
C.~Rowley, I.~Mezi{\'c}, S.~Bagheri, P.~Schlatter, and D.~Henningson,
  ``Reduced-order models for flow control: balanced models and koopman modes,''
  in {\em Seventh IUTAM Symposium on Laminar-Turbulent Transition}, pp.~43--50,
  Springer, 2010.

\bibitem{kutz2016dynamic}
J.~N. Kutz, S.~L. Brunton, B.~W. Brunton, and J.~L. Proctor, {\em Dynamic mode
  decomposition: data-driven modeling of complex systems}.
\newblock SIAM, 2016.

\bibitem{noack_2016}
B.~R. Noack, ``From snapshots to modal expansions – bridging low residuals
  and pure frequencies,'' {\em Journal of Fluid Mechanics}, vol.~802, p.~1–4,
  2016.

\bibitem{PWG18}
B.~Peherstorfer, K.~Willcox, and M.~Gunzburger, ``Survey of multifidelity
  methods in uncertainty propagation, inference, and optimization,'' {\em SIAM
  Review}, vol.~60, pp.~550--591, 2019/04/26 2018.

\bibitem{Perdikaris:2015aa}
P.~Perdikaris, D.~Venturi, J.~O. Royset, and G.~E. Karniadakis,
  ``Multi-fidelity modelling via recursive co-kriging and gaussian--markov
  random fields,'' 2015.

\bibitem{babaee2016}
H.~Babaee, P.~Perdikaris, C.~Chryssostomidis, and G.~E. Karniadakis,
  ``Multi-fidelity modelling of mixed convection based on experimental
  correlations and numerical simulations,'' {\em Journal of Fluid Mechanics},
  vol.~809, pp.~895--917, 12 2016.

\bibitem{cheng2013dynamicallyII}
M.~Cheng, T.~Y. Hou, and Z.~Zhang, ``A dynamically bi-orthogonal method for
  time-dependent stochastic partial differential equations {II: A}daptivity and
  generalizations,'' {\em Journal of Computational Physics}, vol.~242,
  pp.~753--776, 2013.

\bibitem{2019arXiv190409846B}
H.~{Babaee}, ``{A Scalable Observation-Driven Time-Dependent Basis for a
  Reduced Description of Transient Systems},'' {\em arXiv e-prints},
  p.~arXiv:1904.09846, Apr 2019.

\bibitem{MNZ15}
E.~Musharbash, F.~Nobile, and T.~Zhou, ``Error analysis of the dynamically
  orthogonal approximation of time dependent random pdes,'' {\em SIAM Journal
  on Scientific Computing}, vol.~37, pp.~A776--A810, 2018/01/17 2015.

\bibitem{beck2000multiconfiguration}
M.~H. Beck, A.~J{\"a}ckle, G.~A. Worth, and H.-D. Meyer, ``The
  multiconfiguration time-dependent hartree (mctdh) method: a highly efficient
  algorithm for propagating wavepackets,'' {\em Physics reports}, vol.~324,
  no.~1, pp.~1--105, 2000.

\bibitem{bardos2003mean}
C.~Bardos, F.~Golse, A.~D. Gottlieb, and N.~J. Mauser, ``Mean field dynamics of
  fermions and the time-dependent hartree--fock equation,'' {\em Journal de
  math{\'e}matiques pures et appliqu{\'e}es}, vol.~82, no.~6, pp.~665--683,
  2003.

\bibitem{koch2007dynamical}
O.~Koch and C.~Lubich, ``Dynamical low-rank approximation,'' {\em SIAM Journal
  on Matrix Analysis and Applications}, vol.~29, no.~2, pp.~434--454, 2007.

\bibitem{choi2014equivalence}
M.~Choi, T.~P. Sapsis, and G.~E. Karniadakis, ``On the equivalence of
  dynamically orthogonal and bi-orthogonal methods: Theory and numerical
  simulations,'' {\em Journal of Computational Physics}, vol.~270, pp.~1--20,
  2014.

\bibitem{babaee2017reduced}
H.~Babaee, M.~Farazmand, G.~Haller, and T.~P. Sapsis, ``Reduced-order
  description of transient instabilities and computation of finite-time
  lyapunov exponents,'' {\em Chaos: An Interdisciplinary Journal of Nonlinear
  Science}, vol.~27, no.~6, p.~063103, 2017.

\bibitem{babaee2016minimization}
H.~Babaee and T.~Sapsis, ``A minimization principle for the description of
  modes associated with finite-time instabilities,'' {\em Proceedings of the
  Royal Society A: Mathematical, Physical and Engineering Sciences}, vol.~472,
  no.~2186, p.~20150779, 2016.

\bibitem{babaee2017robust}
H.~Babaee, M.~Choi, T.~P. Sapsis, and G.~E. Karniadakis, ``A robust
  bi-orthogonal/dynamically-orthogonal method using the covariance
  pseudo-inverse with application to stochastic flow problems,'' {\em Journal
  of Computational Physics}, vol.~344, pp.~303--319, 2017.

\bibitem{battles2004extension}
Z.~Battles and L.~N. Trefethen, ``An extension of matlab to continuous
  functions and operators,'' {\em SIAM Journal on Scientific Computing},
  vol.~25, no.~5, pp.~1743--1770, 2004.

\bibitem{MN18}
E.~Musharbash and F.~Nobile, ``Dual dynamically orthogonal approximation of
  incompressible navier stokes equations with random boundary conditions,''
  {\em Journal of Computational Physics}, vol.~354, pp.~135--162, 2018.

\bibitem{choi2013convergence}
M.~Choi, T.~P. Sapsis, and G.~E. Karniadakis, ``A convergence study for {SPDE}s
  using combined polynomial chaos and dynamically-orthogonal schemes,'' {\em
  Journal of Computational Physics}, vol.~245, pp.~281--301, 2013.

\end{thebibliography}
\end{document}